\documentclass[a4paper,11pt]{article}
\pdfoutput=1
\usepackage{subcaption}
\usepackage{jheppub}
\usepackage{hyperref}
\usepackage{mathdots}
\usepackage{amsthm}
\usepackage{amsmath}
\usepackage{amsfonts}
\usepackage{graphicx}
\usepackage{placeins}
\usepackage{tikz}
\usepackage{svg}
\usepackage{subfiles}
\usepackage{float}
\usepackage{extarrows}
\usepackage{cancel}
\usepackage[english]{babel}
\usepackage{multicol,multirow,bigdelim}
\usepackage{graphbox}
\usepackage{CJK}

\usepackage{tabularray}

\usepackage{array}
\usepackage{mathtools}
\def\be{\begin{equation}}
\def\ee{\end{equation}}
\def\ba{\begin{eqnarray}}
\def\ea{\end{eqnarray}}

\makeatletter
\newcommand{\subalign}[1]{
    \vcenter{
    \Let@ \restore@math@cr \default@tag
    \baselineskip\fontdimen10 \scriptfont\tw@
    \advance\baselineskip\fontdimen12 \scriptfont\tw@
    \lineskip\thr@@\fontdimen8 \scriptfont\thr@@
    \lineskiplimit\lineskip
    \ialign{\hfil$\m@th\scriptstyle##$&$\m@th\scriptstyle{}##$\hfil\crcr
    #1\crcr}}
}

\newcommand{\raisemath}[1]{\mathpalette{\raisem@th{#1}}}
\newcommand{\raisem@th}[3]{\raisebox{#1}{$#2#3$}}
\makeatother

\newcommand{\eqtitleref}[1]{\texorpdfstring{\eqref{#1}}{(\protect\ref{#1})}}

\newcommand{\zp}[1]{\texttt{zp[$#1$]}}

\newcommand{\residue}[1]{\mathop{\mathrm{Res}}_{\raisemath{-2pt}{#1}}}

\theoremstyle{definition}
\newtheorem{theorem}{Theorem}[section]
\newtheorem{lemma}[theorem]{Lemma}

\newtheorem{conjecture}[theorem]{Conjecture}

\title{
Supergluon scattering in AdS: constructibility, spinning amplitudes, and new structures} 

\author[a,c]{Qu Cao (曹趣),}\author[a,b,g]{Song He (何颂),}\author[a,d]{Xiang Li (李想),}\author[a,d]{Yichao Tang (唐一朝)}

\affiliation[a]{CAS Key Laboratory of Theoretical Physics, Institute of Theoretical Physics, Chinese Academy of Sciences, Beijing 100190, China}
\affiliation[b]{School of Fundamental Physics and Mathematical Sciences, Hangzhou Institute for Advanced Study \& ICTP-AP, UCAS, Hangzhou 310024, China}
\affiliation[c]{Zhejiang Institute of Modern Physics, Department of Physics, Zhejiang University, Hangzhou, 310027, China}
\affiliation[d]{School of Physical Sciences, University of Chinese Academy of Sciences, No.19A Yuquan Road, Beijing 100049, China}
\affiliation[g]{Peng Huanwu Center for Fundamental Theory, Hefei, Anhui 230026, P. R. China}
\emailAdd{qucao@zju.edu.cn}
\emailAdd{songhe@itp.ac.cn}
\emailAdd{lixiang@itp.ac.cn}
\emailAdd{tangyichao@itp.ac.cn}

\abstract{We elaborate on a new recursive method proposed in~\cite{Cao:2023cwa} for computing tree-level $n$-point supergluon amplitudes as well as those with one gluon, {\it i.e.} spinning amplitudes, in ${\rm AdS}_5 \times S^3$. We present an improved proof for the so-called ``constructibility" of supergluon and spinning amplitudes based on their factorizations and flat-space limit, which allows us to determine these amplitudes in Mellin space to all $n$. We present explicit and remarkably simple expressions for up to $n=7$ supergluon amplitudes and $n=6$ spinning amplitudes, which can be viewed as AdS generalizations of the scalar-scaffolded gluon amplitudes proposed recently. We then reveal a series of hidden structures of these AdS amplitudes including (1). an understanding of general pole structures especially the precise truncation on descendent poles (2). a derivation of simple ``Feynman rules" for the all-$n$ amplitudes with the simplest R-symmetry structures, and (3). certain universal behavior analogous to the soft/collinear limit of flat-space amplitudes. }
                    
\begin{document}
\begin{CJK*}{UTF8}{}
\CJKfamily{gbsn}
\maketitle
\end{CJK*}
\section{Introduction}
Recent years have witnessed remarkable progress in computing and revealing new structures of holographic correlators, or ``scattering amplitudes'' in AdS space in supersymmetric models, at both tree~\cite{Rastelli:2016nze, Rastelli:2017udc,Rastelli:2017ymc,Zhou:2017zaw,Goncalves:2019znr,Alday:2020lbp,Alday:2020dtb,Zhou:2021gnu,Goncalves:2023oyx} and loop~\cite{Alday:2017xua,Aprile:2017bgs,Aprile:2017qoy,Aprile:2019rep,Alday:2019nin,Huang:2021xws,Drummond:2022dxw} level.
Although more focus has been on supergravity amplitudes in AdS, explicit results have also been obtained for ``supergluon" tree amplitudes up to $n=6$~\cite{Zhou:2018ofp,Alday:2021odx,Alday:2022lkk,Bissi:2022wuh,Alday:2023kfm} in AdS super-Yang-Mills (sYM) theories (see~\cite{Alday:2021ajh,Huang:2023oxf,Huang:2023ppy} for loop-level results). Very recently in~\cite{Cao:2023cwa}, we initiated a new method for constructing higher-point supergluon amplitudes purely from lower-point ones, and along the way we have started revealing new structures for these amplitudes. 

The natural language for holographic correlators is the Mellin representation~\cite{Mack:2009mi,Penedones:2010ue,Fitzpatrick:2011ia}. Mellin tree amplitudes are rational functions of Mellin variables. They can be determined by the residues at all physical poles (and pole at infinity encoded in the flat-space limit), which are given by factorization/OPE with scalar and gluon exchanges~\cite{Goncalves:2014rfa}. Since spinning amplitudes needed for gluon factorization are more difficult to obtain than scalar amplitudes, in~\cite{Cao:2023cwa} we have proposed to exploit scalar factorization as much as possible. It was very nice to see how powerful these constraints are: purely from scalar factorizations and only imposing ``gauge invariance" of spinning amplitudes, together with flat-space limit, we fully determine
supergluon amplitudes to all multiplicities with no other input. We will elaborate on our method and discuss some higher-point results in this paper. 

The key idea for organizing such amplitudes, which also becomes a source for simplifications, comes from a natural R-symmetry basis built from $SU(2)_R$ traces compatible with color ordering, and we decompose any color-ordered amplitude into partial amplitudes with up to $\lfloor \frac{n}{2} \rfloor$ traces. Since gluon exchanges cannot contribute in any channel incompatible with the trace structure, these channels are determined by scalar factorization. For compatible channels where gluon exchanges are generically allowed, at special no-gluon kinematics closely related to ``gauge invariance'', gluon exchanges are forbidden and only scalar exchanges survive.  As we have shown in~\cite{Cao:2023cwa}, such constraints suffice for single-, double- and triple-trace partial amplitudes to all $n$ (thus for all $n\leq 7$ amplitudes). Moreover, as we have outlined in~\cite{Cao:2023cwa}, by bootstrapping the $n=5$ single-gluon amplitude using the same method allows us to fix the remaining quadruple-trace partial amplitudes at $n=8$. We will review all these basic ingredients in sec.~\ref{sec:review}, and we will then push the frontier significantly: not only will we present an improved proof for the ``constructibility" to all $n$, but we will also unveil various new structures regarding these supergluon and spinning amplitudes with a single gluon. 

Another important motivation for us is to set up an ideal playground for studying such scattering amplitudes in $AdS_5 \times S^3$; since we have focused on tree-level supergluon and spinning amplitudes, these are simply rational functions in Mellin variables. The {\it planar variables} automatically solve all linear constraints, thus for each independent R-symmetry structure, we have a rational function of $n(n{-}3)/2$ linearly independent variables ${\cal X}_{i,j}$ for $i<j{-}1$. Note that all the locations of the poles for any color-ordered amplitude are simply given by planar variables at integer values, ${\cal X}_{i,j}=0, 2, 4, \cdots$. Moreover, one of the most basic observations we have is the {\it truncation} of descendent poles, which can be summarized as ``half-circle rule": we will see that it is only possible to have poles at ${\cal X}_{1,3}=0$, ${\cal X}_{1,4}=0,2$ for $n\geq 6$, and ${\cal X}_{1,5}=0,2$ for $n\geq8$, {\it etc.}, which is a prior not obvious at all! Our investigations will also indicate that both supergluon and spinning amplitudes can be computed by certain simple {\it ``Feynman rules"} in AdS space, and they exhibit some nice behaviors similar to collinear/soft limits of flat-space amplitudes. In addition, these amplitudes can be viewed as AdS generalizations of Yang-Mills-scalar amplitudes~\cite{Cachazo:2014xea} and the so-called scalar-scaffolded gluon amplitudes recently proposed in~\cite{Arkani-Hamed:2023jry} based on curve integrals~\cite{Arkani-Hamed:2023lbd, Arkani-Hamed:2023mvg} and positive geometries~\cite{Arkani-Hamed:2017mur, Arkani-Hamed:2017tmz}. We believe that we have only scratched the surface of numerous beautiful structures underlying these AdS amplitudes. 

Let us summarize the main results of this paper, or the organization of the paper after the review in~sec.~\ref{sec:review}.
\begin{itemize}
    \item  In sec.~\ref{sec:construct} We give an improved recursive algorithm for computing supergluon amplitudes to all multiplicities, which gives a new proof for the ``constructibility'' as outlined in~\cite{Cao:2023cwa}. Our algorithm starts with extracting $(n{-}1)$-point spinning amplitudes (with $n{-}2$ supergluon and one gluon) from the $n$-point supergluon amplitude, and remarkably we can use such information of lower-point amplitudes to determine not only the $(n{+}1)$-point,  but actually even the $(n{+}2)$-point supergluon amplitude! We provide a detailed example for how this works for the $n=6$ case.
    
    \item As a byproduct, our method also amounts to a recursive algorithm for computing the spinning amplitudes to all multiplicities. Hence, we present explicit compact formulas for $n\leq7$ supergluon amplitude and $n\leq6$ spinning amplitudes in sec.~\ref{sec:result}, which demonstrate the power of our new method and provide valuable theoretical data (we will also present illuminating examples for their flat-space limits). 
    
    \item In sec.~\ref{sec:structure} we study various new structures of this data, which include:
    \begin{itemize}
        \item A conjecture about the most general pole structures of supergluon/spinning amplitudes, especially the truncation for descendant poles mentioned above.
        \item Interesting Feynman rules for the simplest R-symmetry structures: the single-trace supergluon amplitude can be simply described by ``Feynman rules'' of the form $\phi^3+ \phi^4$ and similarly for the spinning case with an additional gluon-scalar-scalar vertex.
        \item Some interesting universal behavior, which can be viewed as certain generalization of soft/collinear behavior of flat-space scattering amplitudes.
    \end{itemize}  
\end{itemize}

Finally we conclude with some immediate next steps in sec.~\ref{sec:outlook}, and we collect some useful material in the appendices~\ref{sec:Witten-st-Coefficients} and~\ref{sec:rule_derivation}.

 \section{Symmetries, poles and factorizations for Mellin amplitudes}\label{sec:review}
In this section, we will provide an overview of the fundamental background. The most famous example of the ${\rm AdS}/{\rm CFT}$ duality establishes a correspondence between the type IIB superstring living in the ${\rm AdS}_{5}\times \rm{S}^{5}$ bulk spacetime and the $\mathcal{N}=4$ supersymmetric Yang-Mills (sYM) conformal field theory on the boundary. In bulk spacetime, the low-energy limit of the superstring theory yields the supergravity field theory. However, calculating the tree-level amplitudes in the supergravity theory is challenging. As an alternative, there exists a related theory, the super Yang-Mills theory, which resides in the bulk spacetime.

This theory arises as the low energy description of many different theories~\cite{Fayyazuddin:1998fb,Aharony:1998xz,Karch:2002sh,Alday:2021odx}. One way to obtain the theory is through the low-energy description of D3-D7-brane system in Type IIB string theory in the probe limit (number $N_f$ of D7-branes much less than number $N_c$ of D3-branes)~\cite{Karch:2002sh}. On the worldvolume of D3-branes, we have an $\mathcal N=2$ SCFT, while on the worldvolume of D7-branes, gravity decouples at tree level and we have $\mathcal N=1$ sYM on ${\rm AdS}_5\times S^3$~\footnote{The gravitational coupling is proportional to $1/N_c$, which is much smaller than the (super)gluon self coupling proportional to $1/\sqrt{N_c}$. Hence, gravity decouples at tree level, i.e., leading $1/N_c$ order.}. The system has a symmetry $G_F={SU}(N_f)$~\footnote{Other constructions such as those arising in F-theories~\cite{Fayyazuddin:1998fb,Aharony:1998xz} lead to different $G_F$, but otherwise the effective descriptions are the same.} which is global on the boundary and local in the bulk. The dictionary reads:

\begin{center}
    \begin{tabular}{rc|cl}
        \hline\hline
        \multicolumn{2}{c|}{\textbf{Boundary CFT$_{\boldsymbol4}$}} & \multicolumn{2}{|c}{\textbf{Bulk AdS$_{\boldsymbol5}\boldsymbol{\times S^3}$}} \\
        \hline
        \multicolumn{2}{c|}{R-symmetry $SU(2)_R\times SU(2)_L$} & \multicolumn{2}{|c}{isometry $SO(4)$ of $S^3$} \\
        \multicolumn{2}{c|}{global symmetry $G_F$} & \multicolumn{2}{|c}{gauge symmetry $G_F$} \\
        \ldelim\{{2}{*}[$\substack{\displaystyle\text{half-BPS}\\[0.5ex]\displaystyle\text{supermultiplet}}$\ ]\hspace{-0.5ex} & Noether current $\mathcal J_\mu^a$ & ${\rm AdS}_5$ YM field $A_{\bar\mu}^a$ & \rdelim\}{2}{*}[\ $\substack{\displaystyle\text{lowest KK of}\\[0.5ex]\displaystyle\text{8d YM field}}$] \\
        & scalar superprimary $\mathcal O^a_{\alpha\beta}$ & ${\rm AdS}_5$ scalar field $\phi^a_m$ \\
        \hline\hline
    \end{tabular}
\end{center}


We are interested in the $n$-point ``supergluon'' amplitudes, i.e., the connected correlator of $n$ half-BPS operators $\mathcal O^a(x,v)$ with conformal dimension $\Delta=2$:
\begin{gather}
    G_n^{(s)a_1\cdots a_n}=\langle\mathcal O^{a_1}(x_1,v_1)\cdots\mathcal O^{a_n}(x_n,v_n)\rangle,\quad
    \mathcal O^a(x,v)=\mathcal O^{a;\alpha_1\alpha_2}(x)v^{\beta_1}v^{\beta_2}\epsilon_{\alpha_1\beta_1}\epsilon_{\alpha_2\beta_2}.
\end{gather}
Here, $a_i=1,\cdots,\dim G_F$ are adjoint indices of $G_F$,
and $v^\beta$ ($\alpha_i,\beta_i=1,2$) are auxiliary $SU(2)_R$-spinors which extracts the R-spin-1 part of $\mathcal O^{a;\alpha_1\alpha_2}(x)$. By antisymmetry, the auxiliary spinors are null: $v^{2}:=v^{\beta_1}v^{\beta_2}\epsilon_{\beta_1\beta_2}=0$. The superscript $^{(s)}$ denotes that $G_n^{(s)}$ is a correlator of scalar operators.

For convenience, we also introduce the single-gluon correlators $G_{n,\mu}^{(v)}$ involving a Noether current $\mathcal J_\mu^a(x)$ of $G_F$, an $SU(2)_R$-singlet with conformal dimension $\Delta=3$:
\begin{equation}
    G_{n,\mu}^{(v)a_1\cdots a_n}=\langle\mathcal O^{a_1}(x_1)\cdots\mathcal O^{a_{n-1}}(x_{n-1})\mathcal J_\mu^{a_n}(x_n)\rangle\,,
\end{equation}
where the superscript $^{(v)}$ denotes that $G_n^{(v)}$ is a correlator consisting of a vector operator.

Since the bulk dual of $\mathcal J_\mu^a$ is the gauge field $A_\mu^a$ (``gluon''), its superprimary partner $\mathcal O^a$/$\phi_m^a$ ($m=1,2,3$) is dubbed the ``supergluon''. Together, they compose the lowest Kaluza-Klein (KK) mode of the $G_F$ gauge field on ${\rm AdS}_5\times S^3$. 

The operators $\mathcal O^a$ and $\mathcal J_\mu^a$ are all we need to compute $G_n^{(s)}$ at the tree level. This can be seen as follows. Within the supermultiplet containing $\mathcal O^a$ and $\mathcal J_\mu^a$, all other primaries are charged under $U(1)_r$, the Abelian part of the $\mathcal N=2$ R-symmetry. Other half-BPS supermultiplets are dual to higher KK modes, and they are charged under ${SU}(2)_L$ which is part of the isometry group ${SO}(4)={SU}(2)_L\times{SU}(4)_R$ of $S^3$. The operators $\mathcal O^a$ and $\mathcal J_\mu^a$ are special in that they are neutral under $U(1)_r$ and $SU(2)_L$. Hence, all other fields can only appear in pairs and contribute at loop level.

Now, we will use three tools to simplify the supergluon amplitudes in AdS: the color decomposition, the Mellin representation, and the R-structure decomposition. Then we will describe constraints on the color-ordered Mellin amplitudes, such as the flat-space limit and factorization.

\paragraph{The color decomposition.}The color decomposition for tree amplitudes in AdS space is identical to that for flat-space amplitudes~\cite{DelDuca:1999rs}: we have color-ordered amplitudes as coefficients in front of traces of generators $T^a$ in the adjoint representation:
\begin{equation}
    G_n^{a_1\cdots a_n}=\sum_{\mathclap{\sigma\in S_{n-1}}}{\rm tr}(T^{a_1}T^{a_2^{\sigma}}\cdots T^{a_{n-1}^{\sigma}}T^{a_n^{\sigma}})G_{1\sigma}\,,
\end{equation}
where ${\sigma}$ denotes a permutation of $\{2,\cdots, n\}$. Cyclic and reflection symmetry of the traces implies
\begin{equation}\label{eq:dihe}
    G_{12\cdots n}=G_{2\cdots n1}=(-)^nG_{n\cdots21}\,.
\end{equation}
We will focus on $G_{12\cdots n}$ since other color-ordered amplitude can then be obtained by relabeling.

\paragraph{The Mellin representation.}The natural language to describe holographic CFT correlators is the Mellin representation~\cite{Mack:2009mi}. For scalar amplitudes,
\begin{equation}
    G_{12\cdots n}^{(s)}=\int[{\rm d}\delta]\mathcal M_n^{(s)}(\{\delta_{ij}\},\{v_i\})\prod_{i<j}\frac{\Gamma(\delta_{ij})}{(-2P_i\cdot P_j)^{\delta_{ij}}},
\end{equation}
and for single-gluon amplitudes~\cite{Goncalves:2014rfa}:
\begin{gather}
    G_{12\cdots n}^{(v)}=\int[{\rm d}\delta]\sum_{\ell=1}^{n-1}(Z_n\cdot P_\ell)\mathcal M_n^{(v)\ell}\prod_{i<j}\frac{\Gamma(\delta_{ij}+\delta_i^\ell\delta_j^n)}{(-2P_i\cdot P_j)^{\delta_{ij}+\delta_i^\ell\delta_j^n}},\\
    \text{where }\sum_{\ell=1}^{n-1}\delta_{\ell n}\mathcal M_n^{(v)\ell}=0.\label{eq:gaugeinv}
\end{gather}
Here $\delta_{i}^{l}$ is the Kronecker delta, not to be confused with the Mellin variables $\delta_{ij}$. We have used the embedding formalism following~\cite{Goncalves:2014rfa} to put $G_{n,\mu}^{(v)a_1\cdots a_n}$ into embedding space  contract with an auxliary polarization vector to obtain $G_{n}^{(v)a_1\cdots a_n}=Z_n^MG_{n,M}^{(v)a_1\cdots a_n}$. The embedding vectors $P_i^M$ encode the positions $x_i^\mu$, which satisfies $(x_i-x_j)^2=-2P_i\cdot P_j:=P_{ij}$. The Mellin variables are constrained as if $\delta_{ij}=p_i\cdot p_j$ for auxiliary momenta satisfying $\sum_ip_i=0$ and $p_i^2=-\tau_i=-2$, with conformal twist $\tau_i:=\Delta_i-J_i$ ($J$ is the spin of an operator). Since $\mathcal J$ and $\mathcal O$ have the same twist, they are described by the same ``kinematics''.

For later convenience, we will pack the information $\{\mathcal M_n^{(v)\ell}\}_{\ell=1}^{n-1}$ into
\begin{equation}
    \mathcal M_n^{(v)}=\sum_{\ell=1}^{n-1}\mathcal M_n^{(v)\ell}\zp{\ell},\quad\text{where }\zp{\ell}:=\frac{(Z_n\cdot P_\ell)}{(-2P_n\cdot P_\ell)}\delta_{\ell n}.
\end{equation}
This way, we can write the single-gluon Mellin amplitude in a form similar to the all-scalar case, which makes it clear that requiring $G^{(v)}_{12\cdots n}(Z_n)=G^{(v)}_{12\cdots n}(Z_n+\alpha P_n)$ implies~\eqref{eq:gaugeinv}:
\begin{equation}
    G_{12\cdots n}^{(v)}=\int[{\rm d}\delta]\mathcal M_n^{(v)}\prod_{i<j}\frac{\Gamma(\delta_{ij})}{(-2P_i\cdot P_j)^{\delta_{ij}}}.
\end{equation}
Let us also introduce the compact notation
\begin{equation}
    \zp{ij\cdots k}:=\zp{i}+\zp{j}+\cdots+\zp{k}.
\end{equation}
At this stage, the definition of $\zp{\cdots}$ seems completely whimsical. However, we will later find that $\zp{i}$ behaves similarly to $\epsilon_n\cdot p_i$ in flat-space.

\begin{figure}[ht]
    \vspace{-1em}
    \centering
    \includegraphics[scale=0.7]{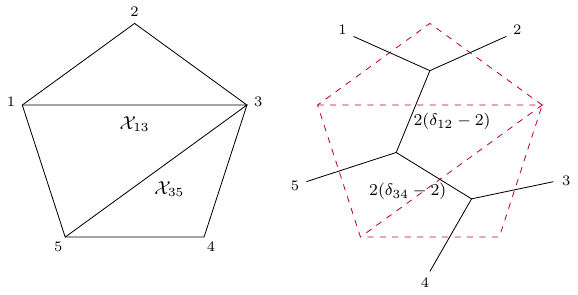}
    \vspace{-1em}
    \caption{Planar variables and dual skeleton graph for $n=5$.}
    \label{fig:n-gonXij}
\end{figure}

Only $\frac12n(n-3)$ of the $\delta_{ij}$'s are independent. Inspired by flat space~\cite{Arkani-Hamed:2017mur}, it proves convenient to introduce $\frac12n(n-3)$ planar variables induced by the color ordering (with $\delta_{ii}\equiv-2$)\footnote{Here, we define ${\cal X}_{ij}$ differently from \cite{Cao:2023cwa} by a sign: $\mathcal X_{ij}^{\text{here}}=-\mathcal X_{ij}^{\text{there}}$.},
\begin{equation}
{\cal X}_{ij}:=-2-\sum_{i\leq k, l<j} \delta_{kl}=-2-\Bigg(\sum_{i\leq k<j}p_k\Bigg)^2,
\end{equation}
where we have ${\cal X}_{i,j}={\cal X}_{j,i}$ with special cases ${\cal X}_{i,i{+}1}=0$ and ${\cal X}_{i,i}\equiv -2$. The inverse transform which motivated the associahedron in~\cite{Arkani-Hamed:2017mur, Arkani-Hamed:2019vag} reads:
\begin{equation}
    2\delta_{ij}=\mathcal X_{i,j}+\mathcal X_{i+1,j+1}-\mathcal X_{i,j+1}-\mathcal X_{i+1,j}.
\end{equation}
Planar variables correspond to $n$-gon chords (Figure~\ref{fig:n-gonXij}).

The planar variables are particularly suited for factorization~\cite{Goncalves:2014rfa} of color-ordered amplitudes. Schematically,
\begin{equation}\label{eq:facschem}
    \mathcal M_{12\cdots n}\sim\frac{\mathcal M_{1\cdots(k-1)I}^{(m)}\mathcal M_{k\cdots nI}^{(m)}}{\mathcal X_{1k}-2m},\quad m=0,1,2,\cdots
\end{equation}
where a pole at $\mathcal X_{1k}=2m$ corresponds to the exchange of a level-$m$ descendant of $\mathcal O$ or $\mathcal J$. But we will see in section~\ref{sec:truncation}, the $m$ can not be the infinity in our story, which indicates a truncation.
By induction, all simultaneous poles of $\mathcal M_n$ consist of \emph{compatible} planar variables (non-intersecting chords), which gives a partial triangulation of the $n$-gon dual to  planar skeleton graphs (Figure~\ref{fig:n-gonXij}).

\begin{figure}[ht]
    \centering
    \includegraphics[scale=0.65]{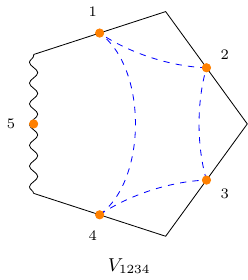}\hspace{1ex}\includegraphics[scale=0.65]{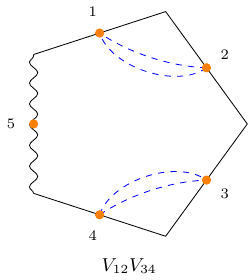}\hspace{1ex}\includegraphics[scale=0.65]{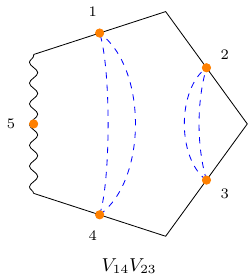}
    \caption{$\mathcal M_5^{(v)}$ R-structures.}
    \label{fig:gluonR}
\end{figure}

\paragraph{The R-structure decomposition.}Another advantage of working with color-ordered amplitude is a natural basis for the R-charge structures. Let us define a $SU(2)_R$ trace as $V_{i_1i_2\cdots i_r}:=\langle i_1i_2\rangle\langle i_2i_3\rangle\cdots\langle i_ri_1\rangle$, where $\langle ij\rangle:=v_i^\alpha v_j^\beta\epsilon_{\alpha\beta}\equiv v_{i}\cdot v_{j}$. The Schouten identity $\langle ik\rangle\langle jl\rangle=\langle ij\rangle\langle kl\rangle+\langle il\rangle\langle jk\rangle$ enables us to expand any R-structure to products of non-crossing cycles or $SU(2)_R$ traces:
\begin{align}
{\cal M}_n^{(s)}&=\sum_{\substack{\text{non-crossing}\\\text{partition }\pi\\\text{of }\{1,\cdots,n\}}}\Bigg(\prod_{\text{cycle }\tau\,\in\,\pi}V_\tau\Bigg)M^{(s)}_n(\pi),\\
{\cal M}_n^{(v)\ell}&=i \sqrt{2/3}\sum_{\substack{\text{non-crossing}\\\text{partition }\pi\\\text{of }\{1,\cdots,n{-}1\}}}\Bigg(\prod_{\text{cycle }\tau\,\in\,\pi}V_\tau\Bigg)M^{(v)\ell}_n(\pi).
\end{align}
For example, (Figure~\ref{fig:gluonR})
\begin{align*}
    \mathcal M_4^{(s)}&=M_4^{(s)}(1234)V_{1234}\nonumber\\
    &+M_4^{(s)}(12;34)V_{12}V_{34}+M_4^{(s)}(14;23)V_{14}V_{23},\\
    \mathcal M_4^{(v)\ell}&=i \sqrt{2/3} M_4^{(v)\ell}(123)V_{123},\\
    \mathcal M_5^{(s)}&=M_5^{(s)}(12345)V_{12345}\nonumber\\
    &+M_5^{(s)}(12;345)V_{12}V_{345}+\text{cyclic},\\
    \mathcal M_5^{(v)\ell}&=i \sqrt{2/3} \left(M_5^{(v)\ell}(1234)V_{1234}\right. \nonumber\\
    &+\left.M_5^{(v)\ell}(12;34)V_{12}V_{34}+M_5^{(v)\ell}(14;23)V_{14}V_{23}\right).
\end{align*}
Because a length-$L$ trace picks up $(-)^L$ under reflection, for scalar amplitudes this cancels the sign in~\eqref{eq:dihe} while for single-gluon amplitudes the net result is a minus sign:
\begin{align*}
    M_4^{(s)}(12;34)&\xlongequal{\text{ref}}M_4^{(s)}(21;43)\xlongequal{\text{cyc}}M_4^{(s)}(14;23),\\
    M_5^{(v)}(12;34)&\xlongequal{\text{ref}}-M_5^{(v)}(21;43),\\
    M_5^{(v)}(12;34)&\text{ unrelated to }M_5^{(v)}(14;23).
\end{align*}

\paragraph{The flat-space limit.}In~\cite{Alday:2021odx}, the authors showed that with $\delta_{ij}=R^2s_{ij}$, the leading terms of $\mathcal M_4^{(s)}$ in the limit $R\to\infty$ is closely related to the color-ordered 4-gluon amplitude. Specifically, setting $\epsilon_i\cdot p_j=0$ and $\epsilon_i\cdot\epsilon_j=\langle ij\rangle^2=-V_{ij}$,
\begin{equation}
    \mathcal A_4^\text{flat gluon}\leadsto V_{12}V_{34}\frac{s_{12}+s_{23}}{s_{12}}+(1\leftrightarrow3)-V_{13}V_{24}.
\end{equation}
Using $V_{13}V_{24}=V_{12}V_{34}+V_{14}V_{23}-2V_{1234}$, this matches the leading terms of $\mathcal M_4^{(s)}$ (up to overall normalization):
\begin{equation}\label{eq:m4}
    \begin{aligned}
        \mathcal M_4^{(s)}&=-2 \left(\frac1{\mathcal X_{13}}+\frac1{\mathcal X_{24}}+1\right)V_{1234}-\frac{\mathcal X_{24}-2}{\mathcal X_{13}}V_{12}V_{34}-\frac{\mathcal X_{13}-2}{\mathcal X_{24}}V_{14}V_{23}\\
        &=-2V_{1234}-\frac{\mathcal X_{24}}{\mathcal X_{13}}V_{12}V_{34}-\frac{\mathcal X_{13}}{\mathcal X_{24}}+O(R^{-2}).
    \end{aligned}
\end{equation}
Here, we argue that this is a general property which holds for any multiplicity, justifying the manipulations in~\cite{Alday:2023kfm}. Recall that, from the bulk perspective, we have a $G_F$ gauge theory with adjoint scalars. The scalars are additionally charged under $SU(2)_R$, which can be viewed as a flavor symmetry. Importantly, at tree level, the R-symmetry current (which lives in the stress-tensor supermultiplet) decouples, so the flavored scalars must form R-neutral pairs before interacting with the $G_F$ gauge field. It is well-known~\cite{Cachazo:2014xea} that, in flat space, flavor-stripped amplitudes of paired scalars in Yang-Mills-flavored-scalar theory can be extracted from pure gluon amplitudes by considering the coefficient of $\epsilon_i\cdot\epsilon_j$. The flavor-dressed amplitudes are then obtained by dressing each pair with the R-neutral factor $\langle ij\rangle^2$. The net result is that we are setting $\epsilon_i\cdot p_j=0$ and $\epsilon_i\cdot\epsilon_j=\langle ij\rangle^2$ in the pure gluon amplitudes. The result of paired scalar amplitudes have been computed using the CHY formula~\cite{Cachazo:2013iea} explicitly through $n=12$.

The flat-space limit constrains the $R\to\infty$ power-counting of $\mathcal M_n^{(s)}$. For even $n$, everything is clear, and $\mathcal M_n^{(s)}\sim\delta^{2-\frac n2}$. For odd $n$, the flat space amplitude vanishes due to the prescription $\epsilon_i\cdot p_j=0$. The power counting $s^{2-\frac n2}$ means that the order $\delta^{2-\lfloor\frac n2\rfloor}$ vanishes, and $\mathcal M_n^{(s)}\sim\delta^{2-\lceil\frac n2\rceil}$.

\subsection{Factorization of the Mellin amplitudes}

\begin{figure}[ht]
    \centering
    \includegraphics[scale=0.65]{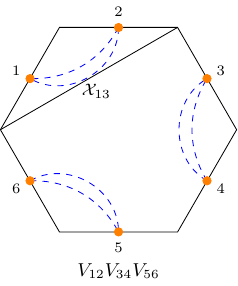}\hspace{1em}\includegraphics[scale=0.65]{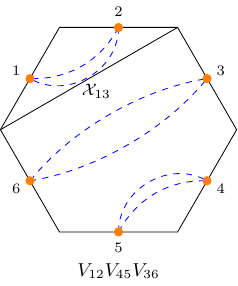}
    
    \includegraphics[scale=0.65]{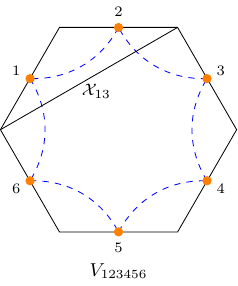}\hspace{1em}\includegraphics[scale=0.65]{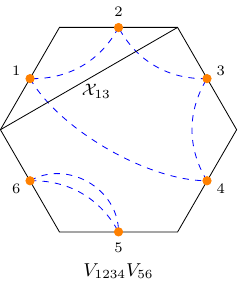}\hspace{1em}\includegraphics[scale=0.65]{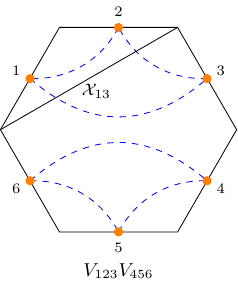}
    \caption{$\mathcal M_6^{(s)}$ R-structures compatible (above) and incompatible (below) with $\mathcal X_{13}$.}
    \label{fig:compatR}
\end{figure}

\begin{figure}[H]
    \centering
    \includegraphics[scale=0.65]{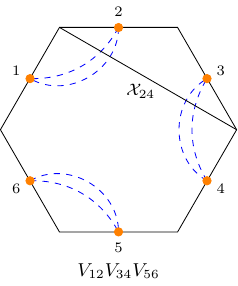}\hspace{1em}\includegraphics[scale=0.65]{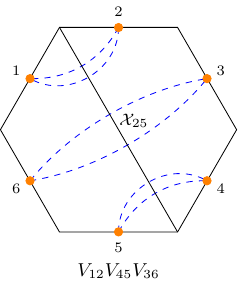}
    \caption{Vanishing R-structures.}
    \label{fig:vanishR}
\end{figure}

Different exchanged fields contribute to different R-structures. For a given channel, say $\mathcal X_{1k}$, we distinguish the compatible R-structures (none of the cycles intersect $\mathcal X_{1k}$) from the incompatible ones (Figure~\ref{fig:compatR}). For scalar exchanges, \eqref{eq:facschem} reads
\begin{equation}\label{eq:scalar-fac}
    \mathop{\rm Res}_{\mathcal X_{1k}=2m}^{(s)}\mathcal M_n^{(s)}=\mathcal N_s^{(m)}\texttt{glueR}\left(\mathcal M_{1\cdots(k-1)I}^{(s)(m)}\mathcal M_{k\cdots nI}^{(s)(m)}\right).
\end{equation}
Here, $\mathcal N_s^{(m)}=-2$, and $\mathcal M_{1\cdots(k-1)I}^{(s)(m)}$ is a shifted version of the scalar amplitude $\mathcal M_{1\cdots(k-1)I}^{(s)}$:
\begin{equation}\label{eq:shift}
    \mathcal M_{1\cdots(k-1)I}^{(s)(m)}=\sum_{\substack{n_{ab}\geq0\\\sum n_{ab}=m}}\mathcal M_{1\cdots(k-1)I}^{(s)}(\delta_{ab}+n_{ab})\prod_{1\leq a<b<k}\frac{(\delta_{ab})_{n_{ab}}}{n_{ab}!}.
\end{equation}
$\mathcal M_{k\cdots nI}^{(s)(m)}$ is defined similarly.
The operation $\texttt{glueR}$ glues together the traces. Note that there is the 1-1 correspondence of R-structures in amplitudes and OPE:
\begin{gather*}
    \langle\mathcal O(v_I)\mathcal O\cdots\mathcal O\rangle\supset\text{something}\times V_{ia\cdots bjI}\\
    \Updownarrow\\
    \mathcal O\cdots\mathcal O\supset\text{something}\times\langle ia\rangle\cdots\langle bj\rangle v_i^{(\alpha}v_j^{\beta)}\mathcal O_{\alpha\beta}
\end{gather*}
Since $\langle\mathcal O_{\alpha\beta}\mathcal O_{\gamma\delta}\rangle=\frac12(\epsilon_{\alpha\gamma}\epsilon_{\beta\delta}+\epsilon_{\alpha\delta}\epsilon_{\beta\gamma})$, we have
\begin{align}
v_i^{(\alpha}v_j^{\beta)}v_k^{(\gamma}v_l^{\delta)}\langle\mathcal O_{\alpha\beta}\mathcal O_{\gamma\delta}\rangle
    =\langle il\rangle\langle jk\rangle-\frac12\langle ij\rangle\langle lk\rangle.
\end{align}
which implies the following gluing rule:
\begin{equation}\label{eq:glueR}
    \texttt{glueR}:\ V_{i\cdots jI}\otimes V_{Ik\cdots l}\mapsto V_{i\cdots jk\cdots l}-\frac12V_{i\cdots j}V_{k\cdots l}.
\end{equation}
We see that scalar exchanges contribute to both compatible and incompatible R-structures. R-structures with more than one cycle intersecting $\mathcal X_{1k}$ vanish (Figure~\ref{fig:vanishR}).

For gluon exchanges, \eqref{eq:facschem} reads
\begin{equation}\label{eq:gluonfac}
    \mathop{\rm Res}_{\mathcal X_{1k}=2m}^{(v)}\mathcal M_n^{(s)}=\mathcal N_v^{(m)}\sum_{a=1}^{k-1}\sum_{i=k}^n\delta_{ai}\mathcal M_{1\cdots(k-1)I}^{(v)(m)a}\mathcal M_{k\cdots nI}^{(v)(m)i}.
\end{equation}
Here, $\mathcal N_v^{(m)}=\frac3{1+m}$, and $\mathcal M_{1\cdots(k-1)I}^{(v)(m)a}$ is $\mathcal M_{1\cdots(k-1)I}^{(v)a}$ shifted according to \eqref{eq:shift}. We no longer need $\texttt{glueR}$ because $\mathcal J$ is R-neutral; gluon exchanges contribute to compatible R-structures only.

An important consequence of ``gauge invariance'' \eqref{eq:gaugeinv} is that, at certain \emph{no-gluon kinematics}, gluon exchanges are forbidden completely. To see this, let us denote $\mathcal M_{1\cdots(k-1)I}^{(v)(m)a}\equiv\mathcal L^{(m)a}$ and $\mathcal M_{k\cdots nI}^{(v)(m)i}\equiv\mathcal R^{(m)i}$, and solve $\mathcal L^{(m)1},\mathcal R^{(m)k}$ using \eqref{eq:gaugeinv}. The double sum in \eqref{eq:gluonfac} becomes
\begin{equation*}
    \sum_{a=2}^{k-1}\sum_{i=k+1}^{n}\left(\delta_{ai}-\frac{\delta_{aI}}{\delta_{1I}}\delta_{1i}-\frac{\delta_{iI}}{\delta_{kI}}\delta_{ak}+\frac{\delta_{aI}\delta_{iI}}{\delta_{1I}\delta_{kI}}\delta_{1k}\right)\mathcal L^{(m)a} \mathcal R^{(m)i}.
\end{equation*}
If all $(k-2)(n-k)$ coefficients vanish on the support of $\mathcal X_{1k}=2m$, gluon exchanges are forbidden, regardless of the detailed form of $\mathcal L^{(m)}$ and $\mathcal R^{(m)}$. The number of conditions equals the number of chords $\mathcal X_{ai}$ ($2\leq a\leq k-1$ and $k+1\leq i\leq n$) crossing $\mathcal X_{1k}$. Hence, the no-gluon conditions translate to $\mathcal X_{ai}$ taking special values $\mathcal X_{ai}^*$:
\begin{align}\label{eq:no-gluon kin}
    \mathcal E_{ai}^{(m)}&:=\mathcal X_{ai}-\mathcal X_{ai}^*=0,\\
    X_{ai}^*&=1-m+\frac{\mathcal X_{1a}+\mathcal X_{1i}+\mathcal X_{ak}+\mathcal X_{ik}}2-\frac{(\mathcal X_{1a}-\mathcal X_{ak})(\mathcal X_{1i}-\mathcal X_{ik})}{4(m+1)}.
\end{align}
Since gluon exchanges are forbidden at no-gluon kinematics, scalar exchanges alone fix the residue up to polynomials of $\mathcal E$'s:
\begin{equation}\label{eq:NGfix}
    \mathop{\rm Res}_{\mathcal X_{1k}=2m}\mathcal M_n =\left.\mathop{\rm Res}_{\mathcal X_{1k}=2m}^{(s)}\mathcal M_n\right|_{\mathcal X_{ai}=\mathcal X_{ai}^*}+\text{poly}(\mathcal E_{ai}^{(m)}).
\end{equation}

The special case of \eqref{eq:gluonfac} where $k=n-1$ is particularly important. From the 3-point single-gluon amplitude~\footnote{We can easily compute it as follows. Firstly, the R-structure is unique. Secondly, there are no free Mellin variables, so the amplitude must be a constant. Thirdly, \eqref{eq:gaugeinv} fixes the relative sign to be $-1$. Lastly, we can fix the overall normalization by comparing to the gluon-exchange contribution to $\mathcal M_4^{(s)}$.}:
\begin{equation}
    \mathcal M_{n-1,n,I}^{(v)(0)n-1}=\frac i{\sqrt6}V_{n-1,n},\ \mathcal M_{n-1,n,I}^{(v)(0)n}=-\frac i{\sqrt6}V_{n-1,n},
\end{equation}
we see that
\begin{equation}
    \mathop{\rm Res}_{\mathcal X_{1,n-1}=0}^{(v)}\mathcal M_n^{(s)}=\frac{3i}{\sqrt6}V_{n-1,n}\sum_{a=1}^{n-2}(\delta_{a,n-1}-\delta_{a,n})\mathcal M_{n-1}^{(v)a}.
\end{equation}
This is similar to the scaffolding relation in~\cite{Arkani-Hamed:2023jry}. If we write the $\delta$'s in terms of $\mathcal X$'s, one can show that for each $2\leq a\leq n-2$,
\begin{equation}\label{eq:extract}
    \mathcal M_{n-1}^{(v)a}-\mathcal M_{n-1}^{(v)a-1}=\frac\partial{\partial\mathcal X_{an}}\left(\frac{i\sqrt{2/3}}{V_{n-1,n}}\mathop{\rm Res}_{\mathcal X_{1,n-1}=0}^{(v)}\mathcal M_n^{(s)}\right).
\end{equation}
Together with \eqref{eq:gaugeinv}, these $(n-3)+1$ equations completely determine $\{\mathcal M_{n-1}^{(v)a}\}_{a=1}^{n-2}$. In other words, $(n-1)$-point single-gluon amplitudes can be extracted from the $n$-point scalar amplitude!

 \section{Constructibility: proof and examples}\label{sec:construct}

In~\cite{Cao:2023cwa}, we showed that supergluon amplitudes of any multiplicity can be recursively constructed by alternatingly extracting single-gluon amplitudes and bootstrapping supergluon amplitudes:
\begin{equation}\label{eq:zigzagold}
    \includegraphics[align=c]{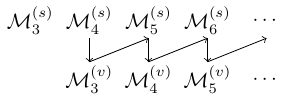}
\end{equation}
However, we observed in practice that we only need the explicit result of $\mathcal M_5^{(v)}$ instead of $\mathcal M_6^{(v)}$ to construct $\mathcal M_8^{(s)}$. In other words, the above zigzag course can be strengthened:
\begin{equation}\label{eq:zigzagnew}
    \includegraphics[align=c]{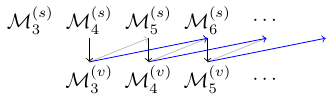}
\end{equation}
In this section, we review and generalize the proof in~\cite{Cao:2023cwa} to explain eq.~\eqref{eq:zigzagnew}, before demonstrating the algorithm with a $\mathcal M_6^{(s)}$ example.

\subsection{Proof of eq.~\eqtitleref{eq:zigzagold}}

The vertical arrows in eq.~\eqref{eq:zigzagold} have already been explained in~\eqref{eq:extract}. Here, we need only show the validity of the diagonal arrows.

Suppose we have already obtained the explicit results of $\mathcal M_{\leq n-1}^{(s)}$ and $\mathcal M_{\leq n-2}^{(v)}$, and we would like to construct $\mathcal M_n^{(s)}$. We already have enough input to fully constrain the residue of $\mathcal M_n^{(s)}$ at $\mathcal X_{ij}=2m$ where $\|i-j\|\geq3$ (length$\geq3$ poles)\footnote{Distance is measured on the circle: $\|i-j\|:=\min\{|i-j|,n-|i-j|\}$.}, because both half amplitudes only require $\mathcal M_{\leq n-2}$. Note that this determines not just terms with only length$\geq3$ poles, but all terms with \emph{at least one} length$\geq3$ pole. For instance, a term in $\mathcal M_6^{(s)}$ of the form
\begin{equation*}
    \frac{\text{something}}{\mathcal X_{13}\mathcal X_{14}}
\end{equation*}
is visible upon taking the residue at $\mathcal X_{14}=0$ and is fixed by factorization.

\begin{figure}[ht]
    \centering
    \begin{subfigure}{0.3\linewidth}
        \centering
        \includegraphics{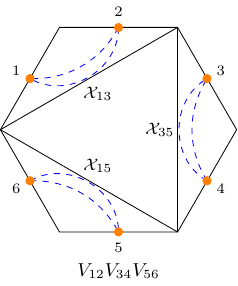}
        \caption{}
    \end{subfigure}
    \hfill
    \begin{subfigure}{0.3\linewidth}
        \centering
        \includegraphics{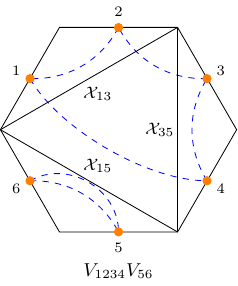}
        \caption{}
    \end{subfigure}
    \hfill
    \begin{subfigure}{0.3\linewidth}
        \centering
        \includegraphics{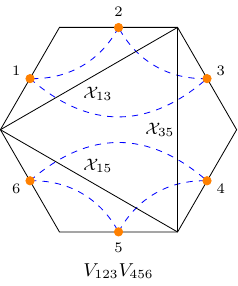}
        \caption{}
    \end{subfigure}
    \caption{(a) Not determined by scalar factorization. (b) Determined by scalar factorization at $\mathcal X_{13}=0$ or $\mathcal X_{35}=0$. (c) Vanish because $\mathcal X_{35}$ intersects 2 R-cycles $V_{123}$ and $V_{456}$.}
    \label{fig:shortchannels}
\end{figure}

Terms with only $\mathcal X_{i,i+2}=2m$ poles (length-2 poles) are not fixed by factorization, because the gluon-exchange contributions involve $\mathcal M_{n-1}^{(v)}$ which is not known by assumption. However, any incompatible R-structure is fixed by scalar factorization, because gluon-exchanges do not contribute (Figure~\ref{fig:shortchannels}). So the remaining problem is to determine terms with only short poles and fully-compatible R-structures (Figure~\ref{fig:shortchannels}(a)). Luckily, no-gluon kinematics enables us to constrain the gluon-exchange contribution without knowing it precisely. It turns out that, together with the flat-space limit, these are sufficient to fully constrain such terms.

\begin{figure}[ht]
    \centering
    \begin{subfigure}{0.3\linewidth}
        \centering
        \includegraphics[scale=0.8]{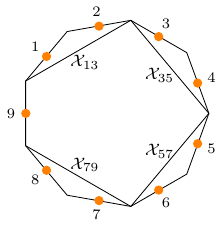}
        \caption{}
    \end{subfigure}
    \quad
    \begin{subfigure}{0.6\linewidth}
        \centering
        \includegraphics[scale=0.8]{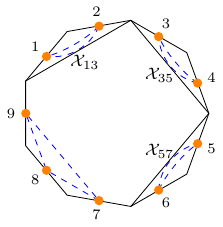}\includegraphics[scale=0.8]{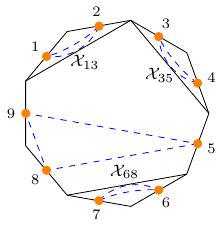}\includegraphics[scale=0.8]{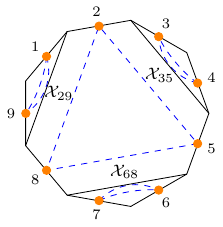}
        \caption{}
    \end{subfigure}
    \caption{Odd $n=9$ example. (a) With $n'=4$ length-2 poles, any R-cycle involving the 9-th particle necessarily intersect one of the poles. (b) With $n'-1=3$ length-2 poles, there is a unique fully compatible R-structure.}
    \label{fig:fixodd}
\end{figure}

Let us try to construct a term (``survivor'') that cannot be fixed by these conditions. We will soon see that this is impossible. As mentioned above, such survivor terms can only have length-2 poles and fully compatible R-structures.

Suppose $n=2n'+1$ is odd. The flat-space limit imposes power-counting $\mathcal X^{n'-1}$, so that each term contains at least $(n'-1)$ poles. On the other hand, since each $\mathcal X_{ij}$ pole has $\|i-j\|\geq2$ and incompatible poles are not allowed, there can be at most $n'$ poles (all length-2 poles). However, none of the R-structures is simultaneously compatible with $n'$ length-2 poles (Figure~\ref{fig:fixodd}(a)), while a unique R-structure exists for $(n'-1)$ length-2 poles (Figure~\ref{fig:fixodd}(b)). Therefore, a survivor term must have exactly $(n'-1)$ length-2 poles, with a constant numerator. But, according to~\eqref{eq:NGfix}, no-gluon kinematics in any one channel (say, $\mathcal X_{13}=0$) forces the amplitude to take the form
\begin{equation}
    \mathop{\rm Res}_{\mathcal X_{13}=0}\mathcal M_n^{(s)}=\frac{\text{known}+\text{poly}(\mathcal E_{2i}^{(0)})}{\text{other denominator factors}},\quad\text{poly}(\mathcal E_{2i}^{(0)})\sim\mathcal X^{\geq1}.
\end{equation}
In order to survive the no-gluon kinematics constraints, a survivor term must be proportional to $\text{poly}(\mathcal E)\sim\mathcal X^{\geq1}$, which is impossible with only a constant numerator. As a result, we have just shown that no survivor term exists for odd $n$.

Now, suppose $n=2n'$ is even. The flat-space limit imposes power-counting $\mathcal X^{n'-2}$, so that each term contains at least $(n'-2)$ poles. Furthermore, the leading order terms are determined by the flat-space result; in particular, terms with exactly $(n'-2)$ poles do not survive the flat-space limit constraint. On the other hand, since each $\mathcal X_{ij}$ pole has $\|i-j\|\geq2$ and incompatible poles are not allowed, there can be at most $n'$ poles (all length-2 poles). Hence, multiplying both numerator and denominator by some $\mathcal X_{i,i+2}$ and creating spurious poles if necessary, we can put survivor terms into the form
\begin{equation*}
    \frac{\mathcal X^{\leq1}}{\mathcal X^{n'}}.
\end{equation*}
For the interesting case of $n'\geq3$, there are at least two (perhaps spurious) poles. It turns out that no such linear numerator can survive no-gluon kinematics constraints in both channels. Let us demonstrate this with an example, and show that no linear numerator $f(\mathcal X)$ can survive the no-gluon kinematics constraints in both $\mathcal X_{13}=0$ and $\mathcal X_{35}=0$. To survive the no-gluon kinematics constraint in $\mathcal X_{13}=0$\footnote{We use $\oplus$ to denote some unknown linear combination of individual terms.},
\begin{equation}\label{eq:bothngk1}
    f(\mathcal X)=\mathcal X_{13}\oplus\bigoplus_{i\neq1,2,3}\mathcal E_{2i}^{(0)}=\mathcal X_{13}\oplus\bigoplus_{i\neq1,2,3}\left(\mathcal X_{2i}-1-\frac{\mathcal X_{1i}+\mathcal X_{3i}}2\right).
\end{equation}
Similarly, to survive the no-gluon kinematics constraint in $\mathcal X_{35}=0$,
\begin{equation}\label{eq:bothngk2}
    f(\mathcal X)=\mathcal X_{35}\oplus\bigoplus_{i\neq3,4,5}\mathcal E_{4i}^{(0)}=\mathcal X_{35}\oplus\bigoplus_{i\neq3,4,5}\left(\mathcal X_{4i}-1-\frac{\mathcal X_{3i}+\mathcal X_{5i}}2\right).
\end{equation}
In eq.~\eqref{eq:bothngk1} there are no $\mathcal X_{4i}$ except for $i=2$, while in eq.~\eqref{eq:bothngk2} there are no $\mathcal X_{2i}$ except for $i=4$. Comparing the terms,
\begin{equation}
    f(\mathcal X)=\mathcal X_{13}\oplus\left(\mathcal X_{24}-1-\frac{\mathcal X_{14}}2\right)=\mathcal X_{35}\oplus\left(\mathcal X_{24}-1-\frac{\mathcal X_{25}}2\right).
\end{equation}
For the interesting case of $n\geq6$, we have $\mathcal X_{1i}\neq\mathcal X_{5i}$, so that $f(\mathcal X)=0$. As a result, we have just shown that no survivor term exists for even $n$.

In summary, we have ruled out the existence of a survivor term not determined by our constraints, and the validity of the diagonal arrows in eq.~\eqref{eq:zigzagold} is established.

\subsection{Proof of eq.~\eqtitleref{eq:zigzagnew}}

The validity of the blue arrows in eq.~\eqref{eq:zigzagnew} can be shown in a similar fashion. Since only $\mathcal M_{\leq n-3}^{(v)}$ is presumed to be known, survivor terms shoule be constructed with length-2 poles or length-3 poles and fully compatible R-structures.

\begin{figure}[h]
    \centering
    \includegraphics[scale=0.8]{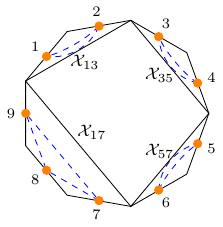}
    \caption{With length-3 poles available, it is possible to have $n'$ poles and fully compatible R-structures.}
    \label{fig:fixoddlong}
\end{figure}

Suppose $n=2n'+1$ is odd. Again, flat-space limit implies that survivor terms have at least $(n'-1)$ poles. However, since length-3 poles are allowed, we cannot rule out terms with $n'$ poles by compatibility of R-structures (Figure~\ref{fig:fixoddlong}). However, to have $n'$ poles means to have numerators $\mathcal X^{\leq1}$, and one of the $n'$ poles must have length-3. The no-gluon kinematics constraint in the length-3 channel reads
\begin{equation}
    \mathop{\rm Res}_{\mathcal X_{ii+3}=2m}\mathcal M_n^{(s)}=\frac{\text{known}+\text{poly}(\mathcal E)}{\text{other denominator factors}},\quad\text{poly}(\mathcal E)\sim\mathcal X^{\geq2}.
\end{equation}
Importantly, the $\mathcal E$ factors are now quadratic, making it impossible to construct a survivor with $\mathcal X^{\leq1}$ numerators. Like before, terms with $(n'-1)$ poles and of the form
\begin{equation*}
    \frac{\text{constant}}{\mathcal X^{n'-1}}
\end{equation*}
are ruled out by the no-gluon kinematics constraint in any one channel.

Suppose $n=2n'$ is even. Again, flat-space limit implies that survivor terms have at least $(n'-1)$ poles. Creating spurious poles as before if necessary, we can put survivor terms into the form
\begin{equation*}
    \frac{\mathcal X^{\leq1}}{\mathcal X^{n'}}.
\end{equation*}
Now, if one of the $n'$ (perhaps spurious) poles has length 3, the no-gluon kinematics constraint in that channel forces the numerator to be $\text{poly}(\mathcal E)\sim\mathcal X^{\geq2}$, killing any survivor. On the other hand, if all $n'$ (perhaps spurious) poles have length 2, the situation reduces to that discussed in the previous subsection, and we know that no survivors exist.

Therefore, we see that it is not much harder to prove eq.\eqref{eq:zigzagnew} than to prove eq.\eqref{eq:zigzagold}.

\subsection{Worked example}

Let us illustrate how to use the constructing algorithm in practice with a $\mathcal M_6^{(s)}$ example. As the arrows in \eqref{eq:zigzagnew} indicates, we will only need the explicit results of $\mathcal M_3^{(s)},\mathcal M_4^{(s)},\mathcal M_5^{(s)}$ and $\mathcal M_3^{(v)}$.

In the first step, we construct a rational function that has all the correct residues for length$\geq4$ poles (none for $\mathcal M_6^{(s)}$) and the correct incompatible-channel contributions for length-2 and length-3 poles, by gluing lower-point amplitudes together using~\eqref{eq:scalar-fac}. This is achieved by the following observation: suppose we seek a function $F(x,y)$ with only simple poles at $x=0$ and $y=0$ and require
\begin{align*}
    \mathop{\rm Res}_{x=0}F(x,y)&=F_x(y)=\frac ry+f_x(y),\\
    \mathop{\rm Res}_{y=0}F(x,y)&=F_y(x)=\frac{r'}x+f_y(x).
\end{align*}
As long as the input is consistent, i.e., $\mathop{\rm Res}_{x=0}F_y(x)=\mathop{\rm Res}_{y=0}F_x(y)$ or $r=r'$, we can simply take
\begin{equation*}
    F(x,y)=\frac{F_x(y)}x+\frac{F_y(x)-\frac{\mathop{\rm Res}_{x=0}F_y(x)}x}y\xlongequal{\text{Mathematica}}\frac{F_x(y)}x+\frac{F_y(x)\texttt{/.\{$\frac1x\to0$\}}}y.
\end{equation*}
Iterating this leads to a simple algorithm that constructs a multi-variate function $\mathcal S_6^{(s)}$\footnote{The letter $\mathcal S$ stands for scalar, as $\mathcal S_6^{(s)}$ correctly captures all scalar-exchange contributions, but may have missed gluon-exchange contributions (compatible channel poles) for length-2 and length-3 channels.} with correct codim-1 residues. In our particular case, we obtain
\begin{align*}
    \frac{S_6^{(s)}(123456)}{-8}&=\frac1{\mathcal X_{14}}+\frac1{\mathcal X_{14}-2}+\frac1{\mathcal X_{13}\mathcal X_{14}}+\frac1{\mathcal X_{13}\mathcal X_{15}}+\frac1{\mathcal X_{14}\mathcal X_{15}}+\frac1{\mathcal X_{13}\mathcal X_{46}}\\
    &+\frac1{\mathcal X_{13}\mathcal X_{14}\mathcal X_{15}}+\frac1{\mathcal X_{13}\mathcal X_{35}\mathcal X_{15}}+\frac1{\mathcal X_{14}\mathcal X_{15}\mathcal X_{24}}+\frac1{\mathcal X_{15}\mathcal X_{24}\mathcal X_{25}}\\
    &+\text{their inequivalent cyclic images},\\
    \frac{S_6^{(s)}(123;456)}{4}&=\frac2{\mathcal X_{13}\mathcal X_{14}}+\frac2{\mathcal X_{14}\mathcal X_{15}}-\frac{\mathcal X_{35}-2}{\mathcal X_{13}\mathcal X_{14}\mathcal X_{15}}-\frac{\mathcal X_{25}-2}{\mathcal X_{14}\mathcal X_{15}\mathcal X_{24}}-\frac{\mathcal X_{36}-2}{\mathcal X_{13}\mathcal X_{14}\mathcal X_{46}}\\
    &+\text{their inequivalent cyclic-by-3 images},\\
    \frac{S_6^{(s)}(12;34;56)}{-2}&=\frac{(\mathcal X_{24}-2)(\mathcal X_{46}-2)}{\mathcal X_{13}\mathcal X_{14}\mathcal X_{15}}+\frac{(\mathcal X_{24}-2)(\mathcal X_{46}-2)}{\mathcal X_{13}(\mathcal X_{14}-2)\mathcal X_{15}}\\
    &+\text{their inequivalent cyclic-by-2 images},\\
    &\vdots
\end{align*}
Notice that $S_6^{(s)}(123456)=M_6^{(s)}(123456)$ because it does not have compatible channels and gluon-exchange contributions(recall this in Fig.~\ref{fig:compatR}).

In the second step, we write down all possible monomials with (a) the correct power-counting, (b) only length-2 and length-3 poles, and (c) fully compatible R-structures. In our case, we have the following list of possible monomials:
\begin{align*}
    V_{123456}:&\quad\text{none},\\
    V_{12}V_{3456}:&\quad\frac1{\mathcal X_{13}},\\
    V_{123}V_{456}:&\quad\frac1{\mathcal X_{14}},\ \frac1{\mathcal X_{14}-2},\\
    V_{14}V_{23}V_{56}:&\quad\frac1{\mathcal X_{15}},\ \frac1{\mathcal X_{24}},\ \frac1{\mathcal X_{15}\mathcal X_{24}},\ \frac{\mathcal X_{13}}{\mathcal X_{15}\mathcal X_{24}},\ \frac{\mathcal X_{14}}{\mathcal X_{15}\mathcal X_{24}},\ \frac{\mathcal X_{25}}{\mathcal X_{15}\mathcal X_{24}},\ \frac{\mathcal X_{26}}{\mathcal X_{15}\mathcal X_{24}},\\
    &\quad\frac{\mathcal X_{35}}{\mathcal X_{15}\mathcal X_{24}},\ \frac{\mathcal X_{36}}{\mathcal X_{15}\mathcal X_{24}},\ \frac{\mathcal X_{46}}{\mathcal X_{15}\mathcal X_{24}},\\
    V_{12}V_{34}V_{56}:&\quad\frac1{\mathcal X_{13}},\ \frac1{\mathcal X_{13}\mathcal X_{35}},\ \frac{\mathcal X_{14}}{\mathcal X_{13}\mathcal X_{35}},\ \frac{\mathcal X_{15}}{\mathcal X_{13}\mathcal X_{35}},\ \frac{\mathcal X_{24}}{\mathcal X_{13}\mathcal X_{35}},\ \frac{\mathcal X_{25}}{\mathcal X_{13}\mathcal X_{35}},\ \frac{\mathcal X_{26}}{\mathcal X_{13}\mathcal X_{35}},\\
    &\quad\frac{\mathcal X_{36}}{\mathcal X_{13}\mathcal X_{35}},\ \frac{\mathcal X_{46}}{\mathcal X_{13}\mathcal X_{35}},\ \frac1{\mathcal X_{13}\mathcal X_{35}\mathcal X_{15}},\ \frac{\mathcal X_{14}}{\mathcal X_{13}\mathcal X_{35}\mathcal X_{15}},\ \frac{\mathcal X_{24}}{\mathcal X_{13}\mathcal X_{35}\mathcal X_{15}},\ \frac{\mathcal X_{14}\mathcal X_{24}}{\mathcal X_{13}\mathcal X_{35}\mathcal X_{15}},\\
    &\quad\frac{\mathcal X_{14}\mathcal X_{25}}{\mathcal X_{13}\mathcal X_{35}\mathcal X_{15}},\ \frac{\mathcal X_{14}\mathcal X_{26}}{\mathcal X_{13}\mathcal X_{35}\mathcal X_{15}},\ \frac{\mathcal X_{24}\mathcal X_{25}}{\mathcal X_{13}\mathcal X_{35}\mathcal X_{15}},\ \frac{\mathcal X_{24}\mathcal X_{26}}{\mathcal X_{13}\mathcal X_{35}\mathcal X_{15}},\\
    &\quad\text{and their inequivalent cyclic-by-2 images}.
\end{align*}

In the third step, we write down the most general linear combination $\sum\text{coef}(\text{term})\times\text{term}$ of the above monomials, and add it to $\mathcal S_6^{(s)}$. Demanding the result has the correct flat-space limit fixes a bunch of terms. In our particular case, flat-space limit fixes
\begin{gather*}
    \text{coef}\left(V_{12}V_{34}V_{56}\frac{\mathcal X_{14}}{\mathcal X_{13}\mathcal X_{35}}\right)=0,\quad \text{coef}\left(V_{12}V_{34}V_{56}\frac{\mathcal X_{24}}{\mathcal X_{13}\mathcal X_{35}}\right)=-4,\\
    \text{coef}\left(V_{123}V_{456}\frac1{\mathcal X_{14}}\right)+\text{coef}\left(V_{123}V_{456}\frac1{\mathcal X_{14}-2}\right)=8,\quad\text{etc.}
\end{gather*}
In other words, the unfixed terms are those subleading in the large-$\mathcal X$ limit:
\begin{align*}
    V_{123456}:&\quad\text{none},\\
    V_{12}V_{3456}:&\quad\text{none},\\
    V_{123}V_{456}:&\quad\frac1{\mathcal X_{14}}-\frac1{\mathcal X_{14}-2},\\
    V_{14}V_{23}V_{56}:&\quad\frac1{\mathcal X_{15}\mathcal X_{24}},\\
    V_{12}V_{34}V_{56}:&\quad\frac1{\mathcal X_{13}\mathcal X_{35}},\ \frac1{\mathcal X_{13}\mathcal X_{35}\mathcal X_{15}},\ \frac{\mathcal X_{14}}{\mathcal X_{13}\mathcal X_{35}\mathcal X_{15}},\ \frac{\mathcal X_{24}}{\mathcal X_{13}\mathcal X_{35}\mathcal X_{15}},\\
    &\quad\text{and their inequivalent cyclic-by-2 images}.
\end{align*}

In the fourth and final step, we consider the no-gluon kinematics in each factorization channel. For instance, consider the ansatz for $M_6^{(s)}(12;34;56)$ which now reads:
\begin{align*}
    M_6^{(s)}(12;34;56)&=\frac{-2(\mathcal X_{24}-2)(\mathcal X_{46}-2)}{\mathcal X_{13}\mathcal X_{14}\mathcal X_{15}}+\frac{-2(\mathcal X_{24}-2)(\mathcal X_{46}-2)}{\mathcal X_{13}(\mathcal X_{14}-2)\mathcal X_{15}}\\
    &+\frac{c_1-4\mathcal X_{14}+4\mathcal X_{24}-4\mathcal X_{26}+4\mathcal X_{46}}{\mathcal X_{13}\mathcal X_{15}}+\frac{c_2}{\mathcal X_{13}\mathcal X_{35}\mathcal X_{15}}\\
    &+\frac{c_3\mathcal X_{14}+c_4\mathcal X_{24}+4\mathcal X_{14}\mathcal X_{25}-4\mathcal X_{14}\mathcal X_{26}}{\mathcal X_{13}\mathcal X_{35}\mathcal X_{15}}\\
    &+\text{their inequivalent cyclic-by-2 images}.
\end{align*}
The no-gluon kinematics for the compatible channel $\mathcal X_{13}=0$ reads~\eqref{eq:no-gluon kin},
\begin{align*}
    \mathcal X_{24}&=1+\frac{\mathcal X_{14}}2,\\
    \mathcal X_{25}&=1+\frac{\mathcal X_{15}+\mathcal X_{35}}2,\\
    \mathcal X_{26}&=1+\frac{\mathcal X_{36}}2.
\end{align*}
Taking the residue and evaluating at the no-gluon kinematics,
\begin{align*}
    \mathop{\rm Res}_{\mathcal X_{13}=0}M_6^{(s)}(12;34;56)\Big|_{\text{no-gluon}}&=\frac{4+c_1+\frac12c_3}{\mathcal X_{15}}+\frac{4+c_1+\frac12c_3}{\mathcal X_{35}}+\frac{2(\mathcal X_{46}-2)}{\mathcal X_{14}\mathcal X_{15}}+\frac{2(\mathcal X_{46}-2)}{\mathcal X_{35}\mathcal X_{36}}\\
    &+\frac{c_2+c_3+2c_4+(c_3+\frac12c_4)\mathcal X_{14}+(c_3+\frac12c_4)\mathcal X_{36}+(c_4-4)\mathcal X_{46}}{\mathcal X_{15}\mathcal X_{35}}.
\end{align*}
On the other hand, we know that the correct result should match the scalar-exchange contribution obtained through gluing $\mathcal M_3^{(s)}\otimes\mathcal M_5^{(s)}$:
\begin{align*}
    \mathop{\rm Res}_{\mathcal X_{13}=0}\mathcal M_6^{(s)}\Big|_{\text{no-gluon}}&=-2\texttt{glueR}\left(\mathcal M_{12I}^{(s)(0)}\mathcal M_{3456I}^{(s)(0)}\right)\Big|_{\text{no-gluon}}\\
    &=V_{12}V_{34}V_{56}\left(\frac{-4}{\mathcal X_{15}}+\frac{-4}{\mathcal X_{35}}+\frac{2(\mathcal X_{46}-2)}{\mathcal X_{14}\mathcal X_{15}}+\frac{2(\mathcal X_{46}-2)}{\mathcal X_{35}\mathcal X_{36}}-\frac{2(4-\mathcal X_{14}-\mathcal X_{36})}{\mathcal X_{15}\mathcal X_{35}}\right)\\
    &+\text{other R-structures}.
\end{align*}
Comparing the coefficients, we see that
\begin{equation*}
    c_1=-8,\quad c_2=-16,\quad c_3=0,\quad c_4=4.
\end{equation*}
For this particular example, one channel suffices. In general, we need to consider every factorization channel to completely fix the answer. In the end, we arrive at
\begin{align*}
    M_6^{(s)}(12;34;56)&=\frac{-2(\mathcal X_{24}-2)(\mathcal X_{46}-2)}{\mathcal X_{13}\mathcal X_{14}\mathcal X_{15}}+\frac{-2(\mathcal X_{24}-2)(\mathcal X_{46}-2)}{\mathcal X_{13}(\mathcal X_{14}-2)\mathcal X_{15}}\\
    &+\frac{-8-4\mathcal X_{14}+4\mathcal X_{24}-4\mathcal X_{26}+4\mathcal X_{46}}{\mathcal X_{13}\mathcal X_{15}}+\frac{-16}{\mathcal X_{13}\mathcal X_{35}\mathcal X_{15}}\\
    &+\frac{4\mathcal X_{24}+4\mathcal X_{14}\mathcal X_{25}-4\mathcal X_{14}\mathcal X_{26}}{\mathcal X_{13}\mathcal X_{35}\mathcal X_{15}}\\
    &+\text{their inequivalent cyclic-by-2 images}.
\end{align*}
 \section{Explicit results for supergluon and spinning amplitudes}\label{sec:result}
For convenience, we record explicit, human-readable results for supergluon amplitudes up to $n=7$ and spinning amplitudes up to $n=6$. All these and some higher-point results can be found in the ancillary file of~\cite{Cao:2023cwa}, but we present these compact formula mainly to show how simple these amplitudes actually are and to make their interesting mathematical structures more visible. The $n=4,5$ supergluon amplitudes can be found in~\cite{Alday:2021odx,Alday:2022lkk}, so we start with $n=6$ supergluon amplitudes:
\begin{align}
    \frac{M_6(123456)}{-8}&=\left[\left(\frac{1}{{\cal X}_{14}}+\frac{1}{{\cal X}_{14}-2}\right)+ 2~{\rm cyclic}\right]\nonumber\\
    &+\left[\left(\frac{1}{{\cal X}_{13}{\cal X}_{14}} + \frac{1}{{\cal X}_{13} {\cal X}_{15}} + \frac{1}{{\cal X}_{14} {\cal X}_{15}}\right)+ 5~{\rm cyclic} + \frac{1}{{\cal X}_{13} {\cal X}_{46}} + 2~ {\rm cyclic}\right]\nonumber \\
    &+\left[\left(\frac{1}{{\cal X}_{13}{\cal X}_{14} {\cal X}_{15}}+ \frac{1}{{\cal X}_{13} {\cal X}_{14} {\cal X}_{46}}\right) + 5~{\rm cyclic} + \frac{1}{{\cal X}_{13} {\cal X}_{15} {\cal X}_{35}} + 1~{\rm cyclic}\right]\!,\!\label{6pt-single}\\[5pt]
    \frac{M_6(12;3456)}{4}&=\frac{1}{{\cal X}_{13}} \left[\,2+\frac{2}{{\cal X}_{15}}+ \frac{2}{{\cal X}_{35}}+\frac{2}{{\cal X}_{46}}-\frac{{\cal X}_{25}-2}{{\cal X}_{15} {\cal X}_{35}}\right.\nonumber\\
    &\hspace{3.7em}\left.-\,({\cal X}_{24}-2) \left(\frac{1}{{\cal X}_{14}}+\frac{1}{{\cal X}_{14}-2} +\frac{1}{{\cal X}_{14}{\cal X}_{15}}+\frac{1}{{\cal X}_{14} {\cal X}_{46}}\right) \right.\nonumber \\
    &\hspace{3.7em}\left.-\,({\cal X}_{26}-2) \left(\frac{1}{{\cal X}_{36}}+\frac{1}{{\cal X}_{36}-2}+\frac{1}{{\cal X}_{36}{\cal X}_{46}}+\frac{1}{{\cal X}_{35} {\cal X}_{36}}\right)\right],
\end{align}

\begin{align}
    \frac{M_6(123;456)}{4}&=\frac{1}{{\cal X}_{14}-2}+\frac{1}{{\cal X}_{14}}\left(1+\frac{2}{{\cal X}_{13}}+ \frac{2}{{\cal X}_{15}}+\frac{2}{{\cal X}_{24}}+\frac{2}{{\cal X}_{46}}\right.\nonumber\\
    &\hspace{8.5em}-\left.\frac{{\cal X}_{25}-2}{{\cal X}_{15}{\cal X}_{24}}-\frac{{\cal X}_{35}-2}{{\cal X}_{13}{\cal X}_{15}} -\frac{{\cal X}_{26}-2}{{\cal X}_{24}{\cal X}_{46}} -\frac{{\cal X}_{36}-2}{{\cal X}_{13}{\cal X}_{46}} \right),\\[5pt]
    \frac{M_6(12;36;45)}{-2}&=\frac{1}{{\cal X}_{13} {\cal X}_{46}}\left(8-2 {\cal X}_{25}+\frac{\left({\cal X}_{15}-2\right) \left({\cal X}_{24}-2\right)}{{\cal X}_{14}}+\frac{\left({\cal X}_{15}-2\right) \left({\cal X}_{24}-2\right)}{{\cal X}_{14}-2}\right.\nonumber\\
    &\hspace{5.5em} \left.+\,\frac{\left({\cal X}_{26}-2\right) \left({\cal X}_{35}-2\right)}{{\cal X}_{36}}+\frac{\left({\cal X}_{26}-2\right) \left({\cal X}_{35}-2\right)}{{\cal X}_{36}-2}\right),\\[2pt]
    \frac{M_6(12;34;56)}{-2}&=\left[\frac{1}{{\cal X}_{13} {\cal X}_{15}}\left(2\left({\cal X}_{14}-{\cal X}_{24}+{\cal X}_{26}-{\cal X}_{46}+2\right) \phantom{\frac{\left({\cal X}_{24}-2\right) \left({\cal X}_{46}-2\right)}{ {\cal X}_{14}}}\right.\right.\nonumber\\
    &\hspace{6em}\left.\left.+\,\frac{\left({\cal X}_{24}-2\right) \left({\cal X}_{46}-2\right)}{ {\cal X}_{14}}+\frac{\left({\cal X}_{24}-2\right) \left({\cal X}_{46}-2\right)}{ {\cal X}_{14}-2}\right)+2\text{ cyclic}\right]\nonumber\\
    &+\frac{2}{{\cal X}_{13} {\cal X}_{15} {\cal X}_{35}}\left(
    \begin{aligned}
    &4-{\cal X}_{46}-{\cal X}_{24}-{\cal X}_{26}+{\cal X}_{36} {\cal X}_{24}-{\cal X}_{14} {\cal X}_{25}\\
    &+{\cal X}_{14} {\cal X}_{26}-{\cal X}_{14} {\cal X}_{36}-{\cal X}_{25} {\cal X}_{36}+{\cal X}_{25} {\cal X}_{46}
    \end{aligned}\right).
\end{align}

\noindent Next we give $n=7$ supergluon amplitudes:
\begin{align}
    \frac{M_7(1234567)}{16}&=\left[\left(\frac{1}{\mathcal{X}_{1 3}\mathcal{X}_{1 5}}+\frac{1}{\mathcal{X}_{1 3}\left( \mathcal{X}_{1 5}-2 \right)}+\frac{1}{\mathcal{X}_{1 3}\mathcal{X}_{1 4}\mathcal{X}_{1 5}}+\frac{1}{\mathcal{X}_{1 3}\mathcal{X}_{1 4}\mathcal{X}_{1 6}}\right.\right.\nonumber\\
    &\hspace{2em}\left.\left.+\frac{1}{\mathcal{X}_{1 4}\mathcal{X}_{1 5}\mathcal{X}_{2 4}}+\frac{1}{\mathcal{X}_{1 4}\mathcal{X}_{1 6}\mathcal{X}_{2 4}}+\frac{1}{\mathcal{X}_{1 5}\mathcal{X}_{1 6}\mathcal{X}_{2 4}}\right.\right.\nonumber\\
    &\hspace{2em}\left.\left.+\frac{1}{\mathcal{X}_{1 4}\mathcal{X}_{1 5}\mathcal{X}_{1 6}\mathcal{X}_{2 4}}+\frac{1}{\mathcal{X}_{1 3}\mathcal{X}_{1 5}\mathcal{X}_{1 6}\mathcal{X}_{3 5}}\right)+14~\text{dihedral}\right],\\[5pt]
    \frac{M_7(123;4567)}{-8}&=\left\{ \frac{1}{\mathcal{X}_{14}}\left[ \frac{2}{\mathcal{X}_{15}}+\frac{1}{\mathcal{X}_{15}-2}+\frac{1}{\mathcal{X}_{16}}+\frac{1/2}{\mathcal{X}_{57}}+\frac{2}{\mathcal{X}_{13}}-\frac{\mathcal{X}_{26}-2}{\mathcal{X}_{16}\mathcal{X}_{24}\mathcal{X}_{46}} \right. \right. \nonumber\\
    &\hspace{3.3em}-\left( \frac{\mathcal{X}_{25}-2}{\mathcal{X}_{24}}+\frac{\mathcal{X}_{35}-2}{\mathcal{X}_{13}} \right) \left( \frac{1}{\mathcal{X}_{15}}+\frac{1}{\mathcal{X}_{15}-2}+\frac{1}{\mathcal{X}_{15}\mathcal{X}_{16}}+\frac{1}{\mathcal{X}_{15}\mathcal{X}_{57}} \right) \nonumber\\
    &\hspace{3.6em}+\left. \frac{2}{\mathcal{X}_{15}\mathcal{X}_{16}}+\frac{2}{\mathcal{X}_{16}\mathcal{X}_{24}}+\frac{2}{\mathcal{X}_{13}\mathcal{X}_{57}}+\frac{2}{\mathcal{X}_{15}\mathcal{X}_{57}}+\frac{2}{\mathcal{X}_{13}\mathcal{X}_{16}}+\frac{1}{\mathcal{X}_{16}\mathcal{X}_{46}} \right] \nonumber\\[2pt]
    &\hspace{1.5em}+\left. \frac{1}{\mathcal{X}_{14}-2}\left( \frac{1}{\mathcal{X}_{16}}+\frac{1}{\mathcal{X}_{15}-2}+\frac{1/2}{\mathcal{X}_{57}} \right) \right\} +(1\leftrightarrow 4,2\leftrightarrow 3,5\leftrightarrow 7),\\[5pt]
    \frac{M_7(14;23;567)}{4}&=\frac{1}{\mathcal{X}_{24}}\left[ \frac{4}{\mathcal{X}_{15}}+\frac{2}{\mathcal{X}_{15}-2}+\frac{\mathcal{X}_{13}-2}{\mathcal{X}_{14}\mathcal{X}_{15}}\left( \frac{\mathcal{X}_{46}-2}{\mathcal{X}_{16}}+\frac{\mathcal{X}_{47}-2}{\mathcal{X}_{57}}-2 \right) \right. \nonumber\\
    &\hspace{3.3em}-\frac{\mathcal{X}_{13}-2}{\mathcal{X}_{14}-2}\left( \frac{1}{\mathcal{X}_{15}}+\frac{1}{\mathcal{X}_{15}-2}-\frac{\mathcal{X}_{46}-2}{\mathcal{X}_{15}\mathcal{X}_{16}}-\frac{\mathcal{X}_{47}-2}{\mathcal{X}_{15}\mathcal{X}_{57}} \right) \nonumber\\
    &\hspace{3.3em}-\frac{2\mathcal{X}_{37}-8}{\mathcal{X}_{15}\mathcal{X}_{57}}-\frac{2\mathcal{X}_{36}-8}{\mathcal{X}_{15}\mathcal{X}_{16}}+\frac{\mathcal{X}_{35}-2}{\mathcal{X}_{15}\mathcal{X}_{25}}\left( \frac{\mathcal{X}_{26}-2}{\mathcal{X}_{16}}+\frac{\mathcal{X}_{27}-2}{\mathcal{X}_{57}}-2 \right)\nonumber\\
    &\hspace{3.3em}\left. -\frac{\mathcal{X}_{35}-2}{\mathcal{X}_{25}-2}\left( \frac{1}{\mathcal{X}_{15}}+\frac{1}{\mathcal{X}_{15}-2}+\frac{\mathcal{X}_{26}-2}{\mathcal{X}_{15}\mathcal{X}_{16}}+\frac{\mathcal{X}_{27}-2}{\mathcal{X}_{15}\mathcal{X}_{57}} \right) \right] .
\end{align}

\begin{align}
    \frac{M_7(12;34567)}{-8}&=\frac{1}{\mathcal{X}_{13}}\left[\frac{2}{\mathcal X_{15}}+\frac{2}{\mathcal X_{15}-2}+\frac{2}{\mathcal{X}_{16}}+\frac{2}{\mathcal{X}_{46}}+\frac1{\mathcal X_{47}}+\frac1{\mathcal X_{47}-2}-\frac{\mathcal{X}_{2 5}-2}{\left(\mathcal{X}_{1 5}-2\right) \mathcal{X}_{3 5}} \right.\nonumber\\
    &\hspace{3.7em}+\frac1{\mathcal X_{16}\mathcal X_{35}}+\frac2{\mathcal X_{15}\mathcal X_{16}}+\frac2{\mathcal X_{16}\mathcal X_{46}}+\frac2{\mathcal X_{36}\mathcal X_{46}}+\frac2{\mathcal X_{47}\mathcal X_{57}}\nonumber\\
    &\hspace{3.7em}-\frac{\mathcal{X}_{2 4}-2}{\mathcal{X}_{1 4}}\left(\frac{1}{\mathcal{X}_{15}}+\frac{1}{\mathcal{X}_{16}}+\frac{1}{\mathcal{X}_{46}}+\frac{1}{\mathcal{X}_{47}}+\frac{1}{\mathcal{X}_{57}}\right.\nonumber\\
    &\hspace{7.5em}\left.+\frac{1}{\mathcal{X}_{15}\mathcal{X}_{16}}+\frac{1}{\mathcal{X}_{16}\mathcal{X}_{46}}+\frac{1}{\mathcal{X}_{46}\mathcal{X}_{47}}+\frac{1}{\mathcal{X}_{47}\mathcal{X}_{57}}+\frac{1}{\mathcal{X}_{57}\mathcal{X}_{15}}\right)\nonumber\\[2pt]
    &\hspace{3.7em}-\frac{\mathcal{X}_{2 4}-2}{\mathcal{X}_{1 4}-2} \left( \frac{1}{\mathcal{X}_{1 5}-2}+\frac{1}{\mathcal{X}_{1 6}}+\frac{1}{\mathcal{X}_{4 6}}+\frac{1}{\mathcal{X}_{4 7}-2}+\frac{1}{\mathcal{X}_{5 7}}\right)\nonumber\\[2pt]
    &\hspace{3.7em}\left.-\frac{\mathcal{X}_{2 5}-2}{\mathcal{X}_{1 5}\mathcal{X}_{3 5}}\left(\frac{1}{\mathcal{X}_{5 7}}+\frac{1}{\mathcal{X}_{1 6}}+1 \right)\right]+(1\leftrightarrow 3,4\leftrightarrow 7,5\leftrightarrow 6),\\[5pt]
    \frac{M_7(12;34;567)}{4}&=\left\{ \frac{1}{\mathcal{X}_{13}}\left[ \frac{2}{\mathcal{X}_{15}}+\frac{2}{\mathcal{X}_{15}-2}-\frac{\mathcal{X}_{24}-2}{\mathcal{X}_{14}-2}\left( \frac{1}{\mathcal{X}_{15}}+\frac{1}{\mathcal{X}_{15}-2} \right) -\frac{2\left( \mathcal{X}_{24}-2 \right)}{\mathcal{X}_{14}\mathcal{X}_{15}} \right. \right.\nonumber\\
    &\hspace{3.3em}-\frac{\mathcal{X}_{27}-2}{\mathcal{X}_{35}}\left( \frac{2}{\mathcal{X}_{37}}+\frac{2}{\mathcal{X}_{37}-2}-\frac{\mathcal{X}_{46}-2}{2\mathcal{X}_{36}\mathcal{X}_{37}}-\frac{\mathcal{X}_{46}-2}{2\left( \mathcal{X}_{36}-2 \right) \left( \mathcal{X}_{37}-2 \right)} \right)\nonumber\\
    &\hspace{3.3em}+\frac{3}{\mathcal{X}_{35}}-\frac{\mathcal{X}_{14}-\mathcal{X}_{24}+\mathcal{X}_{26}-\mathcal{X}_{46}+2}{\mathcal{X}_{15}\mathcal{X}_{16}}-\frac{\mathcal{X}_{14}+\mathcal{X}_{25}-4}{2\left( \mathcal{X}_{15}-2 \right) \mathcal{X}_{35}}\nonumber\\
    &\hspace{3.3em}\left. +\frac{\mathcal{X}_{14}-\mathcal{X}_{24}+\mathcal{X}_{27}-\mathcal{X}_{47}+2}{\mathcal{X}_{15}\mathcal{X}_{57}}-\frac{\mathcal{X}_{14}+2\mathcal{X}_{24}+\mathcal{X}_{25}-8}{2\mathcal{X}_{15}\mathcal{X}_{35}} \right]\nonumber\\
    &\hspace{1.5em}+\frac{\mathcal{X}_{24}-\mathcal{X}_{25}+\mathcal{X}_{26}-\mathcal{X}_{46}-2}{\mathcal{X}_{15}\mathcal{X}_{16}\mathcal{X}_{35}}-\frac{\mathcal{X}_{24}-\mathcal{X}_{25}+\mathcal{X}_{27}-\mathcal{X}_{47}-2}{\mathcal{X}_{15}\mathcal{X}_{35}\mathcal{X}_{57}}\nonumber\\
    &\hspace{1.5em}\left. -\frac{M_6(12;34;56)}{2\mathcal{X}_{16}} \right\}+(1\leftrightarrow 5,2\leftrightarrow 4,6\leftrightarrow 7).\\[5pt]
    \frac{M_7(12;56;347)}{4}&=\frac{1}{\mathcal{X}_{13}\mathcal{X}_{57}}\left[ 6+\frac{\mathcal{X}_{16}-2}{\mathcal{X}_{15}}\left( \frac{\mathcal{X}_{24}-2}{\mathcal{X}_{14}}+\frac{\mathcal{X}_{25}-2}{\mathcal{X}_{35}}-2 \right) -\frac{2(\mathcal{X}_{46}-2)}{\mathcal{X}_{47}} \right. \nonumber\\
    &\hspace{5em}+\frac{\mathcal{X}_{16}-2}{\mathcal{X}_{15}-2}\left( \frac{\mathcal{X}_{24}-2}{\mathcal{X}_{14}-2}+\frac{\mathcal{X}_{25}-2}{\mathcal{X}_{35}}-2 \right) +\frac{\mathcal{X}_{24}-2}{\mathcal{X}_{14}}\left( \frac{\mathcal{X}_{46}-2}{\mathcal{X}_{47}}-2 \right) \nonumber\\
    &\hspace{5em}+\frac{\mathcal{X}_{46}-2}{\mathcal{X}_{47}-2}\left( \frac{\mathcal{X}_{24}-2}{\mathcal{X}_{14}-2}+\frac{\mathcal{X}_{27}-2}{\mathcal{X}_{37}-2}-2 \right) +\frac{\mathcal{X}_{27}-2}{\mathcal{X}_{37}-2}\left( \frac{\mathcal{X}_{36}-2}{\mathcal{X}_{35}}-2 \right) \nonumber\\
    &\hspace{5em}\left. +\frac{\mathcal{X}_{27}-2}{\mathcal{X}_{37}}\left( \frac{\mathcal{X}_{36}-2}{\mathcal{X}_{35}}+\frac{\mathcal{X}_{46}-2}{\mathcal{X}_{47}}-2 \right) -\frac{2\mathcal{X}_{26}-8}{\mathcal{X}_{35}} \right] .
\end{align}

\noindent We also present all spinning amplitudes for $n=3,4,5,6$. 
\begin{align}
    2M^{(v)}_3(12)&=\zp{1}-\zp{2}, \\[5pt]
    M^{(v)}_4(123)&=-\frac{\zp{1}-\zp{23}}{\mathcal{X}_{2 4}}-\frac{\zp{12}-\zp{3}}{\mathcal{X}_{1 3}},\\[5pt]
    \frac{M^{(v)}_5(1234)}{2}&=\,\frac{2 \zp{1}}{\mathcal{X}_{2 5}-2}+\frac{\zp{1}-\zp{234}}{\mathcal{X}_{2 5}}\left(\frac{1}{\mathcal{X}_{3 5}}+\frac{1}{\mathcal{X}_{2 4}}+1\right)+\frac{\zp{12}-\zp{34}}{\mathcal{X}_{1 3} \mathcal{X}_{3 5}}\nonumber\\
    &\hspace{1em}-\frac{2 \zp{4}}{\mathcal{X}_{1 4}-2}+\frac{\zp{123}-\zp{4}}{\mathcal{X}_{1 4}}\left(\frac{1}{\mathcal{X}_{2 4}}+\frac{1}{\mathcal{X}_{1 3}}+1\right),\\[5pt]
    M^{(v)}_5(12;34)&=\,\frac{2 \zp{1} \left(\mathcal{X}_{2 4}-2\right)}{\left(\mathcal{X}_{2 5}-2\right) \mathcal{X}_{3 5}}-\frac{(\zp{1}-\zp{234}) \left(\mathcal{X}_{2 4}-2\right)}{\mathcal{X}_{2 5} \mathcal{X}_{3 5}} \nonumber\\
    &\hspace{1em}-\frac{2 \zp{4} \left(\mathcal{X}_{2 4}-2\right)}{\mathcal{X}_{1 3} \left(\mathcal{X}_{1 4}-2\right)}+\frac{(\zp{123}-\zp{4}) \left(\mathcal{X}_{2 4}-2\right)}{\mathcal{X}_{1 3} \mathcal{X}_{1 4}} \nonumber\\
    &\hspace{1em}+\frac{1}{\mathcal{X}_{1 3} \mathcal{X}_{3 5}}
    \left(\begin{gathered}
        -4(\zp{1}-\zp{4})+2\mathcal X_{24}(\zp{12}-\zp{34})\\
        +\mathcal X_{25}(3\zp{3}-\zp{124})+\mathcal X_{14}(\zp{134}-3\zp{2})
    \end{gathered}\right) \nonumber\\
    &\hspace{1em}+2(\zp{14}-\zp{23})\left(\frac{1}{\mathcal{X}_{13}}-\frac{1}{\mathcal{X}_{35}}\right),\\[5pt]
    M_5^{(v)}(14;23)&=\frac{2 \zp{1} (\mathcal X_{35}-2)}{\mathcal X_{24} (\mathcal X_{25}-2)}+\frac{(\mathcal X_{35}-2) (\zp{1}-\zp{234})}{\mathcal X_{24} \mathcal X_{25}}-\frac{2 (\zp{12}-\zp{34})}{\mathcal X_{24}}\nonumber\\[2pt]
    &-\frac{2 \zp{4} (\mathcal X_{13}-2)}{(\mathcal X_{14}-2) \mathcal X_{24}}-\frac{(\mathcal X_{13}-2) (\zp{4}-\zp{123})}{\mathcal X_{14} \mathcal X_{24}}.\\[5pt]
    \frac{M^{(v)}_6(12345)}{4}&=-\,\frac{2\zp{1}}{\mathcal{X}_{26}-2}\left( \frac{1}{\mathcal{X}_{25}-2}+\frac{1}{\mathcal{X}_{35}}+\frac{1}{\mathcal{X}_{36}-2}+\frac{1}{\mathcal{X}_{46}}+\frac{1}{\mathcal{X}_{24}} \right)-\frac{2\zp{12}}{\mathcal{X}_{13}\left( \mathcal{X}_{36}-2 \right)} \nonumber\\
    &\hspace{1em}+\frac{2\zp{5}}{\mathcal{X}_{15}-2}\left( \frac{1}{\mathcal{X}_{14}-2}+\frac{1}{\mathcal{X}_{24}}+\frac{1}{\mathcal{X}_{25}-2}+\frac{1}{\mathcal{X}_{35}}+\frac{1}{\mathcal{X}_{13}} \right) +\frac{2\zp{45}}{\left( \mathcal{X}_{14}-2 \right) \mathcal{X}_{46}} \nonumber\\
    &\hspace{1em}-\frac{\zp{1234}-\zp{5}}{\mathcal{X}_{15}}\frac{M_5^{(s)}(12345)}{4} \nonumber-\frac{\zp{1}-\zp{2345}}{\mathcal{X}_{26}}\frac{M_5^{(s)}(23456)}{4} \nonumber\\
    &\hspace{1em}-\frac{\zp{123}-\zp{45}}{\mathcal{X}_{14}\mathcal{X}_{46}}\left( \frac{1}{\mathcal{X}_{24}}+\frac{1}{\mathcal{X}_{13}}+1 \right) \nonumber\\
    &\hspace{1em}-\frac{\zp{12}-\zp{345}}{\mathcal{X}_{13}\mathcal{X}_{36}}\left( \frac{1}{\mathcal{X}_{46}}+\frac{1}{\mathcal{X}_{35}}+1 \right).
\end{align}

\begin{align}
    \frac{M^{(v)}_6(12;345)}{2}&=-\,\frac{2\zp{1}}{\mathcal{X}_{26}-2}\left[ \frac{1}{\mathcal{X}_{36}}\left( \frac{\mathcal{X}_{25}-2}{\mathcal{X}_{35}}+\frac{\mathcal{X}_{24}-2}{\mathcal{X}_{46}}-1 \right) -\frac{1}{\mathcal{X}_{36}-2} \right]  \nonumber\\[2pt]
    &\hspace{1em}+\frac{2\zp{5}}{\mathcal{X}_{13}\left( \mathcal{X}_{15}-2 \right)}\left( \frac{\mathcal{X}_{24}-2}{\mathcal{X}_{14}-2}+\frac{\mathcal{X}_{25}-2}{\mathcal{X}_{35}}-2 \right)+\frac{2(2\zp{1}-\zp{345})}{\mathcal{X}_{13}\mathcal{X}_{36}} \nonumber\\[2pt]
    &\hspace{1em}+\frac{\zp{1}-\zp{2345}}{\mathcal{X}_{26}\mathcal{X}_{36}}\left( 2-\frac{\mathcal{X}_{24}-2}{\mathcal{X}_{46}}-\frac{\mathcal{X}_{25}-2}{\mathcal{X}_{35}} \right) +\frac{2\zp{12}}{\mathcal{X}_{13}\left( \mathcal{X}_{36}-2 \right)} \nonumber\\
    &\hspace{1em}+\frac{\zp{1234}-\zp{5}}{\mathcal{X}_{13}\mathcal{X}_{15}}\left( 2-\frac{\mathcal{X}_{24}-2}{\mathcal{X}_{14}}-\frac{\mathcal{X}_{25}-2}{\mathcal{X}_{35}} \right) +\frac{2\zp{45}\left( \mathcal{X}_{24}-2 \right)}{\mathcal{X}_{13}\left( \mathcal{X}_{14}-2 \right) \mathcal{X}_{46}} \nonumber\\
    &\hspace{1em}+\frac{2(\zp{15}-\zp{234})}{\mathcal{X}_{35}}\left( \frac{1}{\mathcal{X}_{36}}-\frac{1}{\mathcal{X}_{13}} \right)-\frac{(\zp{123}-\zp{45})\left( \mathcal{X}_{24}-2 \right)}{\mathcal{X}_{13}\mathcal{X}_{14}\mathcal{X}_{46}} \nonumber\\
    &\hspace{1em}+\frac{2(\zp{145}-\zp{23})}{\mathcal{X}_{46}}\left( \frac{1}{\mathcal{X}_{36}}-\frac{1}{\mathcal{X}_{13}} \right) \nonumber\\
    &\hspace{1em}-\frac{1}{\mathcal{X}_{13}\mathcal{X}_{36}}\left[\frac{1}{\mathcal{X}_{35}}\left(\begin{aligned}
        \zp{1}\left( \mathcal{X}_{15}+2\mathcal{X}_{25}-\mathcal{X}_{26}-4 \right)&\\
        +\zp{5}\left( \mathcal{X}_{15}-2\mathcal{X}_{25}-\mathcal{X}_{26}+4 \right)&
    \end{aligned}\right.\right.\nonumber\\
    &\hspace{13em}\left.\begin{aligned}
        -\zp{2}\left(3\mathcal{X}_{15}-2\mathcal{X}_{25}
        +\mathcal{X}_{26} \right)&\\
        +\zp{34}\left(\mathcal{X}_{15}
        -2\mathcal{X}_{25}+3\mathcal{X}_{26} \right)&
    \end{aligned}\right)\nonumber\\
    &\hspace{5em}+\frac{1}{\mathcal{X}_{46}}\left(\begin{aligned}
        \zp{1}\left( \mathcal{X}_{14}+2\mathcal{X}_{24}-\mathcal{X}_{26}-4 \right)&\\
        +\zp{45}\left(\mathcal{X}_{14} -2\mathcal{X}_{24}-\mathcal{X}_{26}+4 \right)&
    \end{aligned}\right.\nonumber\\
    &\hspace{13.5em}\left.\left.\begin{aligned}
        -\zp{2}\left( 3\mathcal{X}_{14}-2\mathcal{X}_{24}+\mathcal{X}_{26} \right)&\\
        +\zp{3}\left( \mathcal{X}_{14}-2\mathcal{X}_{24} +3\mathcal{X}_{26} \right)&
    \end{aligned}\right)\right].\\[5pt]
    \frac{M_{6}^{(v)}(23;145)}{2}&=\frac{2\zp{1}}{\mathcal{X}_{24}(\mathcal{X}_{26}-2)}\left( 2-\frac{\mathcal{X}_{35}-2}{\mathcal{X}_{25}-2}-\frac{\mathcal{X}_{36}-2}{\mathcal{X}_{46}} \right) +\frac{(\mathcal{X}_{13}-2)(\zp{45}-\zp{123})}{\mathcal{X}_{14}\mathcal{X}_{24}\mathcal{X}_{46}}\nonumber\\
    &-\frac{2\zp{5}}{(\mathcal{X}_{15}-2)\mathcal{X}_{24}}\left( 2-\frac{\mathcal{X}_{35}-2}{\mathcal{X}_{25}-2}-\frac{\mathcal{X}_{13}-2}{\mathcal{X}_{14}-2} \right) +\frac{2(\zp{12}-\zp{345})}{\mathcal{X}_{24}\mathcal{X}_{46}}\nonumber\\
    &+\frac{\zp{1}-\zp{2345}}{\mathcal{X}_{24}\mathcal{X}_{26}}\left( 2-\frac{\mathcal{X}_{35}-2}{\mathcal{X}_{25}}-\frac{\mathcal{X}_{36}-2}{\mathcal{X}_{46}} \right) +\frac{2(\mathcal{X}_{13}-2)\zp{45}}{(\mathcal{X}_{14}-2)\mathcal{X}_{24}\mathcal{X}_{46}}\nonumber\\
    &-\frac{\zp{5}-\zp{1234}}{\mathcal{X}_{15}\mathcal{X}_{24}}\left( 2-\frac{\mathcal{X}_{35}-2}{\mathcal{X}_{25}}-\frac{\mathcal{X}_{13}-2}{\mathcal{X}_{14}} \right).\\[5pt]
    \frac{M_{6}^{(v)}(15;234)}{2}&=\frac{2\zp{1}}{\mathcal{X}_{26}-2}\left( \frac{1}{\mathcal{X}_{25}}+\frac{1}{\mathcal{X}_{25}-2}-\frac{\mathcal{X}_{46}-2}{\mathcal{X}_{24}\mathcal{X}_{25}}-\frac{\mathcal{X}_{36}-2}{\mathcal{X}_{25}\mathcal{X}_{35}} \right) +\frac{2(\zp{12}-\zp{345})}{\mathcal{X}_{25}\mathcal{X}_{35}}\nonumber\\
    &-\frac{2\zp{5}}{\mathcal{X}_{15}-2}\left( \frac{1}{\mathcal{X}_{25}}+\frac{1}{\mathcal{X}_{25}-2}+\frac{\mathcal{X}_{14}-2}{\mathcal{X}_{24}\mathcal{X}_{25}}+\frac{\mathcal{X}_{13}-2}{\mathcal{X}_{25}\mathcal{X}_{35}} \right) -\frac{2(\zp{45}-\zp{123})}{\mathcal{X}_{24}\mathcal{X}_{25}}\nonumber\\
    &+\frac{\zp{1}-\zp{2345}}{\mathcal{X}_{25}\mathcal{X}_{26}}\left( 2-\frac{\mathcal{X}_{46}-2}{\mathcal{X}_{24}}-\frac{\mathcal{X}_{36}-2}{\mathcal{X}_{35}} \right) \nonumber\\
    &-\frac{\zp{5}-\zp{1234}}{\mathcal{X}_{15}\mathcal{X}_{25}}\left( 2-\frac{\mathcal{X}_{14}-2}{\mathcal{X}_{24}}-\frac{\mathcal{X}_{13}-2}{\mathcal{X}_{35}} \right) .
\end{align}

\subsection{Flat-space limit: Yang-Mills-scalar amplitudes and scalar-scaffolded gluons}
Before proceeding, we comment on the flat-space limit of AdS supergluon amplitudes to arbitrary multiplicity. As pointed out in~\cite{Alday:2021odx,Alday:2023kfm}, the flat space limit corresponds to Yang-Mills amplitudes in the special kinematics where polarizations are orthogonal to momenta, {\it i.e.} $\epsilon_i \cdot p_j=0$, which are non-vanishing only for even $n=2m$. Moreover, the flat-space amplitudes enjoy the same R-symmetry structure as AdS supergluon amplitude with the identification $\epsilon_i \cdot \epsilon_j =\langle i j\rangle^2$, which amounts to a particular way of dimensional reduction of Yang-Mills amplitudes. The upshot is that the flat-space limit for $n=2m$ points is given by the amplitudes in Yang-Mills-scalar (YMS) theory with $m$ pairs of scalars; the Lagrangian is obtained as dimensional reduction of Yang-Mills Lagrangian~\cite{Cachazo:2014xea}:
\begin{equation} \label{eq:LYMS}
    \mathcal{L}_{\mathrm{YMS}}=-\operatorname{Tr}\left(\frac{1}{4} F^{\mu \nu} F_{\mu \nu}+\frac{1}{2} D^\mu \phi^I D_\mu \phi^I-\frac{g_{\rm YM}^2}{4} \sum_{I \neq J}\left[\phi^I, \phi^J\right]^2\right).
\end{equation}
where the flavor group for the scalar is SU$(2)$. The YMS amplitude with $m$ pairs of scalars $(i_1, j_1), \cdots, (i_m, j_m)$ gives the coefficient of $\langle i_1 j_1\rangle^2 \cdots \langle i_m j_m\rangle^2$, and we can compute all these amplitudes using Feynman diagrams from the above Lagrangian: each pair of scalars interact with an internal gluon via three- and four-point vertices (the gluon interact via the usual three- and four-point vertices), and additionally we also have $\phi^4$ vertex. For the special case of $(i_1, j_1)=(12), (i_2, j_2)=(34), \cdots, (i_m, j_m)=(2m{-}1, 2m)$, this amplitude has been studied in~\cite{Arkani-Hamed:2023swr} in the context of scalar-scaffolded gluons: the $m$-gluon YM amplitude can be extracted from this amplitude by taking the residue $X_{13}=X_{35}=\cdots=X_{1, 2m-1}=0$; recall that $X_{1,3}=s_{1,2}$ {\it etc.} are the planar poles correspond to factorization of each scalar pair to produce an on-shell gluon (more details can be found in~\cite{Arkani-Hamed:2023swr}). 

Remarkably, as an alternative to Feynman diagrams, we can also compute all these YMS amplitudes via the following compact Cachazo-He-Yuan (CHY) formula~\cite{Cachazo:2014xea}:
\begin{equation}
{\cal M}_{2m}^{\rm YMS} ( (i_1 j_1), \cdots , (i_m j_m))=\int d\mu_{2m}~{\rm PT}(12 \cdots 2m)~
{\rm Pf}' A_{2m}~\prod_{a=1}^m \frac{1}{z_{i_a, j_a}}
\end{equation}
where the $n$-point CHY measure $d\mu_n$ denotes the $(n{-}3)$-fold integrations over the moduli space ${\cal M}_{0,n}$ localized by the universal scattering equations: $d\mu_n:=\frac{\prod_a d z_a}{{\rm vol~SL}(2)} \prod'_a \delta(\sum_{b\neq a} \frac{s_{a,b}}{z_{a,b}})$ (details can be found in~\cite{Cachazo:2013gna, Cachazo:2013iea}); in addition to the product of $1/z_{i_a, j_a})$ with $z_{i,j}:=z_i-z_j$ for the pairs of scalars, we have the Parke-Taylor factor (indicating the color ordering), and reduced Pfaffian for matrix $A$ defined as
\begin{equation}
{\rm PT}(1,2,\cdots, n):=\frac{1}{z_{1,2} z_{2,3} \cdots z_{n,1}}\,,\quad {\rm Pf}' A:=\frac{(-)^{i+j} |A|_{i,j}}{z_{i,j}}\,,\quad (A_n)_{i,j}=(1-\delta_{i,j}) \frac{s_{i,j}}{z_{i,j}}    
\end{equation}
where the $n\times n$ matrix $A$ has corank $2$ thus the reduced Pfaffian is defined in terms of the matrix with columns and rows $i,j$ deleted (the result is independent of the choice of $i,j$ with the Jacobian). To compare with R-symmetry structures of AdS supergluon amplitudes, we then expand any of these factors $\prod_{a=1}^m \langle i_a, j_a\rangle^2$ into the basis $\{V_\alpha\}$, thus for each $V_\alpha$, the flat-space limit $M^{\rm YMS}_{2m} (V_\alpha)$ becomes a linear combination of ${\cal M}^{\rm YMS}((i_1 j_1), \cdots, (i_m j_m) )$. For example, for $n=4$, ${\cal M}((12), (34))=M(V_{12; 34})$, ${\cal M}((14), (23))=M(V_{14; 23})$ and ${\cal M}((13), (24))=M(V_{12;34})+M(V_{14;23})-2 M(V_{1234})$; by inverting these relations and plug in YMS amplitudes we have
\begin{equation}
M_4^{\rm YMS}(1234)=2\,,\quad M_4^{\rm YMS}(12;34)=\frac{X_{24}}{X_{13}}\,,\quad M_4^{\rm YMS}(14;23)=\frac{X_{13}}{X_{24}}\,,   
\end{equation}
which, up to an overall constant, is the correct flat-space limit (for the rest of the subsection we suppress $V$ for these R-structures) with (flat-space) planar variables defined as
\begin{equation}
X_{a,b}:=(p_a+ p_{a{+}1}+\cdots + p_{b{-}1})^2=\sum_{a\leq i<j\leq b{-}1} s_{i,j}\,. 
\end{equation} Using the automated package in~\cite{He:2021lro}, it is straightforward to compute all these flat-space amplitudes using CHY formulas up to $n=12$. For example, the single-trace result is given by a sum of all planar $\phi^4$ graphs (confirming the $\phi^4$ formula in~\cite{Cachazo:2014xea}), but in general for multi-trace amplitudes we also need interactions with gluons. 

We show some explicit examples to show their simplicity. For $n=6$, we have four inequivalent R-symmetry structure; the single-trace amplitude reads: 
\begin{equation}
M_6^{\rm YMS} (123456)=-4 (\frac{1}{X_{2,5}}+\frac{1}{X_{3,6}}+\frac{1}{X_{1,4}})\,,
\end{equation} 
and for the two inequivalent double-trace cases we have:
\begin{equation}
M_6^{\rm YMS} (12; 3456)=\frac{2}{X_{1,3}} (1-\frac{X_{2,4}}{X_{1,4}}-\frac{X_{2,6}}{X_{3,6}})\,, \quad M_6^{\rm YMS}(123;456)=\frac{2}{X_{1,4}}\,,
\end{equation} 
where the second case has only $\phi^4$ interaction but the first case has gluon propagating between the two traces; the two inequivalent triple-trace cases are
\begin{equation}
M_6^{\rm YMS}(14;23;56)=\frac{1}{X_{1,5} X_{2,4}} (X_{3,6}-\frac{X_{2,6} X_{3,5}}{X_{2,5}}-\frac{X_{1,3} X_{4,6}}{X_{1,4}})\,,
\end{equation}
\begin{equation}
\begin{aligned}
    M_6^{\rm YMS}(12;34;56)=&\left(\frac{X_{2,4}+X_{4,6}-X_{2,6}-X_{1,4}-\frac{X_{2,4} X_{4,6}}{X_{1,4}}}{X_{1,3} X_{1,5}} + i\to i+2) + (i\to i+4)\right)\\
    &+\frac{X_{1,4} (X_{2,5}-X_{2,6}) + X_{2,5} (X_{3,6}-X_{4,6}) + X_{3,6} (X_{1,4}-X_{2,4})}{X_{1,3} X_{1,5} X_{3,5}}
    \end{aligned}
\end{equation}
where for $M(12;34;56)$, we recognize the numerator of the second line, which is just its residue at $X_{13}=X_{35}=X_{15}=0$, as nothing but the scaffolded $3$-point YM amplitude~\cite{Arkani-Hamed:2023swr, Arkani-Hamed:2023jry}:
\begin{equation}
A_3^{\rm sc. YM}=X_{1,4} (X_{2,5}-X_{2,6}) + X_{2,5} (X_{3,6}-X_{4,6}) + X_{3,6} (X_{1,4}-X_{2,4})\,. 
\end{equation}
It is straightforward to check that these are the correct flat-space limit of AdS supergluon amplitudes. Finally, for $n=8$ there are $14$ inequivalent R-symmetry structures, and here are some representative examples. The single-trace case is again the color-ordered $\phi^4$ amplitude, and we present a few double- and triple-trace examples:
\begin{equation}
M_8^{\rm YMS}(123;45678)=\frac{-4}{X_{1,4}}(\frac{1}{X_{1,6}}+\frac{1}{X_{4,7}}+ \frac{1}{X_{5,8}})\,,
\end{equation}
\begin{equation}
M_8^{\rm YMS}(1234; 5678)=\frac{-4}{X_{1,5}}\left(
\frac{X_{1,4}+X_{1,6}-X_{4,6}}{X_{1,4} X_{1,6}}
+\frac{X_{2,5}+X_{5,8}-X_{2,8}}{X_{2,5}X_{5,8}}-\frac{X_{4,8}}{X_{1,4} X_{5,8}}-\frac{X_{2,6}}{X_{1,6} X_{2,5}}\right)\,,
\end{equation}
\begin{equation}
M_8^{\rm YMS}(123; 48; 567)=\frac{2}{X_{1,4} X_{5,8}}    
\end{equation}
\begin{equation}
M_8^{\rm YMS}(12;348; 567)=\frac{2}{X_{1,3} X_{5,8}} (1-\frac{X_{2,4}}{X_{1,4}}-\frac{X_{2,8}}{X_{3,8}}) 
\end{equation}
Other double-, triple and quadruple-trace amplitudes are similar (but longer), and we remark that the most difficult one, with R-symmetry structure $12;34;56;78$ has the residue \begin{equation}
{\rm Res}_{X_{1,3}=\cdots=X_{1,7}=0} M_8^{\rm YMS}(12;34;56;78)=A_4^{\rm sc. YM}    
\end{equation}
given by $4$-point YM amplitude in scaffolded form (explicit results can be found in~\cite{Arkani-Hamed:2023swr}).  

\section{New structures of supergluon and spinning amplitudes}\label{sec:structure}
In this section, we discuss several new structures for supergluon amplitude and spinning amplitudes to all multiplicity. First, we will analyse the most general poles structures of these amplitudes by determining the truncation of descendant poles. Next, we will see that for the simplest cases, {\it i.e.} single-trace supergluon amplitudes and single-trace spinning amplitudes, the amplitudes can be obtained via remarkably simple ``Feynman rules". We will end with derivation of some universal behavior of these amplitudes analogous to the collinear and soft limits of flat-space amplitudes. 

\subsection{Truncation of descendant poles}\label{sec:truncation}
In general, one can not expect a conformal correlator to have truncation, since operators with arbitrarily large  descendant level can be exchanged in the corresponding channel. But in the large $N$ CFT theories, especially the $\mathcal N=2$ SCFT theory considered in this paper and the $\mathcal N=4$ SYM theory, people have found truncation, and even proved it rigorously by the large $N$ power-counting argument at four-point~\cite{Rastelli:2017udc}.

To begin, let us give the definition of truncation for Mellin amplitudes. In general, every $\mathcal X_{ij}$ has an infinite series of poles, each corresponding to a descendant level $m\in\mathbb{N}$~\eqref{eq:facschem}. If all the poles with descendant level $m\geqslant m_{0}$ actually vanish, leaving only the poles with descendant level $m<m_0$ remain, then we call it a truncation at level $m_0$, which reads
\begin{equation}\label{def:truncation}
    \residue{\mathcal X_{ij}=2m_0-2}\mathcal M_n \neq 0\, , \quad \mspace{-6mu} \residue{\mathcal X_{ij}=2m}\mathcal M_n = 0\ \ \text{for}\ \ m\geqslant m_0\,.
\end{equation}

From the explicit results in Section~\ref{sec:result}, we find a pattern of the pole structure, which is taken as a conjecture for general supergluon amplitudes, as stated next.

\begin{conjecture}[half-circle rule]\label{conj:truncation}
For supergluon amplitudes, the poles of $\mathcal{X}_{ij}$ always truncate like
\begin{equation}\label{eq:truncation pattern}
    \residue{\mathcal X_{ij}=2m_0-2}\mathcal M_n \neq 0\, , \quad \mspace{-6mu} \residue{\mathcal X_{ij}=2m}\mathcal M_n = 0\ \ \text{for}\ \ m\geqslant m_0=\lfloor \frac{||i-j||+1}{2}\rfloor.
\end{equation}
\end{conjecture}

\noindent For example, the poles of $\mathcal X_{15}$ truncate at $\lfloor \frac{||1-5||+1}{2}\rfloor=\lfloor \frac{5}{2}\rfloor=2$, which means that the poles of $\mathcal X_{15}$ can have at most descendant level $1$. One can easily check from the results given in Section~\ref{sec:result}, that the poles of $\mathcal X_{15}$ can only locate at $\mathcal X_{15}=0$ or $\mathcal X_{15}=2$, which exactly matches our conjecture. 

\addtolength{\belowdisplayskip}{-0.2\baselineskip}

Due to the complexity of factorization of multi-spin correlator, \eqref{eq:truncation pattern} still remains a conjecture so far. However, with only weak assumptions on multi-spin factorization, we can rigorously prove Conjecture~\ref{conj:truncation}. Recall that the factorization of scalar amplitudes for scalar and vector exchange both behave like 
\begin{equation}\label{eq:general factorizaion}
    \residue{\mathcal X_{1k}=2m}=\mathcal{N}^{(m)} \sum_{\cdots}\sharp^{(m)}_{\,\cdots}(\delta_{ab})\, \mathcal{M}^{(m)\cdots}_{1\cdots(k-1)I}\, \mathcal{M}^{(m)\cdots}_{k\cdots n I}\,,
\end{equation}
where the shifted amplitudes $\mathcal M^{(m)}$ are summations of the following form
\addtolength{\belowdisplayskip}{-0.1\baselineskip}
\begin{equation}\label{eq:general shifted amps}
    \mathcal M_{1\cdots(k-1)I}^{(m)\cdots}=\sum_{\substack{\sum n_{ab}\leqslant m\\ \ldots}}\mathcal M_{1\cdots(k-1)I}(\delta_{ab}+n_{ab})\, \#^{\cdots}_{n_{ab}}(\delta_{ab})\,.
\end{equation}

Here $\cdots$ can be any conditions or indices while $\sharp$ and $\#$ can be any factors without poles. Since the complexity of multi-spin factorization is mainly concentrated in the tensor structure instead of the analytical structure, we have convincing reasons to believe that the multi-spin factorization would share the analytical structure of the above form. 

\addtolength{\belowdisplayskip}{0.3\baselineskip}

With the above weak assumptions~\eqref{eq:general factorizaion} and~\eqref{eq:general shifted amps} on multi-spin factorization, we are now ready to prove Conjecture~\ref{conj:truncation}. Proof of the conjecture can be divided into three steps. First, we notice that the 3- and 4-point contact diagram vanish after shifted by $m\geqslant 1,2$ respectively. That is, $\mathcal{M}^{\raisemath{-3pt}{(m)}}_{\text{3-contact}}=0$ for $m\geqslant 1$ and $\mathcal{M}^{\raisemath{-3pt}{(m)}}_{\text{4-contact}}=0$ for $m\geqslant 2$. Second, we demonstrate that the vanishing behavior of shifted $n$-point contact diagram determine if an $n$-point vertex $V_{m_1,\dots, m_n}$ can exist in a general Witten diagram, as stated in Lemma~\ref{lem: vanishing vertex}. Here $m_1,\dots, m_n$ are the descendant levels of propagators connected to $V_{m_1,\dots, m_n}$, and without lost of generality, we assume $m_n$ to be the maximal one. Finally, we show that the existence conditions of 3- and 4-point vertices imply the expected Conjecture~\ref{conj:truncation}. The first step of the proof can be easily verified, so let us move directly to the second step.

\begin{lemma}\label{lem: vanishing vertex}
If the $n$-point contact diagram shifted by $m$ \textit{i.e.} $\mathcal{M}^{\raisemath{-3pt}{(m)}}_{n\text{-contact}}$ vanishes for all $m \geqslant m_0$, then any Witten diagram containing $n$-point vertex $V_{m_1,\dots, m_n}$ must vanish if $m_n \geqslant m_0+m_1+\cdots +m_{n-1}$. General factorization~\eqref{eq:general factorizaion} and~\eqref{eq:general shifted amps} are assumed to be hold.
\end{lemma}

\begin{proof}
Let us consider an $N$-point Witten diagram $\mathcal{M}_N$ with an $n$-point vertex $V_{m_1,\dots, m_n}$ such that $m_{\flat}:=m_n-m_1-\cdots -m_{n-1}\geqslant 0$, as shown in Figure~\ref{fig:vertex-rule-1}.

\begin{figure}[H]
    \vspace{-1em}
    \centering
    \includegraphics[scale=0.8]{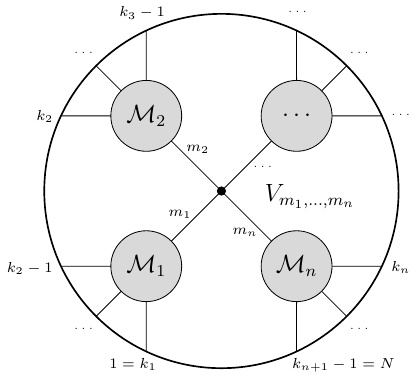}
    \vspace{-1em}
    \caption{$N$-point diagram $\mathcal{M}_N$ with $n$-point vertex $V_{m_1,\dots, m_n}$.}
    \label{fig:vertex-rule-1}
\end{figure}

In order to investigate how the vanishing condition of $\mathcal{M}_N$ is related to $\mathcal{M}^{(m)}_{n\text{-contact}}$, we will factorize out $\mathcal{M}_1,\dots,\mathcal{M}_n$ successively and look into the remaining (shifted) $n$-point contact diagram. Throughout this process we always consider the case that the descendant level of the propagator connecting the $n$-point vertex and $\mathcal{M}_n$ (denoted by $m'$ temporarily) is minimized, since at the final step we will find that $\mathcal{M}_N\neq 0$ only if $\mathcal{M}^{\raisemath{-3pt}{(m')}}_{n\text{-contact}}\neq 0$. Thus in order to make $\mathcal{M}_N\neq 0$, we should strive to minimize $m'$ to be less than $m_0$.

\begin{figure}[ht]
    \centering
    \includegraphics[scale=0.75]{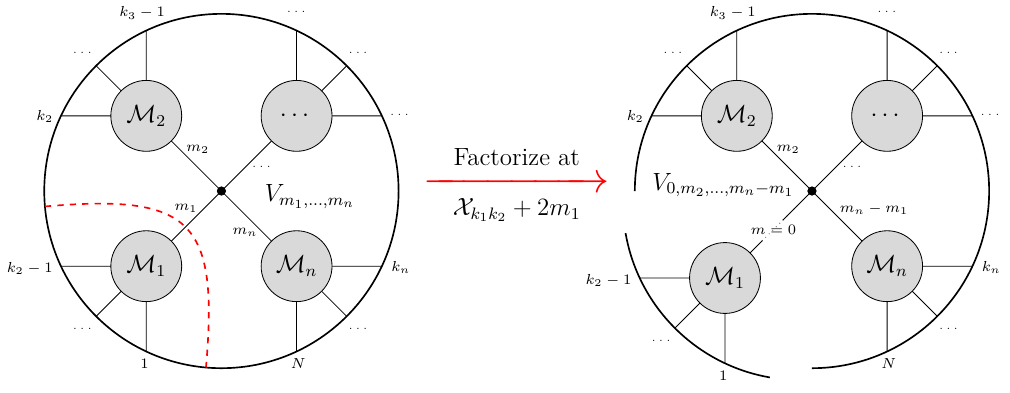}
    \vspace{-1em}
    \caption{Factorize out $\mathcal{M}_1$.}
    \label{fig:vertex-rule-2}
\end{figure}

\addtolength{\belowdisplayskip}{-0.3\baselineskip}

First let us factorize out $\mathcal{M}_1$, as shown in Figure~\ref{fig:vertex-rule-2}. According to assumption~\eqref{eq:general shifted amps}, the half amplitude containing the $n$-point vertex is a linear combination of amplitudes with descendant levels of propagators raised by at most $\sum n_{ab}\leqslant m_1$, which reads
\begin{equation*}
    \mathcal{M}^{(m_1)}_{I k_2\cdots N} =\sum_{\substack{\sum n_{ab}\leqslant m_1\\ \ldots}} \mathcal{M}_{I k_2\cdots N}(\delta_{ab}+n_{ab}) \times \cdots.
\end{equation*}

It would be helpful to imagine that $\sum n_{ab}=m$ are distributed to $\delta_{ab}$, raising them to $\delta_{ab}+n_{ab}$, ascending the descendant levels of some propagators in $\mathcal{M}_{I k_2\cdots N}$. In order to minimize $m'$, we should have all the nonzero $n_{ab}$ distributed to $\delta_{ab}\in \mathcal{X}_{k_n k_{n+1}}$, raising the descendant level of $\mathcal{X}_{k_n k_{n+1}}$ from $m_n-m_1$ to $m_n$. In other words, some terms in $\mathcal{M}_{I k_2\cdots N}$ with $m'\geqslant m_n-m_1$ must exist, otherwise the descendant level of $\mathcal{X}_{k_n k_{n+1}}$ in $\mathcal{M}^{\raisemath{-3pt}{(m_1)}}_{I k_2\cdots N}$ and $\mathcal{M}_N$ would be less than $m_n$.

\begin{figure}[ht]
    \centering
    \includegraphics[scale=0.75]{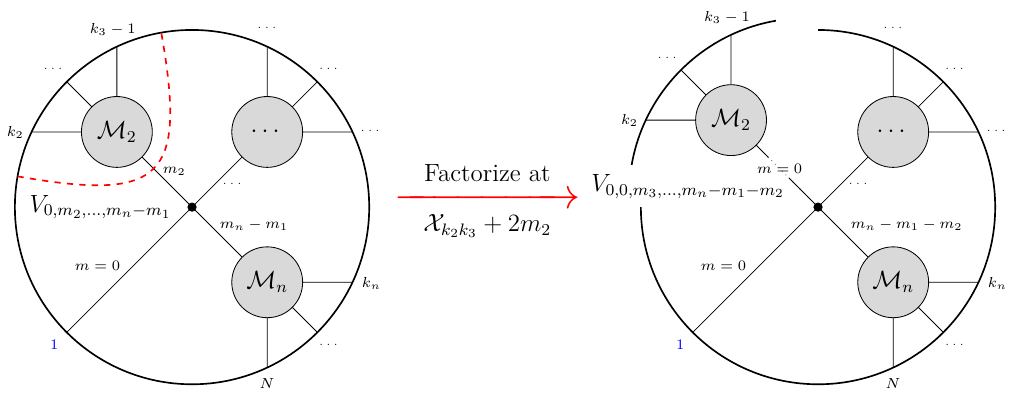}
    \vspace{-0.5em}
    \caption{Factorize out $\mathcal{M}_2$.}
    \label{fig:vertex-rule-3}
\end{figure}

\addtolength{\belowdisplayskip}{0.3\baselineskip}

Labelling the external leg $I$ from factorization by $\color{blue}1$ highlighted in blue, we move on to factorizing out $\mathcal{M}_2$, as shown in Figure~\ref{fig:vertex-rule-3}, and sequentially $\mathcal{M}_3,\dots,\mathcal{M}_{n-1}$. Similar argument shows that some terms in $\mathcal{M}_{{\color{blue}1\cdots n-1}k_n\cdots N}$ (left side of Figure~\ref{fig:vertex-rule-4}) with $m'\geqslant m_{\flat}$ must exist, otherwise the descendant level of $\mathcal{X}_{k_n k_{n+1}}$ in $\mathcal{M}_N$ can not be as large as $m_n$.

\begin{figure}[ht]
    \centering
    \includegraphics[scale=0.75]{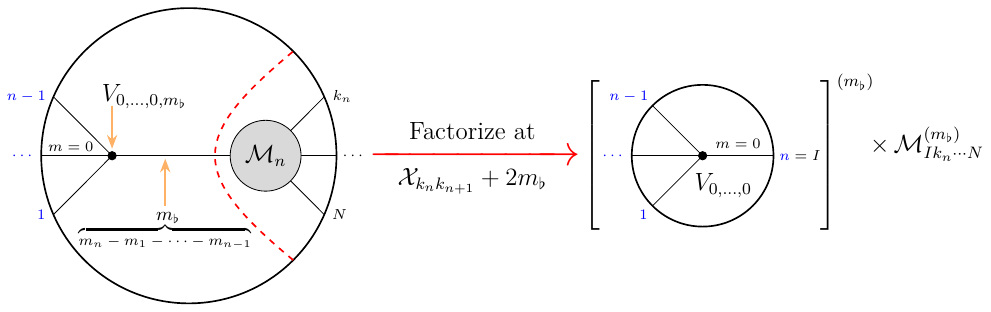}
    \vspace{-1em}
    \caption{Factorize out $\mathcal{M}_n$ to get $\mathcal{M}^{(m_{\flat})}_{n\text{-contact}}$.}
    \label{fig:vertex-rule-4}
\end{figure}

At the last step, we factorize out $\mathcal{M}_n$ from $\mathcal{M}_{{\color{blue}1\cdots n-1}k_n\cdots N}$, as shown in Figure~\ref{fig:vertex-rule-4}, where the external legs $\color{blue}1,\dots,n-1$ highlighted in blue are the legs coming from factorization. As stated before, the minimal $m'$ to have non-vanishing $\mathcal{M}_N$ is $m_{\flat}=m_n-m_1-\cdots -m_{n-1}$. Now from the factorization at $\mathcal{X}_{k_1 k_n}$ we see that, the minimal $m$ by which $\mathcal{M}_{n\text{-contact}}$ need to be shifted is exactly $m_{\flat}$. In other words, all the terms in $\mathcal{M}_{{\color{blue}1\cdots n-1}k_n\cdots N}$ come from some shifted amplitudes $\mathcal{M}^{{\raisemath{-3pt}{(m')}}}_{n\text{-contact}}$ with $m'\geqslant m_{\flat}$. Thus if $\mathcal{M}^{\raisemath{-3pt}{(m)}}_{n\text{-contact}}$ vanishes for all $m\geqslant m_0$ and $m_{\flat}\geqslant m_0$, then all the terms in $\mathcal{M}_N$ would contain a vanishing shifted $n$-point contact diagram, hence $\mathcal{M}_N$ must vanish. This completes the proof.
\end{proof}

Then we can apply Lemma~\ref{lem: vanishing vertex} to the supergluon amplitudes containing only 3- and 4-point vertices in form of $V_{m_1,m_2,m_3}$ and $V_{m_1,m_2,m_3,m_4}$. Without lost of generality, assuming $m_1\leqslant\cdots\leqslant m_n$, we have the following vertices existence conditions
\begin{itemize}
\item $3$-point vertices must have $m_1+m_2\geqslant m_3$.
\item $4$-point vertices must have $m_1+m_2+m_3+1\geqslant m_4$.
\end{itemize}

\begin{figure}[ht]
    \vspace{-0.2em}
    \centering
    \begin{minipage}[b]{0.45\textwidth}
        \centering
        \includegraphics[scale=0.85]{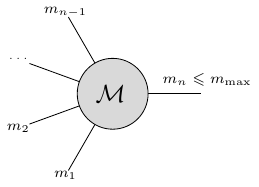}
        \vspace{-0.6em}
        \caption{Witten sub-diagram.}
        \label{fig:sub-diagram-1}
    \end{minipage}\hfill
    \begin{minipage}[b]{0.45\textwidth}
        \centering
        \includegraphics[scale=0.8]{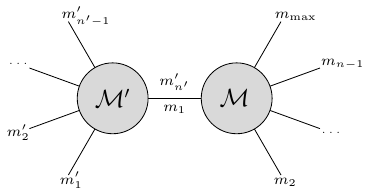}
        \vspace{-0.6em}
        \caption{Induction of sub-diagram.}
        \label{fig:sub-diagram-2}
    \end{minipage}
\end{figure}

Now let us move on to the final step, and investigate how the descendant levels can be restricted by the above vertices existence conditions. With careful observation, we find a helpful property of tree Witten sub-diagram with only 3- and 4-point vertices that, if the descendant levels of propagators connected to it are given by $m_1,m_2,\dots,m_n$, as shown in Figure~\ref{fig:sub-diagram-1}, then the maximal possible value of $m_n$ is given by
\begin{equation}\label{eq:sub-diagram maximal m}
    m_{\text{max}}=m_1+m_2+\cdots+m_{n-1}+(\text{number of 4-point vertex in }\mathcal{M})
\end{equation}

This can be proved by induction. Obviously~\eqref{eq:sub-diagram maximal m} holds for sub-diagrams with a single vertex. Now suppose that~\eqref{eq:sub-diagram maximal m} holds for two tree sub-diagrams $\mathcal{M}$ and $\mathcal{M}'$, then for the sub-diagram $\mathcal{M}'+\mathcal{M}$ constructed by connecting $\mathcal{M}'$ to $\mathcal{M}$ as in Figure~\ref{fig:sub-diagram-2}, we have
\begin{equation}
\begin{aligned}
    m^{\mathcal{M}'+\mathcal{M}}_{\text{max}} &=m^{\mathcal{M}'}_{\text{max}}+m_2+\cdots+m_{n-1}+(\text{number of 4-point vertex in }\mathcal{M})\\
    &=m'_1+m'_2+\cdots+m'_{n'-1}+m_2+\cdots+m_{n-1}\\
    &\phantom{=m'_1}\;+(\text{number of 4-point vertex in }\mathcal{M}'+\mathcal{M})
\end{aligned}
\end{equation}

Now we are just one step away from the desired Conjecture~\ref{conj:truncation}. Note that for the cut $\mathcal X_{ij}$ of a Witten diagram $\mathcal{M}$, the maximal number of 4-point vertices in $\mathcal{M}$ is exactly $\lfloor \frac{||i-j||-1}{2}\rfloor$, we draw the conclusion that the poles of $\mathcal X_{ij}$ must truncate at $\lfloor \frac{||i-j||+1}{2}\rfloor$.

\subsection{Feynman rules for single-trace amplitudes}

Using the algorithm described in section~\ref{sec:construct}, we have explicitly computed supergluon amplitudes up to $n=8$, as well as the single-trace part of $\mathcal M_{\leq12}^{(s)}$ and $\mathcal M_{\leq9}^{(v)}$. Interestingly, the results closely resemble the flat-space amplitudes obtained through simple Feynman rules.

For the single-trace part $M_n^{(s)}(\rm ST)\equiv M_n^{(s)}(12\cdots n)$ and $M_n^{(v)}(\rm ST)\equiv M_n^{(v)}(12\cdots(n-1))$, we are able to make the intuition precise. It turns out that these partial amplitudes can be viewed as arising from the following bulk interaction\footnote{Here, $\lambda$ and $g$ are actually proportional. However, since we will not be careful about the relative normalization of $A^{\bar\mu}$ and $\phi$, we will label the two coupling constants differently. The Feynman rules below are normalized such that $M_n({\rm ST})$ is obtained by putting $g=\lambda=1$.}:
\begin{equation}
    \mathcal L_{\text{int, color-ordered}}\supset g\phi^3-2g^2\phi^4+\lambda A^{\bar\mu}\phi\overleftrightarrow{\nabla}_{\bar\mu}\phi.
\end{equation}
One can try to write down all Witten diagrams needed to compute $\mathcal M_n^{(s)}(12\cdots n)$ and $\mathcal M_n^{(v)}(12\cdots(n-1))$, and then compute their individual contribution to the Mellin amplitude following the Mellin space Feynman rules. The rules for scalar interactions $\phi^3$ and $\phi^4$ were derived in~\cite{Paulos:2011ie,Nandan:2011wc}. On the other hand, the rules for $A^{\bar\mu}\phi\overleftrightarrow{\nabla}_{\bar\mu}\phi$ were not previously known; we will derive the rules following their strategy in the appendix. Here, we merely list the resulting Feynman rules. For a given Witten diagram,
\begin{itemize}
\item Associate auxiliary momenta to each propagator such that momentum is conserved at each vertex. The momentum $p_i$ of a bulk-to-boundary propagator satisfies $-p_i^2=\tau_i$, the conformal twist $\Delta_i-J_i$ of the $i$-th operator. Lorentz products of auxiliary momenta are related to Mellin variables through $\delta_{ij}=p_i\cdot p_j$.
\item To every propagator with auxiliary momentum $p$ corresponding to the exchange of an operator with twist $\tau$, assign a non-negative integer (descendant level) $m$. Bulk-to-boundary propagators have descendant level 0. Each propagator contributes
\begin{equation*}
    \frac{-2}{m!\Gamma(\tau+m+1-h)}\frac1{(-p^2-\tau)-2m}.
\end{equation*}
\item Every $\phi^k$ vertex joining $k$ scalar propagators with descendant levels $m_1,\dots,m_k$ contributes \cite{Paulos:2011ie}
\begin{equation*}
\begin{aligned}
    V^{\Delta_1,\dots,\Delta_k}_{m_1,\dots,m_k}
    &=\sum_{n_1=0}^{m_1} \dots \sum_{n_k=0}^{m_k} \Gamma\!\left[\frac{\Sigma\Delta-d}{2}+{\textstyle \sum_i n_i}\right] \prod_{i=1}^k \frac{(-m_i)_{n_i}}{n_i!(1-h+\Delta_i)_{n_i-m_i}}
\end{aligned}  
\end{equation*}
\item Every $A^{\bar\mu}\phi\overleftrightarrow{\nabla}_{\bar\mu}\phi$ vertex joining scalar propagators with descendant levels $m_L,m_R$ and a vector bulk-to-boundary propagator contributes
\begin{equation*}
    V^{\Delta_L,\Delta_R,\Delta_v+1}_{m_L,m_R,0} \times\left[\texttt{zp[$L$]}\left(\frac{m_R-m_L}2+\frac{\Delta_v-1}4\right)-\texttt{zp[$R$]}\left(\frac{m_L-m_R}2+\frac{\Delta_v-1}4\right)\right],
\end{equation*}
where $L,R$ are the total inflow of auxiliary momenta from the left and right.
\item Sum over all possible assignments of descendant levels associated with bulk-to-bulk propagators.
\end{itemize}

Let us be more concrete. For instance, consider the following terms in $M_8^{(s)}({\rm ST})$:
\begin{equation*}
    M_8^{(s)}({\rm ST})=\frac{-32}{\mathcal X_{14}\mathcal X_{16}}+\frac{-32}{\mathcal X_{14}(\mathcal X_{16}-2)}+\frac{-96}{(\mathcal X_{14}-2)(\mathcal X_{16}-2)}+\frac{-32}{\mathcal X_{13}\mathcal X_{14}\mathcal X_{16}\mathcal X_{17}}+\cdots
\end{equation*}
We can draw their corresponding Witten diagrams (with descendant-level assignments) as follows\footnote{We use blue double-lines to denote level-1 propagators and orange triple-lines to denote level-2 propagators.}:
\begin{equation*}
    \begin{array}{cccc}
        \includegraphics[width=0.2\textwidth]{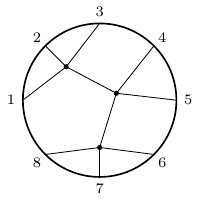}&\includegraphics[width=0.2\textwidth]{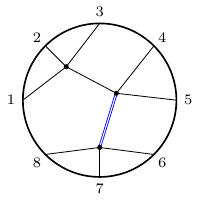}&\includegraphics[width=0.2\textwidth]{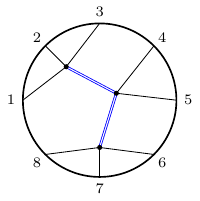}&\includegraphics[width=0.2\textwidth]{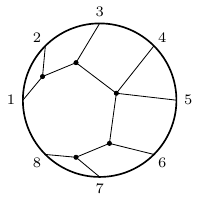} \\
        \downarrow&\downarrow&\downarrow&\downarrow\\
        -32&-32&-96&-32
    \end{array}
\end{equation*}
Using the Feynman rules described above, we indeed obtain the correct coefficient. For example, consider the third diagram above:
\begin{equation*}
    \frac{-2}{(0!)^2}\frac{-2}{(0!)^2}(-2)^3V^{2222}_{0001}V^{2222}_{0011}V^{2222}_{0001}=-96.
\end{equation*}
We can similarly check the following diagram-coefficient pair for higher-point single-trace amplitudes:
\begin{align*}
    \includegraphics[width=0.19\textwidth,align=c]{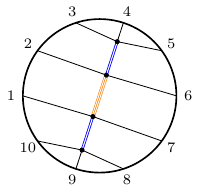}&\to\frac{(-2)^7}{(1!)^2(2!)^2(1!)^2}V^{2222}_{0001}V^{2222}_{0012}V^{2222}_{0012}V^{2222}_{0001}=-512,\\
    \includegraphics[width=0.19\textwidth,align=c]{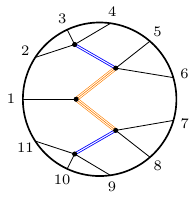}&\to\frac{(-2)^8}{(1!)^2(2!)^2(2!)^2(1!)^2}V^{2222}_{0001}V^{2222}_{0012}V^{222}_{022}V^{2222}_{0012}V^{2222}_{0001}=1024,\\
    \includegraphics[width=0.19\textwidth,align=c]{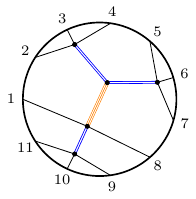}&\to\frac{(-2)^8}{(1!)^2(1!)^2(1!)^2(2!)^2}V^{2222}_{0001}V^{2222}_{0012}V^{222}_{112}V^{2222}_{0001}V^{2222}_{0001}=1024,\\
    \includegraphics[width=0.19\textwidth,align=c]{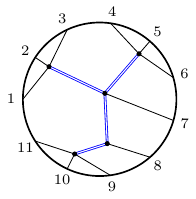}&\to\frac{(-2)^8}{(1!)^2(1!)^2(1!)^2(1!)^2}(V^{2222}_{0001})^3V^{2222}_{0111}V^{222}_{011}=2816,\\
    \includegraphics[width=0.19\textwidth,align=c]{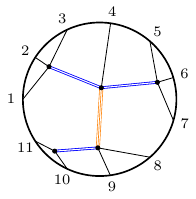}&\to\frac{(-2)^9}{(1!)^6(2!)^2}(V^{2222}_{0001})^3V^{2222}_{0112}V^{2222}_{0012}=-14336,\\
    \includegraphics[width=0.19\textwidth,align=c]{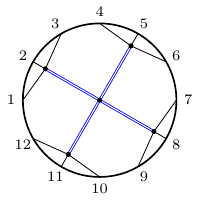}&\to\frac{(-2)^9}{(1!)^8}(V^{2222}_{0001})^4V^{2222}_{1111}=-27136.\\
\end{align*}

For the single-gluon amplitude $M_5^{(v)}({\rm ST})$,
\begin{align*}
    M_5^{(v)}({\rm ST})&=\frac{2(\zp{123}-\zp{4})}{\mathcal X_{14}}+\frac{2(\zp{1}-\zp{234})}{\mathcal X_{25}}+\frac{-4\zp{4}}{\mathcal X_{14}-2}+\frac{4\zp{1}}{\mathcal X_{25}-2}\\
    &+\frac{2(\zp{12}-\zp{34})}{\mathcal X_{13}\mathcal X_{35}}+\frac{2(\zp{123}-\zp{4})}{\mathcal X_{14}\mathcal X_{24}}+\cdots
\end{align*}
\begin{figure}[H]
    \centering
    \includegraphics[width=0.2\textwidth,align=c]{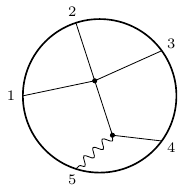}
    \caption{A typical diagram contributing to $M_5^{(v)}({\rm ST})$.}
    \label{fig:5v}
\end{figure}
Consider the diagram Figure~\ref{fig:5v}. The propagator yields a factor of $\frac{-2}{(0!)^2}=-2$, while the $\phi^4$ vertex gives $-2V^{2222}_{0000}=-2$. For the gluon vertex, to its left we have total momentum inflow $p_1+p_2+p_3$, so that $\zp{L}=\zp{123}$ and $\zp{R}=\zp{4}$. Using the Feynman rule described earlier, it gives $V^{224}_{000}=\frac{\zp{123}-\zp{4}}2$ (omitting the overall $i\sqrt{2/3}$ normalization). Putting everything together, we see that this diagram contributes $\frac{2(\zp{123}-\zp{4})}{\mathcal X_{14}}$. For another term, $\frac{-4\zp{4}}{\mathcal X_{14}-2}$, we have $(m_L,m_R)=(1,0)$ so that only the $\zp{L}=\zp{123}$ term have vanishing prefactor. One can similarly check that the Feynman rules give the correct prefactor for all other terms as well.

Some higher-point diagrams contributing to $M_n^{(v)}({\rm ST})$ are:
\begin{equation*}
    \begin{array}{ccc}
        \includegraphics[width=0.2\textwidth,align=c]{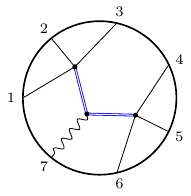} & \includegraphics[width=0.2\textwidth,align=c]{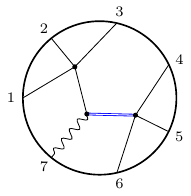} & \includegraphics[width=0.2\textwidth,align=c]{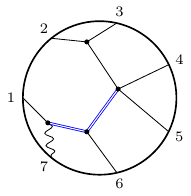} \\
        \downarrow & \downarrow & \downarrow \\
        24(\zp{123}-\zp{456}) & 16\zp{123} & -16\zp{1} \\\\
        \hline\\
        \includegraphics[width=0.2\textwidth,align=c]{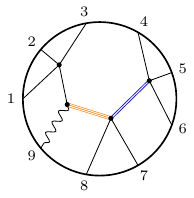} & \includegraphics[width=0.2\textwidth,align=c]{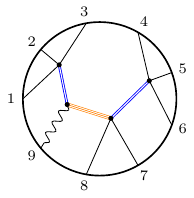} \\
        \downarrow & \downarrow \\
        0 & 256\zp{123}
    \end{array}
\end{equation*}

\subsection{Universal behavior analogous to collinear and soft limits}
Scattering amplitudes of massless particles in flat space exhibit two well-known universal behaviours, namely factorizations  under collinear and soft limits. In this subsection we show that AdS supergluon amplitudes and spinning amplitudes have similar behavior despite that we do not have ``massless particles" here. 

Recall that in the collinear limit, {\it i.e.} two particles become collinear ($s_{ij}=(p_{i}+p_{j})^2\to 0$), the $n$-point scattering amplitude factorizes into $(n{-}1)$-point amplitude and a universal factor called {\it splitting factor}, while in the soft limit, {\it i.e.} the momentum of a particle become soft ($p_{i}\to 0$), the $n$-point scattering amplitude also factorizes into $(n{-}1)$-point amplitude and a universal factor called {\it soft factor}. The collinear/soft behaviours origin from singularities of the scattering amplitudes in the flat space (c.f.~\cite{Dixon:1996wi}). 

In AdS space, the singularities of scattering amplitudes arise from the Operator Product Expansion(OPE) limit, {\it i.e.} the position of two particles becomes close $x_{i,j}^2=(x_{i}-x_{j})^{2}\to 0$. The OPE limit is similar to the collinear limit.  In the Mellin representation of AdS amplitude, the leading twist contribution to the OPE limit ($x_{1}\to x_{2}$) corresponds to the Mellin variable $\mathcal{X}_{1,3}\to 0$, leading to factorization of the Mellin amplitude. For $n$-point supergluon Mellin amplitude, when $\mathcal{X}_{1,3}\to 0$, the 
factorization of Mellin amplitude gives
\begin{equation}
     \mathop{\rm Res}_{\mathcal X_{13}=0}\mathcal M_n^{(s)}=\mathcal N_s~\texttt{glueR}\left(\mathcal M_{12I}^{(s)}\mathcal M_{3\cdots nI}^{(s)}\right)+\mathcal N_v\sum_{a=1}^{2}\sum_{i=3}^n\delta_{ai}\mathcal M_{12I}^{(v)a}\mathcal M_{3\cdots nI}^{(v)i}\,,
\end{equation}
where the residue of the Mellin amplitude at $\mathcal{X}_{13}=0$ has two contributions: scalar exchange and gluon exchange. To distinguish between these two exchanges, we consider a limit for identifying the SU$(2)_R$ spinors $v_{1}$ and $v_{2}$ as $v_{I}$, then $v_{1}\cdot v_{2}= \langle 12\rangle\to v_{I}^2=0$. In this limit, only the incompatible R-structure (i.e., scalar exchange) proportional to $\langle12\rangle$ but not $\langle12\rangle^2$ contributes to the leading order.
In this way we obtain the ``leading collinear limit" for supergluon amplitudes analogous to collinear limit in the flat space.
\begin{equation}
    \mathcal M_n^{(s)}\xrightarrow[\mathcal X_{13}\to0]{\langle 12\rangle\to 0}~\mathcal N_s~\frac{\langle 12\rangle}{\mathcal X_{13}}\times \mathcal M_{3\cdots nI}^{(s)}\,,
\end{equation}
where the $n$-point supergluon amplitude $\mathcal M_n^{(s)}$ factorizes into the $(n{-}1)$-point supergluon amplitude $\mathcal M_{3\cdots nI}^{(s)}$ and the  ``{\it splitting factor}" $\mathcal N_s \frac{\langle 12\rangle}{\mathcal X_{13}}$ for supergluon amplitudes. Compared with the color-ordered scalar amplitude with Tr$(\phi^3)$ interaction in the flat space, the amplitude factors in the collinear limit, $s_{12}\to0$,
\begin{equation}
    \mathcal A_n^{\phi^3}(1,\ldots,n)\xrightarrow[s_{12}\to0]{}\frac{1}{s_{12}}\mathcal A_{n{-}1}^{\phi^3}(3\cdots nI)\,,
\end{equation}
where $ \mathcal A_n^{\phi^3}(1,\ldots,n)$ denotes the $n$-point Tr$(\phi^3)$ tree-level amplitude in the color ordering of $(1,\ldots,n)$.

The derivation for soft limit is very similar. In flat space the soft singularity arises from all poles of the form $s_{i,n}$ when $p_{n}\to 0$. For color-ordered amplitudes, only $s_{n{-}1,n}$ and $s_{1,n}$ contribute. The ``leading  soft " limit of supergluon amplitude in AdS, arises from the limit of $\mathcal X_{1(n{-}1)}/\mathcal X_{2n}\to0$ with identifying $\mathcal X_{jn}$ and  $\mathcal X_{1j}$ for any $j$. Similar to collinear limit, we need to take an operator for the SU$(2)_R$ spinors $v_{n}$, specifically any $V_{\cdots n}\to V_{\cdots}$, in order to suppress the contribution from gluon exchange. In the gluon exchange, the R-structures of $M_{(n{-}1)nI}^{(v)a}/M_{1nI}^{(v)a}$ always vanish ($V_{(n{-}1)n}/V_{1n}\to V_{n{-}1}/V_{1}=0$). The second terms from the $\texttt{glueR}$ of scalar exchange also vanish for the same reason:
\begin{equation}
    \mathcal M_n^{(s)}\xrightarrow[\mathcal X_{1(n{-}1)}/\mathcal X_{2n}\to0]{V_{\cdots n}\to V_{\cdots}}\mathcal N_s \frac{1}{\mathcal X_{1(n{-}1)}}\mathcal M_{1\cdots (n{-}1)}^{(s)}+\mathcal N_s \frac{1}{\mathcal X_{2n}}\mathcal M_{2\cdots n}^{(s)}=\mathcal N_s\left( \frac{1}{\mathcal X_{1(n{-}1)}}+\frac{1}{\mathcal X_{2n}}\right)\mathcal M_{1\cdots (n{-}1)}^{(s)}\,,
\end{equation}
where in the second equality, we identify $\mathcal X_{jn}$ and  $\mathcal X_{1j}$.  In the ``leading soft limit", the $n$-point supergluon amplitude $\mathcal M_n^{(s)}$ factorizes into $(n{-}1)$-point supergluon amplitude $\mathcal M_{1\cdots (n{-}1)}^{(s)}$ and the  ``{\it soft factor}" $\mathcal N_s\left( \frac{1}{\mathcal X_{1(n{-}1)}}+\frac{1}{\mathcal X_{2n}}\right)$ for supergluon amplitudes. This can be compared with the color-ordered scalar amplitude with Tr$(\phi^3)$ interaction in the flat space:
\begin{equation}
\mathcal A_n^{\phi^3}(1,\ldots,n)\xrightarrow[s_{n{-}1, n}\to0]{s_{1n}\to0}(\frac{1}{s_{n{-}1, n}}+\frac{1}{s_{1n}})\mathcal A_{n{-}1}^{\phi^3}(1\cdots (n{-}1))\,.
\end{equation}

Very nicely, similar arguments can be applied for the spinning amplitudes. For collinear limit,

\begin{equation}
     \mathop{\rm Res}_{\mathcal X_{1(n-{1})}=0}\mathcal M_n^{(v)}=\mathcal N_s\texttt{glueR}\left(\mathcal M_{12\cdots (n{-}2)I}^{(s)}\mathcal M_{I(n{-}1)n}^{(v)}\right)\,,
\end{equation}
where the $\texttt{glueR}$~\eqref{eq:glueR} is simplified, because the R-structure of $\mathcal M_{I(n{-}1)n}^{(v)}$ is $V_{I(n{-}1)}$. After the $\texttt{glueR}$, the R-structure of  $\mathcal M_{12\cdots (n{-}2)I}^{(s)}$ becomes the same R-structure of $\mathcal M_{12\cdots (n{-}2)(n{-}1)}^{(s)}$.

\begin{equation}
    \mathcal M_n^{(v)}\xrightarrow[\mathcal X_{1(n{-}1)}\to0]{}\Tilde{\mathcal N}_s \frac{\zp{n{-}1}-\zp{12\cdots n{-}2}}{\mathcal X_{1(n{-}1)}}\mathcal M_{12\cdots (n{-}2)I}^{(s)}\,,
\end{equation}
where $\Tilde{\mathcal N}_s$ absorbs the overall factor from the 3-point spinning amplitude, $\Tilde{\mathcal N}_s=\mathcal N_s \frac{i\sqrt{2/3}}{2}$. The $n$-point spinning amplitude $\mathcal M_n^{(v)}$ factorizes into $(n{-}1)$-point supergluon amplitude $\mathcal M_{12\cdots (n{-}2)I}^{(s)}$ and the  ``{\it splitting factor}" $\Tilde{\mathcal N}_s \frac{\zp{n{-}1}-\zp{12\cdots n{-}2}}{\mathcal X_{1(n{-}1)}}$ for spinning amplitudes. Compared with the color-ordered scalar amplitude minimal coupled with one gluon in the flat space, the amplitude factors in the collinear limit, $s_{(n{-}1)n}\to0$,
\begin{equation}
    \mathcal A_n^{{\rm YM}+\phi^3}(1,\ldots,n{-}1;n)\xrightarrow[s_{(n{-}1)n}\to0]{}\frac{\epsilon_{n}\cdot p_{n{-}1}-\epsilon_{n}\cdot p_{12\ldots n{-}2}}{s_{(n{-}1)n}}\mathcal A_{n{-}1}^{\phi^3}(12\cdots (n{-}2)I)\,,
\end{equation}
where $\mathcal A_n^{{\rm YM}+\phi^3}(1,\ldots,n{-}1;n)$ denotes the $n$-point tree-level amplitude with the $n-1$ scalar in the color ordering $(1,\ldots,n{-}1)$ and one gluon with the polarization vector $\epsilon_{n}$, and here the $p_{12\ldots n{-}2}=\sum_{i=1}^{n{-}2}p_{i}$.

The soft limit of the spinning amplitudes can be similarly derived as,
\begin{equation}
\begin{aligned}
    & \mathcal M_n^{(v)}\xrightarrow[\mathcal X_{1(n{-}1)}\to0]{\mathcal X_{2n}\to0}\Tilde{\mathcal N}_s \frac{\zp{n{-}1}-\zp{12\cdots n{-}2}}{\mathcal X_{1(n{-}1)}}\mathcal M_{12\cdots (n{-}1)}^{(s)}+\Tilde{\mathcal N}_s \frac{\zp{2\cdots n{-}1}-\zp{1}}{\mathcal X_{2n}}\mathcal M_{2\cdots n}^{(s)}\\&=
     \Tilde{\mathcal N}_s \left(\frac{\zp{n{-}1}-\zp{12\cdots n{-}2}}{\mathcal X_{1(n{-}1)}}+\frac{\zp{2\cdots n{-}1}-\zp{1}}{\mathcal X_{2n}}\right) \mathcal M_{1\cdots n{-}1}^{(s)}\,,
\end{aligned}
\end{equation}
where in the second equality, we identify $\mathcal X_{jn}$ and  $\mathcal X_{1j}$.  In the ``leading soft'' limit, the $n$-point spinning amplitude $\mathcal M_n^{(v)}$ factorizes into $(n{-}1)$-point supergluon amplitude $\mathcal M_{1\cdots (n{-}1)}^{(s)}$ and the  ``{\it soft factor}" $\Tilde{\mathcal N}_s \left(\frac{\zp{n{-}1}-\zp{12\cdots n{-}2}}{\mathcal X_{1(n{-}1)}}+\frac{\zp{2\cdots n{-}1}-\zp{1}}{\mathcal X_{2n}}\right) $ for spinning amplitudes. Compared with the color-ordered scalar amplitude minimal coupled with one gluon in the flat space, the amplitude factors in the soft limit, $p_{n}\to0$, which equal to the $s_{1,n}\to0,s_{(n{-}1)n}\to0$ in this color ordering,

\begin{equation}
    \mathcal A_n^{{\rm YM}+\phi^3}\xrightarrow[s_{(n{-}1)n}\to0]{s_{1n}\to0}\left(\frac{\epsilon_{n}\cdot p_{n{-}1}-\epsilon_{n}\cdot p_{12\ldots n{-}2}}{s_{(n{-}1)n}}+\frac{\epsilon_{n}\cdot p_{2\ldots n{-}1}-\epsilon_{n}\cdot p_{1}}{s_{1n}}\right)\mathcal A_{n{-}1}^{\phi^3}(12\cdots (n{-}1))\,.
\end{equation}
 \section{Conclusions and Outlook}~\label{sec:outlook}
In this paper we have elaborated on a new, recursive method proposed in~\cite{Cao:2023cwa} for computing the holographic correlator of $n$ half-BPS operators with conformal dimension $\Delta=2$, or equivalently the $n$-point supergluon amplitudes in AdS${}_5\times S^3$ at tree level. By extracting $(n{-}1)$-point spinning amplitudes from the $n$-point supergluon one, in sec.~\ref{sec:construct} we have shown that these lower-point amplitudes suffice to determine the $(n{+}2)$-point supergluon amplitude from various factorizations and flat-space limit. This amounts to an improved proof for the constructibility of supergluon amplitudes to all multiplicities, which then immediately implies constructibility of spinning amplitudes as a byproduct.  

Apart from providing valuable data for the study of AdS${}_5/$ CFT${}_4$ holography, it is advantageous to have these beautiful expressions of $n$-point Mellin amplitudes (with a given color-ordering, expanded in a basis of R-symmetry structures~\cite{Cao:2023cwa}), which provide an ideal playground for studying hidden structures of AdS amplitudes. One can view our results as the simplest generalizations of the flat-space ``scalar-scaffolded" gluon amplitudes~\cite{Arkani-Hamed:2023jry} to AdS background. In addition to presenting explicit results in sec.~\ref{sec:result}, we have revealed several interesting new structures to all $n$. Most importantly, we have argued that the general structure of descendant poles satisfy 
the ``half-circle rule" (see~\ref{conj:truncation}). For the single-trace case which we have computed to all $n$, we have derived remarkably simple ``Feynman rules": for supergluon case they are given by $\phi^3+ \phi^4$ and for spinning case with additional interactions of gluon with 2 $\phi$s (see app.~\ref{sec:rule_derivation} for more details); we have also shown the analogy of collinear and soft limits for AdS amplitudes. 

Our work has opened up many new avenues for future investigations. 
First of all, within our setting of supergluon tree amplitudes in $AdS_5\times S^3$, there are very interesting questions one can pursue. Although we have proved constructibility for all $n$, it is still important to explicitly run our algorithm for higher points, {\it i.e.} for multi-trace amplitudes with $n\geq 10$, and in particular it would be highly desirable to develop recursion relations for accomplishing this, which already exist for single-trace amplitudes. Relatedly, it would be interesting to derive ``Feynman rules" beyond single-trace cases (both for supergluon and spinning amplitudes), which may be connected with other methods for evaluating Witten diagrams in Mellin space~\cite{Chu:2023kpe, Chu:2023pea, Li:2023azu,Paulos:2011ie}. On the other hand, an important conceptual problem is how to extract multi-gluon spinning amplitudes, since we have yet to develop a formalism with simple factorization properties~\cite{Goncalves:2014rfa} to treat such multi-spinning amplitudes in Mellin space nicely.

From the boundary CFT perspective, the origin of the descendant truncation remains elusive. The large-$N$ argument for $n=4$ in~\cite{Rastelli:2017udc} seems to not generalize straightforwardly to higher multiplicity. Even if a large-$N$ argument exists, one would have to study the multi-operator OPEs in Mellin space more carefully (possibly employing polology techniques~\cite{Yuan:2018qva}) due to the appearance of length$\geq3$ poles such as $\mathcal X_{14}=2m$.

Another important direction is to study all these AdS amplitudes directly in position space, and to see if our program and even the recursion could be formulated there. For this purpose, one may consider the ``differential representation" for these higher-point amplitudes in position space~\cite{Huang:2024dck}, where nice structures have been found in four-point AdS amplitudes in various settings. In particular, the $n=6$ amplitude involves Witten diagrams that evaluate to elliptic integrals~\cite{Paulos:2011ie,Paulos:2012nu}, which certainly deserve further investigations. 

Despite all these open questions, we emphasize that these supergluon amplitudes already provide a rich collection of data and insights for future study of holographic correlator and scattering amplitudes in AdS spacetime. As we have mentioned, it is advantageous to view our results as AdS generalizations of the flat-space program of scalar-scaffolded gluons proposed in~\cite{Arkani-Hamed:2023jry}. Given that the latter is based on combinatorial structures for surfaces~\cite{Arkani-Hamed:2023mvg, Arkani-Hamed:2023lbd}, it is natural to ask if there exists a similar formulation for AdS supergluon amplitudes, and if one could connect it to the exciting progress on worldsheet formulation of AdS Veneziano amplitudes~\cite{Alday:2024yax,Alday:2024ksp} (see also \cite{Alday:2023mvu}). A number of intriguing properties have been uncovered recently for scaffolded gluons in flat space, such as zeros and new factorizations or splitting behaviors~\cite{Arkani-Hamed:2023swr, Cachazo:2021wsz, Cao:2024gln, Arkani-Hamed:2023jry}, and it would be fascinating to look for hints for such structures and properties in our AdS amplitudes. 

Beyond the simplest supergluon amplitudes, it would be very interesting to study scattering of higher Kaluza-Klein modes, which would require some extensions regarding R-symmetry structures and factorization properties. Preliminary studies show that we can at least obtain five-point KK amplitudes in this way, and it is important to develop our method further for constructing higher-point KK amplitudes. Given recent progress on hidden symmetries for four-point (reduced) correlators~\cite{Caron-Huot:2018kta,Rastelli:2019gtj,Alday:2021odx,Abl:2021mxo,Rigatos:2024yjo}, it would be highly desirable to see if such hidden symmetries could be extended to higher points and provide a ``generating function'' of KK amplitudes. Another motivation is that the observed double-copy relation~\cite{Zhou:2021gnu} relates supergluon amplitudes to supergraviton amplitudes as well as conformal scalar amplitudes which can serve as nice toy models. At least for conformal scalars such as $\phi^3$ in ${\rm AdS}_{d+1}\times S^{5-d}$, a version of ``hidden symmetry'' trivially holds.

It would also be extremely interesting to see how this entire program can be extended to supergravity amplitudes in $AdS_5\times S^5$, and especially for higher-point and/or spinning amplitudes. We remark that at least the flat-space limit for such supergravity amplitudes are known: for even points, they are given by generalization of scalar-scaffolded GR which turn out to be pion scattering amplitudes with interactions with gravity; these amplitudes are related to the YMS amplitudes by double copy~\cite{Kawai:1985xq, Bern:2008qj} and can be computed again via nice CHY formulas~\cite{Cachazo:2014xea}. It would be fascinating to explore both these flat-space amplitudes and their AdS counterpart in the future. 

\section*{Acknowledgments}
It is our pleasure to thank Zhongjie Huang, Bo Wang, Ellis Yuan and Xinan Zhou for inspiring discussions or collaboration on related projects. The work of SH is supported by the National Natural Science Foundation of China under Grant No. 12225510, 11935013, 12047503, 12247103, and by the New Cornerstone Science Foundation through the XPLORER PRIZE.

\newpage
\appendix

\section{Coefficients of single-trace Witten diagrams}\label{sec:Witten-st-Coefficients}
Let us summarize some coefficients of single-trace Witten diagrams up to $n=12$ for supergluon amplitudes, both as a check for the Feynman rules and to illustrate how non-trivial these amplitudes become for higher $n$.

\begin{table}[ht]
    \centering
    \begin{tblr}{rowspec={|[1pt]Q[l,m]|[dotted]Q[l,m,abovesep=6pt]|[1pt]}}
        Coefficient $=1$ \\
        \hspace{-8pt} \includegraphics[width=0.98\textwidth]{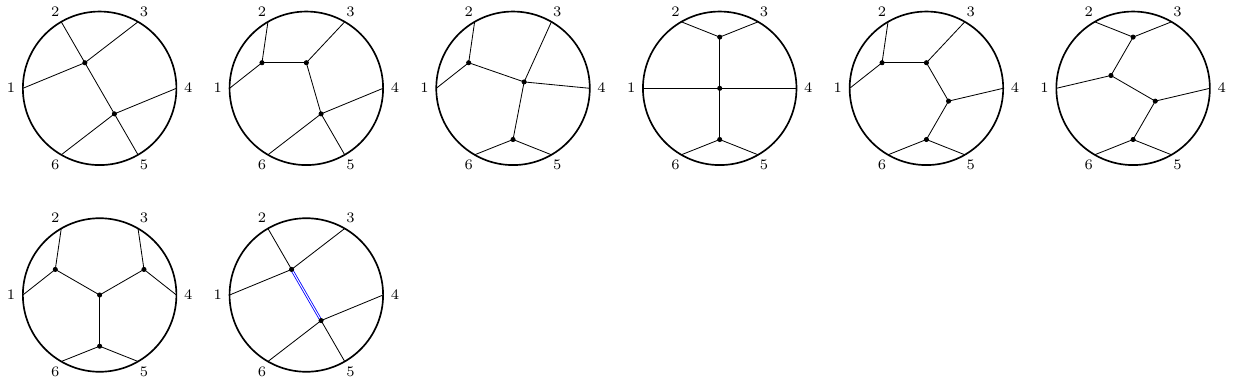}\!\! \\
    \end{tblr}
    \caption{Coefficients of Witten diagrams contributing to $M_6^{(s)}({\rm ST})/(-2)^3$.}
    \label{tab:single-trace-6pt}
\end{table}

\begin{table}[H]
    \centering
    \begin{tblr}{rowspec={|[1pt]Q[l,m]|[dotted]Q[l,m,abovesep=6pt]|[1pt]}}
        Coefficient $=1$ \\
        \hspace{-8pt} \includegraphics[width=0.98\textwidth]{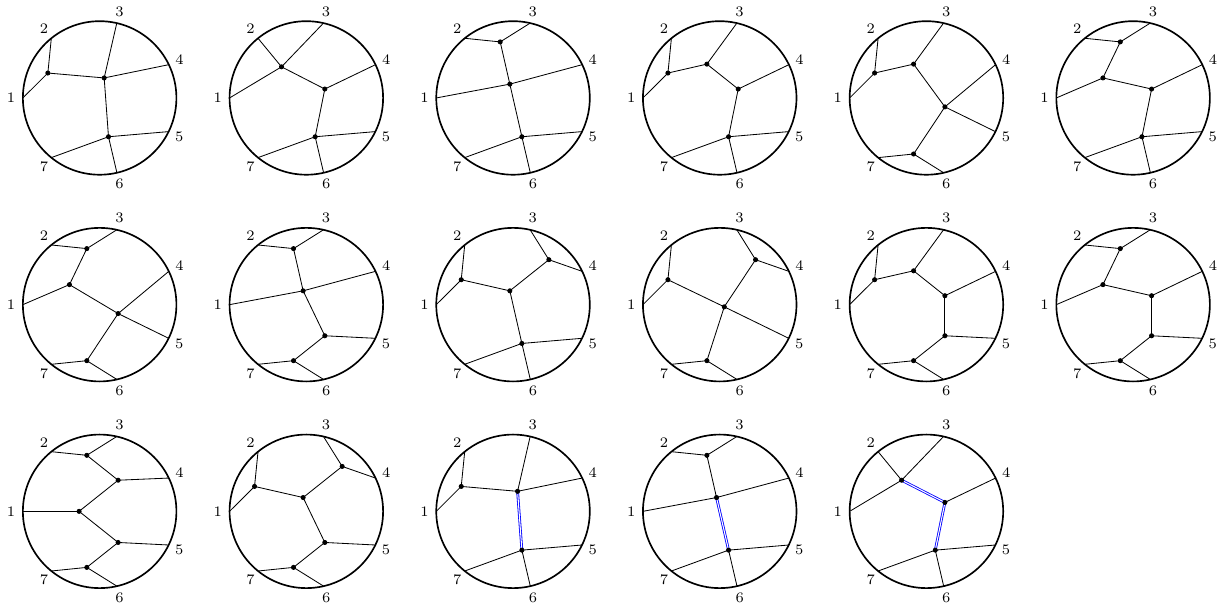}\!\! \\
    \end{tblr}
    \caption{Coefficients of Witten diagrams contributing to $M_7^{(s)}({\rm ST})/(-2)^4$.}
    \label{tab:single-trace-7pt}
\end{table}

\begin{table}[H]
    \centering
    \begin{tblr}{rowspec={|[1pt]Q[l,m]|[dotted]Q[l,m,abovesep=6pt]|Q[l,m]|[dotted]Q[l,m,abovesep=6pt]|[1pt]}}
        Coefficient $=1$ \\
        \hspace{-8pt} \includegraphics[width=0.98\textwidth]{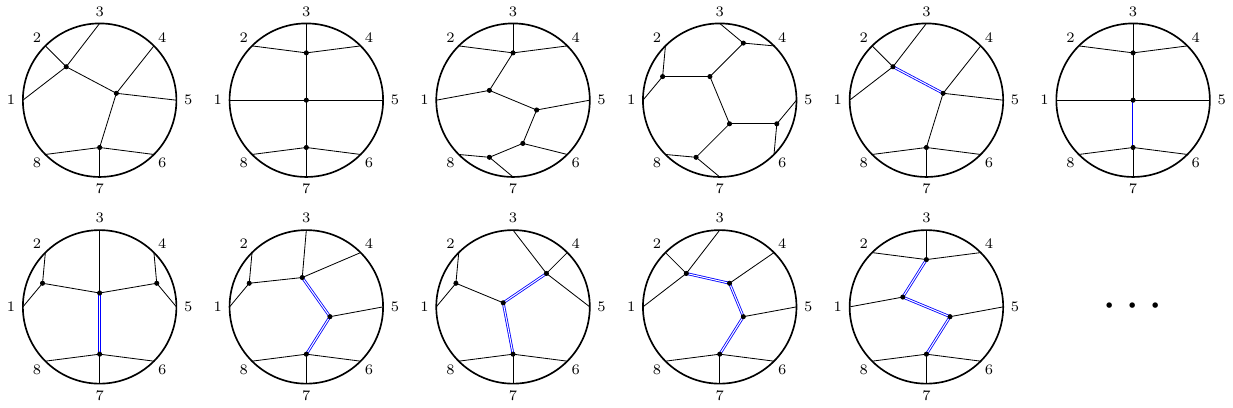}\!\! \\
        Coefficient $=3$ \\
        \hspace{-8pt} \includegraphics[width=0.98\textwidth]{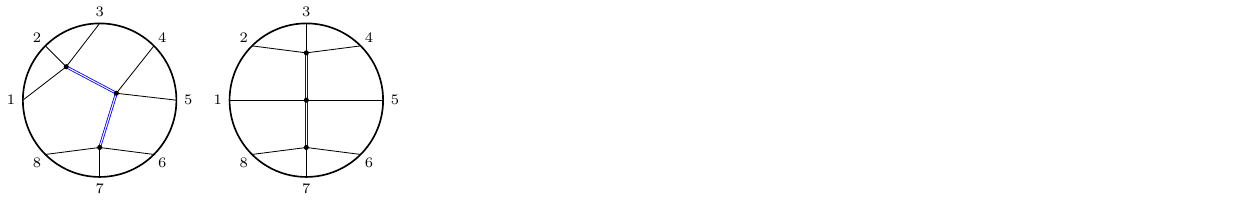}\!\! \\
    \end{tblr}
    \caption{Coefficients of Witten diagrams contributing to $M_8^{(s)}({\rm ST})/(-2)^5$.}
    \label{tab:single-trace-8pt}
\end{table}

\begin{table}[H]
    \centering
    \begin{tblr}{rowspec={|[1pt]Q[l,m]|[dotted]Q[l,m,abovesep=6pt]|Q[l,m]|[dotted]Q[l,m,abovesep=6pt]|Q[l,m]|[dotted]Q[l,m,abovesep=6pt]|[1pt]}}
        Coefficient $=1$ \\
        \hspace{-8pt} \includegraphics[width=0.98\textwidth]{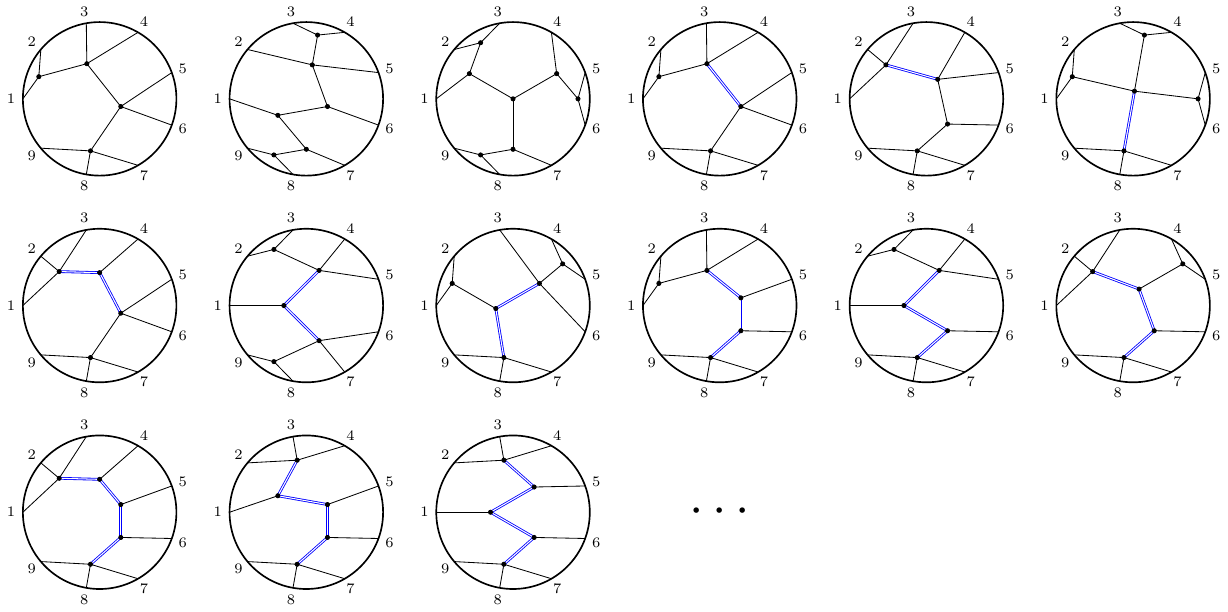}\!\! \\
        Coefficient $=2$ \\
        \hspace{-8pt} \includegraphics[width=0.98\textwidth]{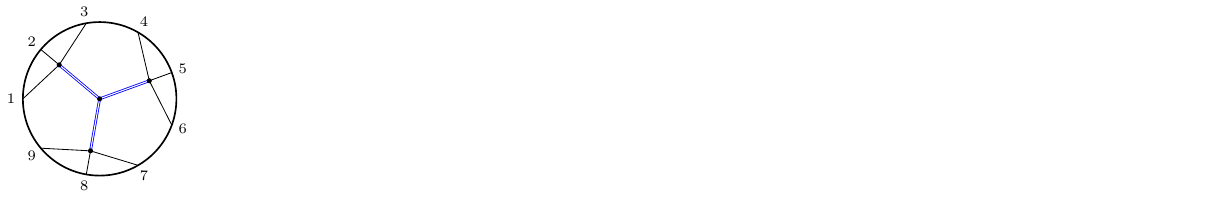}\!\! \\
        Coefficient $=3$ \\
        \hspace{-8pt} \includegraphics[width=0.98\textwidth]{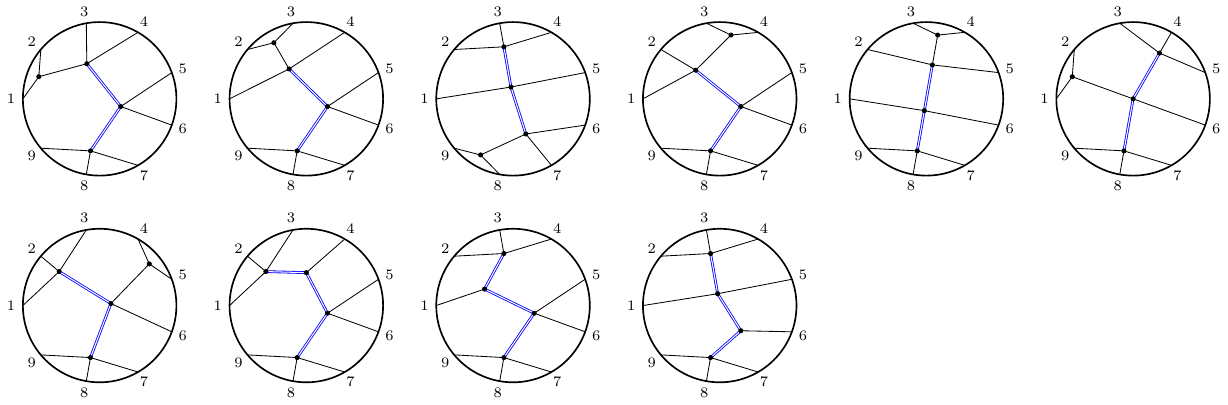}\!\! \\
    \end{tblr}
    \caption{Coefficients of Witten diagrams contributing to $M_9^{(s)}({\rm ST})/(-2)^6$.}
    \label{tab:single-trace-9pt}
\end{table}

\begin{table}[H]
    \centering
    \begin{tblr}{rowspec={|[1pt]Q[l,m]|[dotted]Q[l,m,abovesep=6pt]|Q[l,m]|[dotted]Q[l,m,abovesep=6pt]|Q[l,m]|[dotted]Q[l,m,abovesep=6pt]|Q[l,m]|[dotted]Q[l,m,abovesep=6pt]|Q[l,m]|[dotted]Q[l,m,abovesep=6pt]|Q[l,m]|[dotted]Q[l,m,abovesep=6pt]|[1pt]}}
        Coefficient $=1$ \\
        \hspace{-8pt} \includegraphics[width=0.98\textwidth]{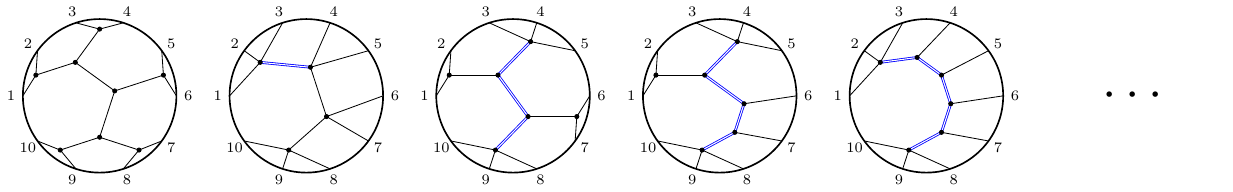}\!\! \\
        Coefficient $=2$ \\
        \hspace{-8pt} \includegraphics[width=0.98\textwidth]{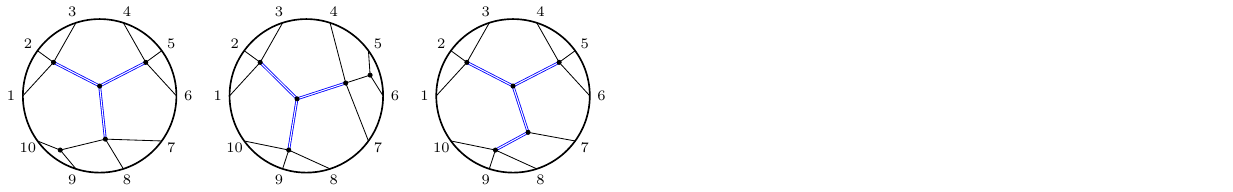}\!\! \\
        Coefficient $=3$ \\
        \hspace{-8pt} \includegraphics[width=0.98\textwidth]{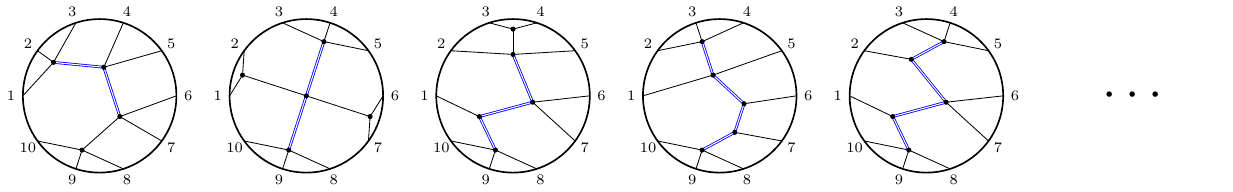}\!\! \\
        Coefficient $=4$ \\
        \hspace{-8pt} \includegraphics[width=0.98\textwidth]{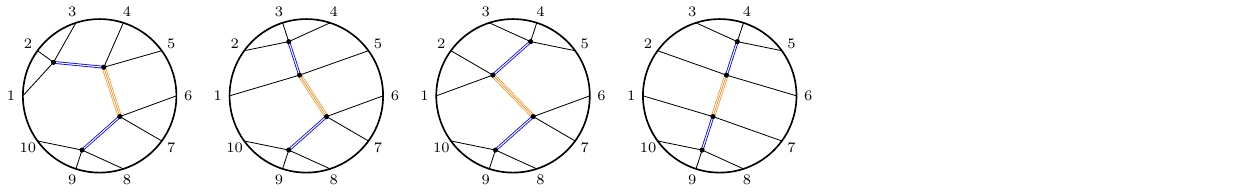}\!\! \\
        Coefficient $=9$ \\
        \hspace{-8pt} \includegraphics[width=0.98\textwidth]{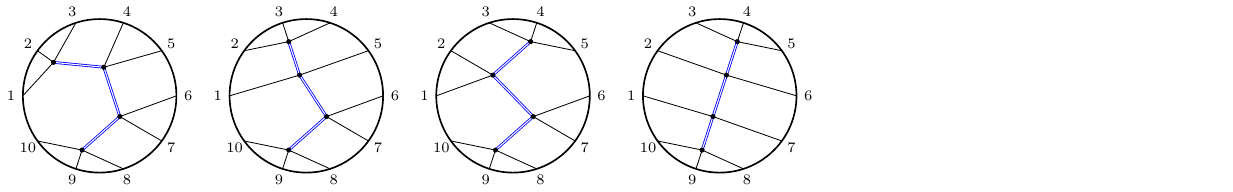}\!\! \\
        Coefficient $=11$ \\
        \hspace{-8pt} \includegraphics[width=0.98\textwidth]{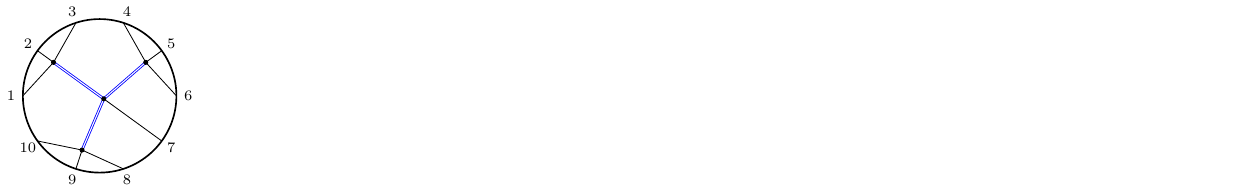}\!\! \\
    \end{tblr}
    \caption{Coefficients of Witten diagrams contributing to $M_{10}^{(s)}({\rm ST})/(-2)^7$.}
    \label{tab:single-trace-10pt}
\end{table}

\begin{table}[H]
    \centering
    \begin{tblr}{rowspec={|[1pt]Q[l,m]|[dotted]Q[l,m]|Q[l,m]|[dotted]Q[l,m,abovesep=6pt]|Q[l,m]|[dotted]Q[l,m,abovesep=6pt]|Q[l,m]|[dotted]Q[l,m,abovesep=6pt]|Q[l,m]|[dotted]Q[l,m,abovesep=6pt]|Q[l,m]|[dotted]Q[l,m,abovesep=6pt]|[1pt]}}
        Coefficient $=1,3$ \\
        $\cdots$ \\
        Coefficient $=2$ \\
        \hspace{-8pt} \includegraphics[width=0.98\textwidth]{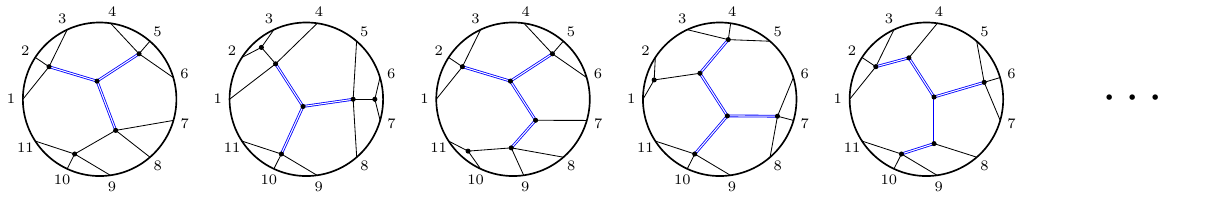}\!\! \\
        Coefficient $=4$ \\
        \hspace{-8pt} \includegraphics[width=0.98\textwidth]{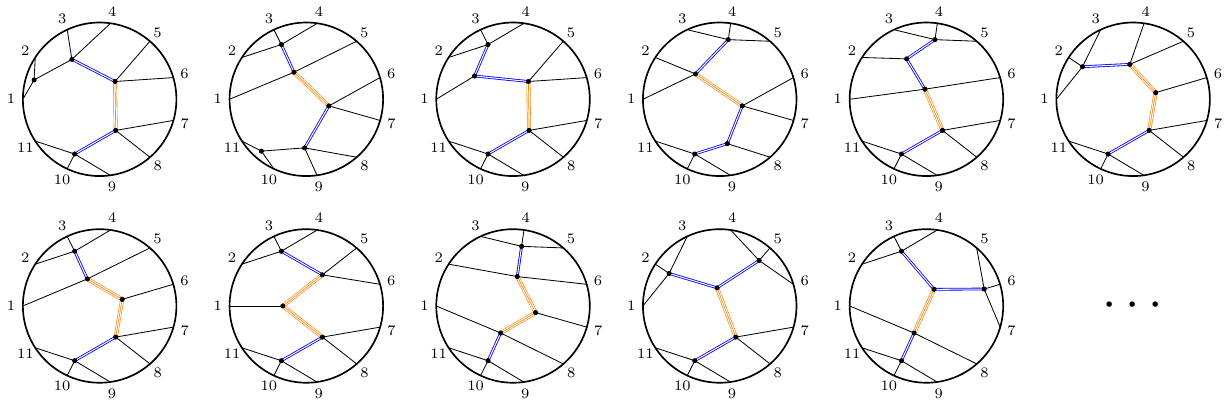}\!\! \\
        Coefficient $=6$ \\
        \hspace{-8pt} \includegraphics[width=0.98\textwidth]{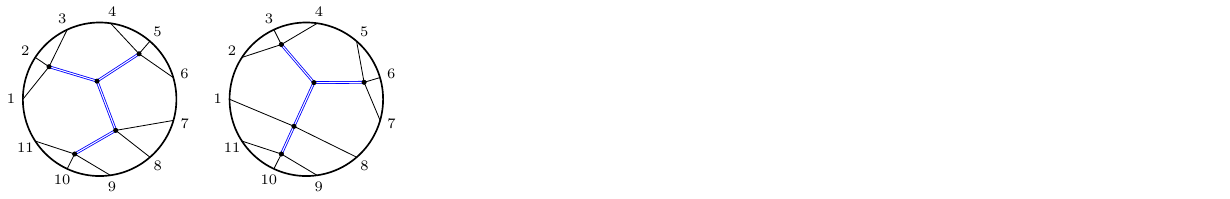}\!\! \\
        Coefficient $=9$ \\
        \hspace{-8pt} \includegraphics[width=0.98\textwidth]{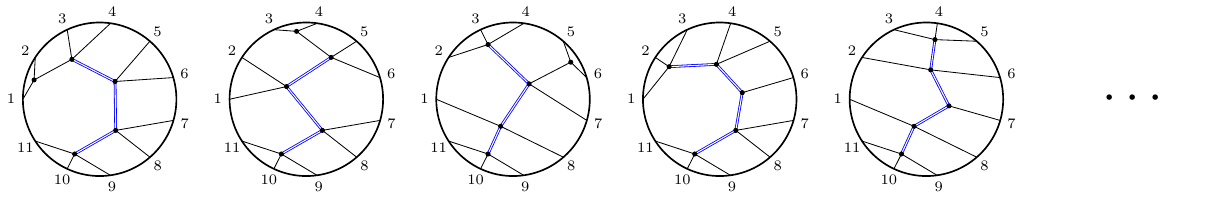}\!\! \\
        Coefficient $=11$ \\
        \hspace{-8pt} \includegraphics[width=0.98\textwidth]{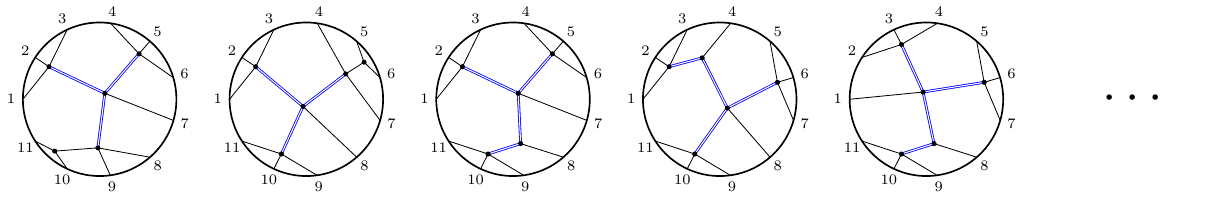}\!\! \\
    \end{tblr}
    \caption{Coefficients of Witten diagrams contributing to $M_{11}^{(s)}({\rm ST})/(-2)^8$.}
    \label{tab:single-trace-11pt}
\end{table}

\begin{table}[H]
    \centering
    \begin{tblr}{rowspec={|[1pt]Q[l,m]|[dotted]Q[l,m]|Q[l,m]|[dotted]Q[l,m,abovesep=3.7pt,belowsep=0pt]|Q[l,m]|[dotted]Q[l,m,abovesep=3.7pt,belowsep=0pt]|Q[l,m]|[dotted]Q[l,m,abovesep=3.7pt,belowsep=0pt]|Q[l,m]|[dotted]Q[l,m,abovesep=3.7pt,belowsep=0pt]|Q[l,m]|[dotted]Q[l,m,abovesep=3.7pt,belowsep=0pt]|Q[l,m]|[dotted]Q[l,m,abovesep=3.7pt,belowsep=0pt]|[1pt]}}
        Coefficient $=1,2,3,4,6,9,11$ \\
        $\cdots$ \\
        Coefficient $=12$ \\
        \hspace{-8pt} \includegraphics[width=0.98\textwidth]{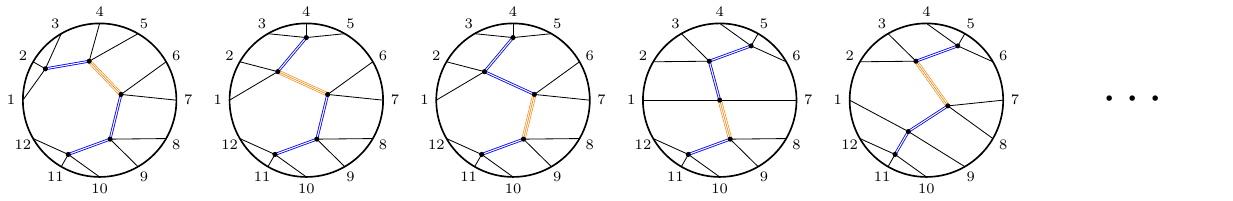}\!\! \\
        Coefficient $=20$ \\
        \hspace{-8pt} \includegraphics[width=0.98\textwidth]{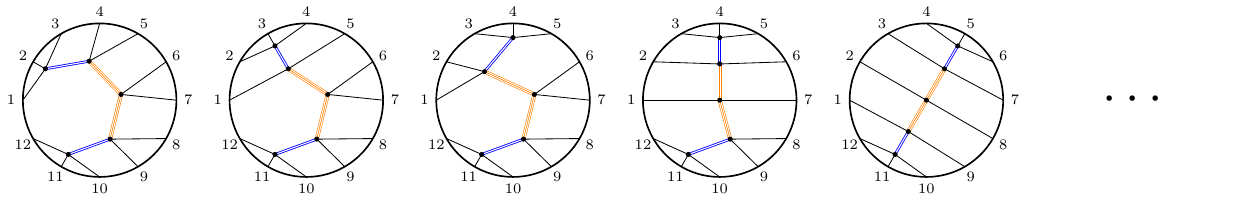}\!\! \\
        Coefficient $=27$ \\
        \hspace{-8pt} \includegraphics[width=0.98\textwidth]{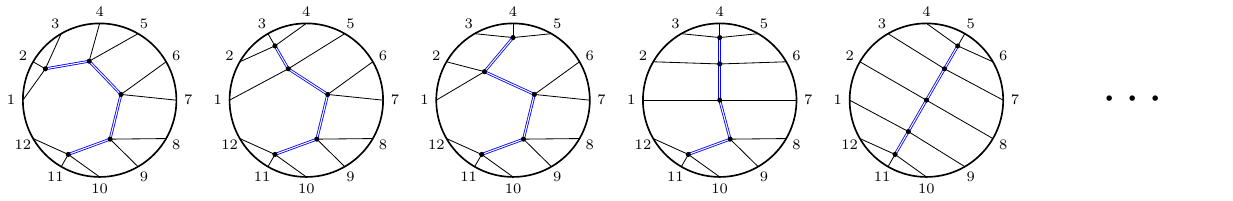}\!\! \\
        Coefficient $=28$ \\
        \hspace{-8pt} \includegraphics[width=0.98\textwidth]{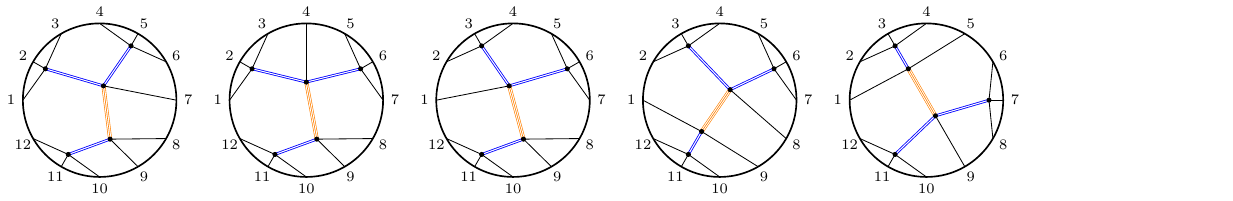}\!\! \\
        Coefficient $=33$ \\
        \hspace{-8pt} \includegraphics[width=0.98\textwidth]{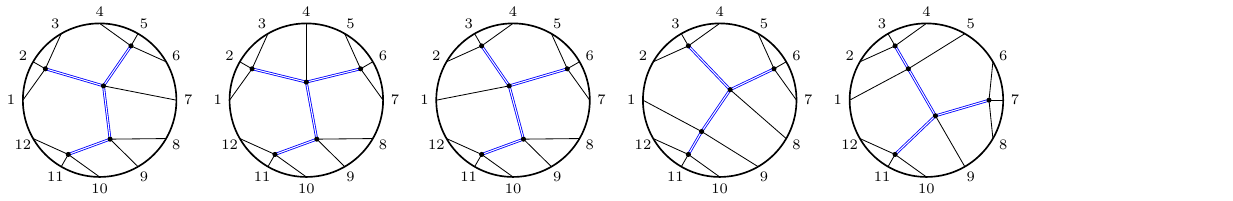}\!\! \\
        Coefficient $=53$ \\
        \hspace{-8pt} \includegraphics[width=0.98\textwidth]{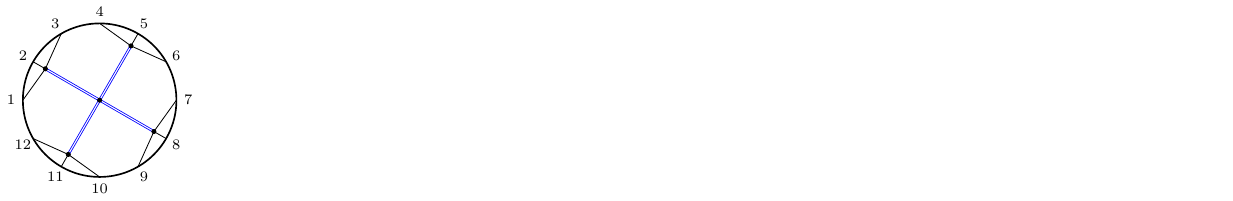}\!\! \\
    \end{tblr}
    \caption{Coefficients of Witten diagrams contributing to $M_{12}^{(s)}({\rm ST})/(-2)^9$.}
    \label{tab:single-trace-12pt}
\end{table}

\begin{table}[H]
    \centering
    \begin{tblr}{rowspec={|[1pt]Q[l,m]|[dotted]Q[l,m]|Q[l,m]|[dotted]Q[l,m,abovesep=3.7pt,belowsep=0pt]|Q[l,m]|[dotted]Q[l,m,abovesep=3.7pt,belowsep=0pt]|Q[l,m]|[dotted]Q[l,m,abovesep=3.7pt,belowsep=0pt]|Q[l,m]|[dotted]Q[l,m,abovesep=3.7pt,belowsep=0pt]|Q[l,m]|[dotted]Q[l,m,abovesep=3.7pt,belowsep=0pt]|Q[l,m]|[1pt]}}
        Coefficient $=2\zp{L}$ \\
        \hspace{-8pt} \includegraphics[width=0.98\textwidth]{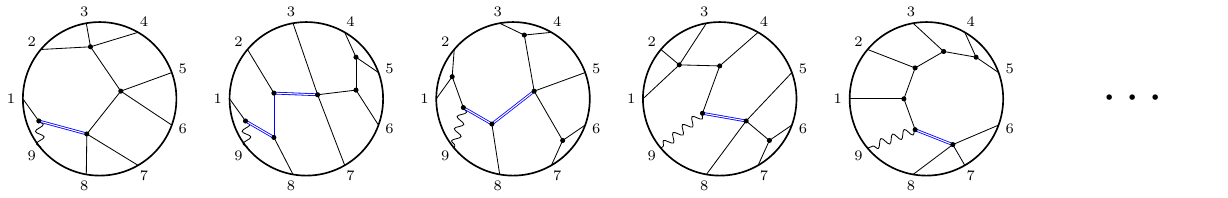}\!\! \\
        Coefficient $=4\zp{L}$ \\
        \hspace{-8pt} \includegraphics[width=0.98\textwidth]{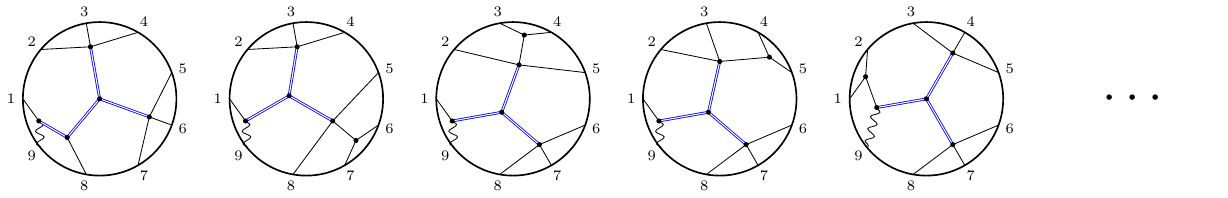}\!\! \\
        Coefficient $=6\zp{L}$ \\
        \hspace{-8pt} \includegraphics[width=0.98\textwidth]{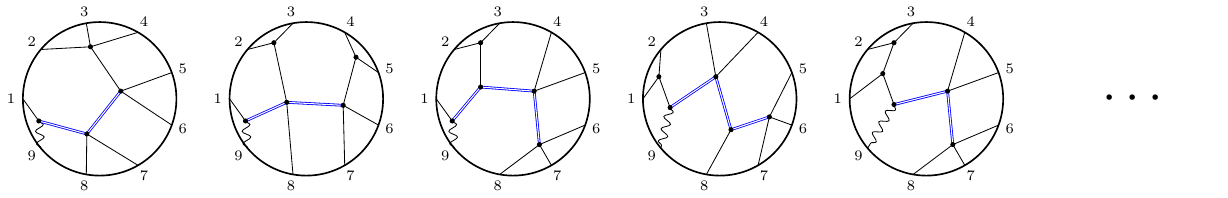}\!\! \\
        Coefficient $=8\zp{L}$ \\
        \hspace{-8pt} \includegraphics[width=0.98\textwidth]{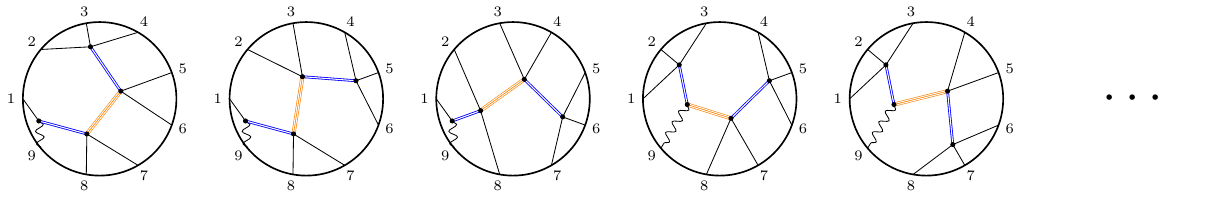}\!\! \\
        Coefficient $=18\zp{L}$ \\
        \hspace{-8pt} \includegraphics[width=0.98\textwidth]{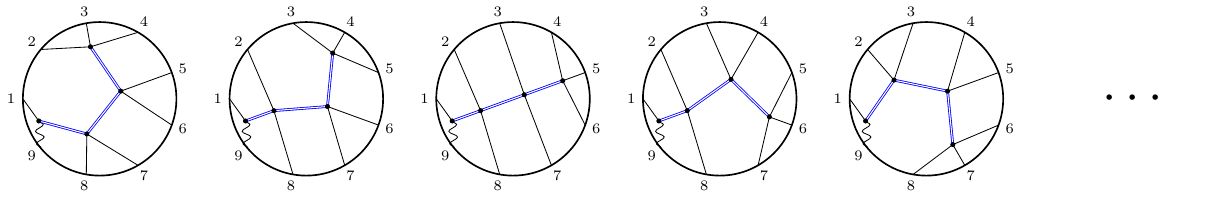}\!\! \\
        Coefficient $=22\zp{L}$ \\
        \hspace{-8pt} \includegraphics[width=0.98\textwidth]{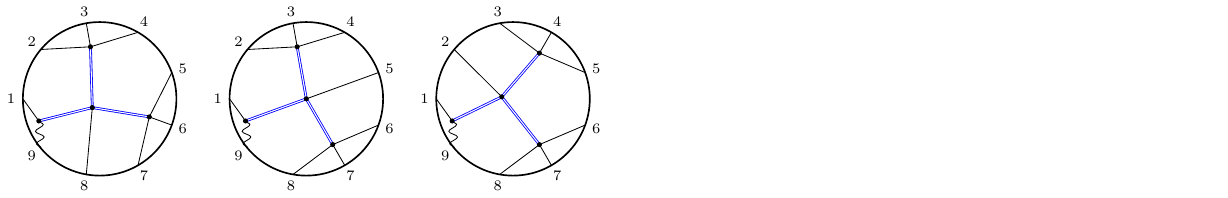}\!\! \\
        Continued on the next page
    \end{tblr}
    \caption{Coefficients of Witten diagrams contributing to $M_{9}^{(v)}({\rm ST})/2^5$.}
    \label{tab:single-trace-gluon-9pt-1}
\end{table}

\begin{table}[H]
    \centering
    \begin{tblr}{rowspec={|[1pt]Q[l,m]|Q[l,m]|[dotted]Q[l,m]|Q[l,m]|[dotted]Q[l,m,abovesep=3.7pt,belowsep=0pt]|Q[l,m]|[dotted]Q[l,m,abovesep=3.7pt,belowsep=0pt]|Q[l,m]|[dotted]Q[l,m,abovesep=3.7pt,belowsep=0pt]|[1pt]}}
        Continued from the previous page\\
        Coefficient $=\zp{L}-\zp{R}$ \\
        \hspace{-8pt} \includegraphics[width=0.98\textwidth]{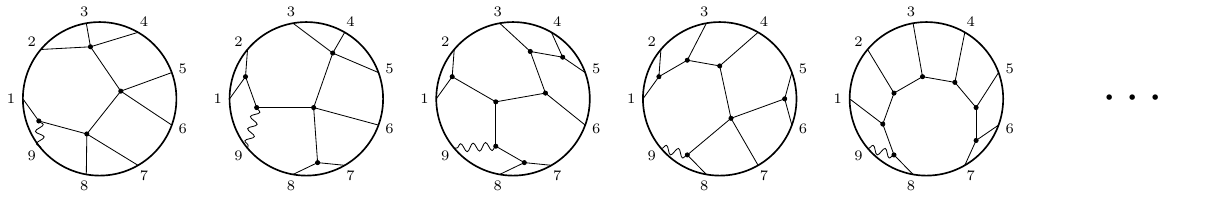}\!\! \\
        Coefficient $=3(\zp{L}-\zp{R})$ \\
        \hspace{-8pt} \includegraphics[width=0.98\textwidth]{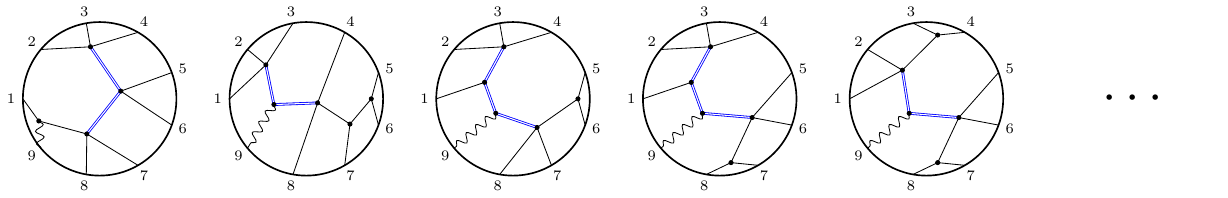}\!\! \\
        Coefficient $=9(\zp{L}-\zp{R})$ \\
        \hspace{-8pt} \includegraphics[width=0.98\textwidth]{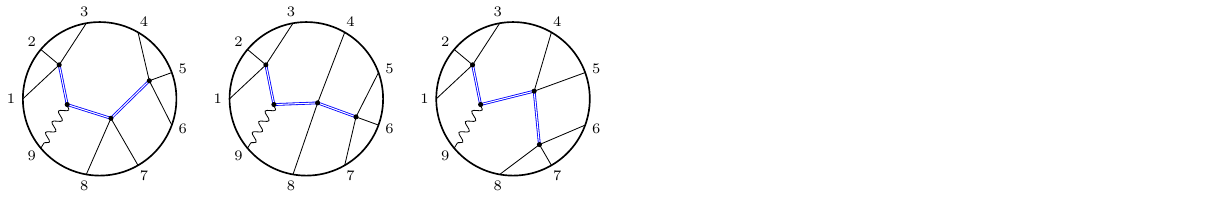}\!\! \\
        Coefficient $\propto\zp{R}$ \\
        $\cdots$ \\
    \end{tblr}
    \caption{Coefficients of Witten diagrams contributing to $M_{9}^{(v)}({\rm ST})/2^5$.}
    \label{tab:single-trace-gluon-9pt-2}
\end{table}


\section{Derivation of \texorpdfstring{$A^{\bar\mu}\phi\overleftrightarrow{\nabla}_{\bar\mu}\phi$}{} Feynman rules}\label{sec:rule_derivation}

\begin{figure}[ht]
    \centering
    \begin{subfigure}{0.45\textwidth}
        \centering
        \includegraphics[width=0.7\textwidth,align=c]{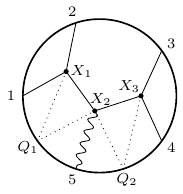}
        \caption{}
    \end{subfigure}
    \hspace{2em}
    \begin{subfigure}{0.45\linewidth}
        \centering
        \includegraphics[width=0.7\textwidth,align=c]{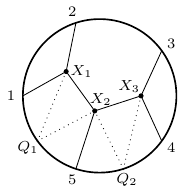}
        \caption{}
    \end{subfigure}
    \caption{(a) A Witten diagram to extract the $A^{\bar\mu}\phi\overleftrightarrow{\nabla}_{\bar\mu}\phi$ vertex. (b) A scalar Witten diagram to compare the result. The dotted line indicates the integration over $Q\in\partial{\rm AdS}$ using the split representation~\eqref{eq:split}.}
    \label{fig:rule_vs}
\end{figure}

In this appendix, we derive the tree-level Mellin space Feynman rule of an on-shell vector and two possibly off-shell scalars. To be completely rigorous, we should follow~\cite{Nandan:2011wc} and show that such a Feynman rule exists and that we can write down an expression for each vertex or edge of a Witten diagram, regardless of the other parts of the diagram. However, we decide to take a short-cut by assuming the existence of Feynman rules. This way, we need only compute a specific Witten diagram (Fig.~\ref{fig:rule_vs}(a)) and extract the corresponding rules.

Using the embedding space formalism, the scalar propagator reads
\begin{align}
    G^\Delta_{B\partial}(X,P)&=\frac1{2\pi^h\Gamma(1+\Delta-h)}\int_0^\infty\frac{{\rm d}t}{t}t^\Delta e^{-2tX\cdot P},\\
    G^\delta_{BB}(X_1,X_2)&=\int[{\rm d}c]f_\delta(c)\int_{\partial{\rm AdS}}{\rm d}Q\int_0^\infty\frac{{\rm d}s}{s}\frac{{\rm d}\bar s}{\bar s}s^{h+c}\bar s^{h-c}e^{-2Q\cdot(sX_1+\bar sX_2)},\label{eq:split}
\end{align}
where
\begin{equation}
    f_\delta(c)=\frac1{2\pi^{2h}}\frac1{((\delta-h)^2-c^2)\Gamma(c)\Gamma(-c)}.
\end{equation}
The gluon bulk-to-boundary propagator can be extracted from the scalar propagator using a differential operator:
\begin{gather}
    D_\Delta^{MA}=\frac{\Delta-1}{\Delta}\eta^{MA}+\frac1\Delta\frac\partial{\partial P_M}P^A,\\
    E_\Delta^{MA}(X,P)=D_\Delta^{MA}G_{B\partial}^\Delta(X,P).
\end{gather}
For the interaction vertex $A^{\bar\mu}\phi\overleftrightarrow\nabla_{\bar\mu}\phi$, the position space Feynman rules can be found in eq.(6.3) of~\cite{Paulos:2011ie}. From these building blocks, the above Witten diagram evaluates to (omitting overall constants)
\begin{align*}
    G^M&=D_{\Delta_5}^{MA}\int_{Q_1,Q_2,X_1,X_2,X_3}\frac{{\rm d}t_1}{t_1}t_1^{\Delta_1}\frac{{\rm d}t_2}{t_2}t_2^{\Delta_2}\frac{{\rm d}t_3}{t_3}t_3^{\Delta_3}\frac{{\rm d}t_4}{t_4}t_4^{\Delta_4}\frac{{\rm d}t_5}{t_5}t_5^{\Delta_5}[{\rm d}c_1][{\rm d}c_2]f_{\delta_1}(c_1)f_{\delta_2}(c_2)\\
    &\times\frac{{\rm d}s_1}{s_1}s_1^{h+c_1}\frac{{\rm d}\bar s_1}{\bar s_1}\bar s_1^{h-c_1}\frac{{\rm d}s_2}{s_2}s_2^{h+c_2}\frac{{\rm d}\bar s_2}{\bar s_2}\bar s_2^{h-c_2}(\bar s_1Q_{1,A}-\bar s_2Q_{2,A}) \\
    &\times e^{-2X_1\cdot(t_1P_1+t_2P_2+s_1Q_1)-2X_3\cdot(t_3P_3+t_4P_4+s_2Q_2)-2X_2\cdot(t_5P_5+\bar s_1Q_1+\bar s_2Q_2)}.
\end{align*}
Without loss of generality, consider the $-\bar s_2Q_{2,A}$ term. Integrating away $X_1,X_2,X_3\in{\rm AdS}$ and $Q_1,Q_2\in\partial{\rm AdS}$, we arrive at
\begin{align*}
    G_2^M&=D_{\Delta_5}^{MA}\int\frac{{\rm d}t_1}{t_1}t_1^{\Delta_1}\frac{{\rm d}t_2}{t_2}t_2^{\Delta_2}\frac{{\rm d}t_3}{t_3}t_3^{\Delta_3}\frac{{\rm d}t_4}{t_4}t_4^{\Delta_4}\frac{{\rm d}t_5}{t_5}t_5^{\Delta_5}[{\rm d}c_1][{\rm d}c_2]f_{\delta_1}(c_1)f_{\delta_2}(c_2)\\
    &\times\frac{{\rm d}s_1}{s_1}s_1^{h+c_1}\frac{{\rm d}\bar s_1}{\bar s_1}\bar s_1^{h-c_1}\frac{{\rm d}s_2}{s_2}s_2^{h+c_2}\frac{{\rm d}\bar s_2}{\bar s_2}\bar s_2^{h-c_2}\\
    &\times{\color{blue}-\bar s_2(s_1\bar s_1\bar s_2(t_1P_1+t_2P_2)+s_2(t_3P_3+t_4P_4)+\cancel{(1+\bar s_1^2)\bar s_2t_5P_5})_A}\\
    &\times e^{-t_1t_2P_{12}-t_3t_4P_{34}+(s_1(t_1P_1+t_2P_2)+\bar s_1t_5P_5)^2+(s_1\bar s_1\bar s_2(t_1P_1+t_2P_2)+s_2(t_3P_3+t_4P_4)+(1+\bar s_1^2)\bar s_2t_5P_5)^2}\\
    &\times\Gamma\left(\frac{\Delta_1+\Delta_2+c_1-h}2\right)\Gamma\left(\frac{\Delta_3+\Delta_4+c_2-h}2\right)\Gamma\left(\frac{\Delta_5+1-c_1-c_2}2\right),
\end{align*}
where we have used the notation $P_{ij}\equiv-2P_i\cdot P_j$. The term proportional to $P_{5A}$ can be dropped because it is annihilated by the differential operator in front. From the form of the exponential, we see that the blue factor can be written as a differential operator:
\begin{equation*}
    {\color{blue}\text{blue}}=\frac{\bar s_2^2}{1+\bar s_2^2(1+\bar s_1^2)}(P_{1,A}\partial_{P_{15}}+P_{2,A}\partial_{P_{25}})+\frac1{1+\bar s_1^2}(P_{3,A}\partial_{P_{35}}+P_{4,A}\partial_{P_{45}}).
\end{equation*}
Now, convert the $t$-integrals into Mellin representation using the Symanzik star formula.
\begin{align*}
    G_2^M&=D_{\Delta_5}^{MA}\int[{\rm d}c_1][{\rm d}c_2]f_{\delta_1}(c_1)f_{\delta_2}(c_2)[{\rm d}\gamma]\frac{{\rm d}s_1}{s_1}s_1^{h+c_1}\frac{{\rm d}\bar s_1}{\bar s_1}\bar s_1^{h-c_1}\frac{{\rm d}s_2}{s_2}s_2^{h+c_2}\frac{{\rm d}\bar s_2}{\bar s_2}\bar s_2^{h-c_2}(s_1\bar s_1)^{-s_{12}}(s_2\bar s_2)^{-s_{34}}\\
    &\times\left[\frac{\bar s_2^2}{1+\bar s_2^2(1+\bar s_1^2)}(P_{1,A}\partial_{P_{15}}+P_{2,A}\partial_{P_{25}})+\frac1{1+\bar s_1^2}(P_{3,A}\partial_{P_{35}}+P_{4,A}\partial_{P_{45}})\right]\prod_{i<j}[P_{ij}]^{\gamma_{ij}}\\
    &\times(1+\bar s_1^2)^{\frac{s_{12}-s_{34}-\tau_5}2}(1+\bar s_2^2(1+\bar s_1^2))^{\frac{s_{34}-s_{12}-\tau_5}2}(1+s_1^2(1+\bar s_1^2\bar s_2^2))^{\frac{s_{12}-\Delta_1-\Delta_2}2}(1+s_2^2)^{\frac{s_{34}-\Delta_3-\Delta_4}2}\\
    &\times\Gamma\left(\frac{\Delta_1+\Delta_2+c_1-h}2\right)\Gamma\left(\frac{\Delta_3+\Delta_4+c_2-h}2\right)\Gamma\left(\frac{\Delta_5+1-c_1-c_2}2\right).
\end{align*}
Here, $s_{ij}=\Delta_i+\Delta_j-2\gamma_{ij}$ and $\tau_5=\Delta_5-1$. Let us now focus on the part that depends on the external kinematics:
\begin{align*}
    &Z_{5M}D_{\Delta_5}^{MA}\left[\frac{\bar s_2^2}{1+\bar s_2^2(1+\bar s_1^2)}(P_{1,A}\partial_{P_{15}}+P_{2,A}\partial_{P_{25}})+\frac1{1+\bar s_1^2}(P_{3,A}\partial_{P_{35}}+P_{4,A}\partial_{P_{45}})\right]\prod_{i<j}[P_{ij}]^{\gamma_{ij}}\\
    &=\Bigg[\frac{\bar s_2^2}{1+\bar s_2^2(1+\bar s_1^2)}\left(-\frac{\tau_5}{\Delta_5}\texttt{zp[$12$]}+\frac{\gamma_{15}+\gamma_{25}}{\Delta_5}\texttt{zp[$1234$]}\right)+\frac1{1+\bar s_1^2}\left(-\frac{\tau_5}{\Delta_5}\texttt{zp[$34$]}+\frac{\gamma_{35}+\gamma_{45}}{\Delta_5}\texttt{zp[$1234$]}\right)\Bigg],
\end{align*}
where $\texttt{zp[$12$]}=\texttt{zp[$1$]}+\texttt{zp[$2$]}$ and so on. Recall that
\begin{equation*}
    \texttt{zp[$i$]}=\frac{(Z_5\cdot P_i)}{(-2P_5\cdot P_i)}\gamma_{5i}.
\end{equation*}
At this stage, we perform the same change of variables as~\cite{Nandan:2011wc}. First, rescale $s_1\mapsto s_1/\sqrt{1+\bar s_1^2\bar s_2^2}$. Then, define $\bar s_1\mapsto\sqrt x$ and $\bar s_2\mapsto\sqrt y$. Finally, redefine $y\mapsto\frac y{1+x}$ and then redefine $x\mapsto\frac x{1+y}$. In the end, we get
\begin{align*}
    Z_{5M}G_2^M&=\int[\rm d\gamma]\prod_{i<j}[P_{ij}]^{\gamma_{ij}}\int[{\rm d}c_1][{\rm d}c_2]f_{\delta_1}(c_1)f_{\delta_2}(c_2)\\
    &\times\Bigg[y\left(-\frac{\tau_5}{\Delta_5}\texttt{zp[$12$]}+\frac{\gamma_{15}+\gamma_{25}}{\Delta_5}\texttt{zp[$1234$]}\right)+(1+y)\left(-\frac{\tau_5}{\Delta_5}\texttt{zp[$34$]}+\frac{\gamma_{35}+\gamma_{45}}{\Delta_5}\texttt{zp[$1234$]}\right)\Bigg]\\
    &\times\frac{{\rm d}s_1}{s_1}s_1^{h+c_1-s_{12}}\frac{{\rm d}s_2}{s_2}s_2^{h+c_2-s_{34}}(1+s_1^2)^{\frac{s_{12}-\Delta_1-\Delta_2}2}(1+s_2^2)^{\frac{s_{34}-\Delta_3-\Delta_4}2}\\
    &\times\frac{{\rm d}x}{x}\frac{{\rm d}y}{y}x^{\frac{h-c_1-s_{12}}2}(1+x)^{\frac{s_{12}-h-c_1}2}y^{\frac{h-c_2-s_{34}}2}(1+y)^{\frac{s_{34}-h-c_2}2}(1+x+y)^{\frac{c_1+c_2-\tau_5-2}2}\\
    &\times\Gamma\left(\frac{\Delta_1+\Delta_2+c_1-h}2\right)\Gamma\left(\frac{\Delta_3+\Delta_4+c_2-h}2\right)\Gamma\left(\frac{\Delta_5+1-c_1-c_2}2\right).
\end{align*}
Together with the $+\bar s_1Q_{1,A}$ part that we ignored from the beginning, the bracket evaluates to the remarkably simple result
\begin{equation*}
    \frac1{\Delta_5}\texttt{zp[$12$]}(s_{34}-s_{12}+\tau_5)+\frac1{\Delta_5}\texttt{zp[$34$]}(s_{34}-s_{12}-\tau_5).
\end{equation*}
Meanwhile, we could repeat the entire calculation for the all-scalar Witten diagram (Fig.~\ref{fig:rule_vs}(b)). If we compare the results, it turns out that the other factors in the vector case precisely match the factors in the scalar case, except in place of $\Delta_5^{\rm scalar}$ we have $\Delta_5^{\rm vector}+1$. Therefore, if the vertex between an on-shell scalar with dimension $\Delta_s$ and two off-shell scalars contributes a factor
\begin{equation*}
    V_{\Delta_L,\Delta_R,\Delta_s}^{m_L,m_R,0},
\end{equation*}
the vertex between an on-shell vector with dimension $\Delta_v$ and two off-shell scalars contributes a factor (up to an overall constant)
\begin{equation*}
    V_{\Delta_L,\Delta_R,\Delta_v{\color{blue}+1}}^{m_L,m_R,0}\times\left[\texttt{zp[$L$]}\left(\frac{m_R-m_L}2+\frac{\Delta_v-1}4\right)-\texttt{zp[$R$]}\left(\frac{m_L-m_R}2+\frac{\Delta_v-1}4\right)\right],
\end{equation*}
where we have used the fact that the above Feynman rules are to be evaluated at the poles $s_{12}=2m_L$ and $s_{34}=2m_R$. Let us remark that, although various conventions enter into the derivation at different stages, the dependence on the descendant levels $m_L,m_R$ is independent of conventions and serves as an important check of the Feynman rules.

\newpage
\bibliographystyle{JHEP}
\bibliography{reference.bib}
\end{document}